%% file: accretion.tex
\documentclass[twocolumn,rmp,amsmath,amssymb]{revtex4-2}

\usepackage{graphicx}
\usepackage{dcolumn}
\usepackage{bm}
\usepackage{hyperref}
\hypersetup{breaklinks=true,colorlinks=true,urlcolor=blue, linkcolor=black,  citecolor=blue}

\newcommand{\mnras}{{\it MNRAS}}
\newcommand{\araa}{{\it ARA\&A}}
\newcommand{\aap}{{\it A\&A}}
\newcommand{\pasj}{{\it PASJ}}
\newcommand{\pasp}{{\it PASP}}
\newcommand{\actaa}{{\it Acta~Astron.}}
\newcommand{\planss}{{\it Planetary and Space Science}}
\newcommand{\apjl}{{\it Astrophys. J. Lett.}}
\newcommand{\nar}{{\it New Astron. Rev.}}
\newcommand{\aj}{{\it AJ}}

\begin{document}

\title{LECTURE NOTES ON ACCRETION DISK PHYSICS}

\author{Philip J. Armitage}
\affiliation{Stony Brook University \& Center for Computational Astrophysics, Flatiron Institute} 

\begin{abstract}
These notes introduce and review some of the physical principles underlying the theory of astrophysical accretion, emphasizing the central roles of angular momentum transport, angular momentum loss, and radiative cooling in determining the structure and evolution of accretion flows. Additional topics covered include the effective viscous theory of thin disks, classical instabilities of disk structure, the evolution of warped or eccentric disks, and the basic properties of waves within disks. 
\end{abstract}

\maketitle
\tableofcontents

\section{Introduction}
Accretion is central to astronomical systems  
as diverse as protoplanetary disks, X-ray binaries, Active Galactic Nuclei, Gamma-Ray Bursts, and Tidal Disruption Events. There are very many important differences between these systems, but what they have in common is that they all involve accretion, usually from a disk, on to a central gravitating object. Because the specific angular momentum in a Keplerian potential increases outward  as $h \propto \sqrt{r}$, the physics problem posed by accretion is to understand how gas in a rotating flow loses angular momentum and moves inward. This problem is generally considered to be solved in principle, though important aspects remain unclear. The astronomical problem is to figure out how the observed manifestations of accretion, in the form of resolved images, spectra and time variability, derive from the underlying physics. Many facets of this problem are certainly not solved.

Indispensable references for the student of accretion are Jim Pringle's 
review of viscous disk theory \citep{pringle81}, Steve Balbus and John Hawley's review of turbulence and angular momentum transport processes 
\citep{balbus98}, and the textbook by Juhan Frank, Andrew King and Derek 
Raine \citep{frank02}. Gordon Ogilvie's lecture notes\footnote{ \url{http://www.damtp.cam.ac.uk/user/gio10/accretion.html}} are extremely useful, and I follow his treatment of several classical problems here. My goal for these notes is to give an accessible introduction that follows a modern point of view, in which the key physical processes of angular momentum transport and radiative efficiency motivate older 
work on effective viscous disk theory. Also included are a smorgasbord of results that are hard to find outside the primary literature, extensive (but still incomplete) references, and some editorializing on my part.

\section{Angular momentum transport in disk accretion}
The angular velocity $\Omega$ of a particle in a circular orbit at distance $r$ from a point mass $M$ in Newtonian 
gravity is,
\begin{equation}
 \Omega = \Omega_{\rm K} = \sqrt{ \frac{GM}{r^3} },
\label{eq_omegaK} 
\end{equation}
where $G$ is the gravitational constant and the subscript ``K" identifies this as being characteristic of 
Keplerian orbits. The specific angular momentum (i.e. the angular momentum per unit mass) is an increasing 
function of orbital radius,
\begin{equation}
 h_{\rm K} = r^2 \Omega_{\rm K} = \sqrt{GMr}.
\end{equation} 
This elementary property of orbits in Newtonian gravity leads immediately to the central problem of accretion physics. 
Very frequently gas that is bound to stars and compact objects has specific angular momentum that 
exceeds that of a circular orbit grazing the object's surface, and notwithstanding the complications wrought 
by general relativity, pressure forces, and so forth, often it orbits in a flattened disk-like 
structure with $h \approx h_{\rm K}$. An individual parcel of gas must then lose angular momentum before it 
can move to an orbit at smaller $r$ and be accreted. There are two logical possibilities for how this 
can happen. One possibility is that the gas parcel {\em exchanges} angular 
momentum with another parcel, so that one moves in while the other moves out to conserve angular 
momentum. Invariably the viscosity of the gas is too small to lead to appreciable angular momentum 
exchange on the time scales inferred for astrophysical systems, so to realize this possibility we require 
that the fluid flow exhibits an instability that leads to turbulence and enhanced transport. The other 
possibility is that the disk is not an isolated system and can {\em lose} angular momentum as a whole. 
This can occur if the disk supports a magnetized outflow that exerts a torque back on the disk to remove angular momentum.

In this Section we will first show that for an important class of disks that are highly flattened, 
known as ``thin disks", the assumption that $h \approx h_{\rm K}$ is easily justified. We justify neglecting the regular fluid viscosity 
that results from particle-particle interactions, which is negligible in most cases of interest. We then consider 
the linear stability of magnetohydrodynamic (MHD) Keplerian shear flows, along with the case where the disk is non-magnetized but self-gravitating. Finally, we will derive a simple condition for when a 
thin disk, threaded by a large-scale magnetic field, can launch an MHD wind.

\subsection{Thin disk structure}
Consider a disk of gas that orbits in the $z=0$ plane in cylindrical polar co-ordinates $(r,\phi,z)$. We assume that the disk is accreting slowly enough that it is in approximate hydrostatic balance, and that the mass of the disk is negligible compared to the central mass $M$.

\begin{figure}
    \centering
    \includegraphics[width=\columnwidth]{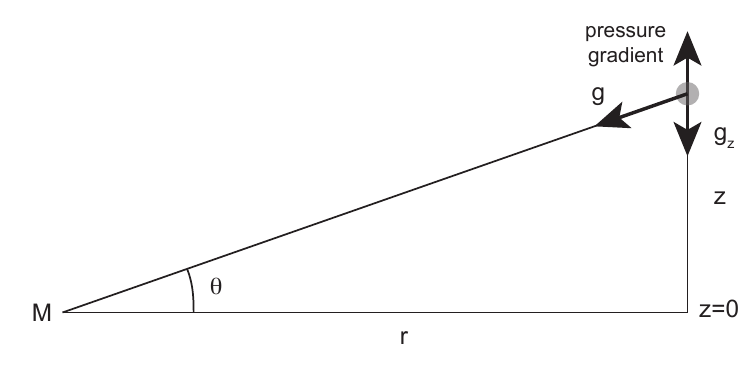}
    \caption{The vertical structure of a thin disk is determined by the hydrostatic balance between the vertical component of gravity from the central source and a pressure gradient.}
    \label{fig_vertical_structure}
\end{figure}

To work out the vertical structure of the disk, we look at the balance of forces in the vertical direction (shown in Figure~\ref{fig_vertical_structure}). The vertical component of gravity,
\begin{equation}
    g_z = g \sin(\theta) = \frac{GM}{ \left( r^2 + z^2 \right)^{3/2} } z,
\end{equation}
is balanced by a pressure gradient. The vertical structure is determined by the equation of vertical hydrostatic equilibrium,
\begin{equation}
    \frac{{\rm d}P}{{\rm d}z} = - \rho g_z.
\end{equation}
In different but reasonable physical circumstances the pressure might have important contributions from gas, radiation, and magnetic fields. The simplest case is when the pressure is dominated by gas pressure, and the vertical {\em temperature} structure is isothermal. This is roughly appropriate when the disk is optically thick and heated from outside. The equation of state is then,
\begin{equation}
    P = \rho c_s^2,
\label{eq_EOS}    
\end{equation}
where $c_s$, the sound speed, is not a function of $z$. If we further assume that $z \ll r$, the equation of hydrostatic equilibrium becomes,
\begin{equation}
    c_s^2 \frac{{\rm d} \rho}{{\rm d}z} = - \Omega_{\rm K}^2 \rho z.
\label{eq_vertical_hydrostatic_eq}    
\end{equation}
Solving this equation we find that,
\begin{equation}
    \rho(z) = \rho_0 \exp \left[ - \frac{\Omega_{\rm K}^2 z^2}{2 c_s^2} \right] = 
    \rho_0 \left[ - \frac{z^2}{2 h^2} \right].
\label{eq_gaussian_profile}    
\end{equation}
The density at the disk mid-plane (at $z=0$) is $\rho_0$, and the second equality serves to define the {\em disk scale height},
\begin{equation}
    h \equiv \frac{c_s}{\Omega_{\rm K}}.
\label{eq_thickness}    
\end{equation}
It is often convenient to work in terms of the surface density $\Sigma$, defined as,
\begin{equation}
    \Sigma \equiv \int_{-\infty}^{\infty} \rho(z) {\rm d}z.
\end{equation}
For the density profile given by equation~(\ref{eq_gaussian_profile}) the relation between the mid-plane and surface densities is,
\begin{equation}
    \rho_0 = \frac{1}{\sqrt{2 \pi}} \frac{\Sigma}{h}.
\label{eq_rho0}    
\end{equation}
The above analysis is justified if $h/r = c_s / v_{\rm K} \ll 1$, i.e. if the disk sound speed is much smaller than the Keplerian orbital velocity. Such disks are described as being geometrically thin.

A geometrically thin disk, by definition, has a mid-plane sound speed that is small compared to the Keplerian orbital velocity. In this limit, radial pressure gradients do not affect the angular velocity profile of the disk at leading order. The radial component of the momentum equation determines the azimuthal velocity $v_\phi$ of the gas,
\begin{equation}
    \frac{v_\phi^2}{r} = \frac{GM}{r^2} + \frac{1}{\rho} \frac{{\rm d}P}{{\rm d}r}.
\end{equation}
Approximating ${\rm d}P / {\rm d}r \sim P/r \sim \rho c_s^2 / r$ we get,
\begin{equation}
    v_\phi^2 = v_{\rm K}^2 \left[ 1 - {\cal O} \left( \frac{h}{r} \right)^2 \right].
\end{equation}
Deviations from Keplerian velocity are thus second order in $h/r$, and we can imagine a toy disk model in which the angular velocity is Keplerian, the mid-plane density and temperature are specified functions of radius, and the vertical density profile is gaussian. In the right circumstances---when the disk is geometrically thin, gas pressure dominated, and isothermal in the vertical direction---such a model will be a decent approximation.

On occasion, one needs a true two-dimensional solution for an axisymmetric disk in hydrostatic equilibrium. Several such solutions are known. \citet{lin90} and \citet{bate02}, for example, quote a solution for a disk with a polytropic equation of state near the mid-plane, and an isothermal atmosphere. This model can be useful as a background state when studying wave propagation in disks. \citet{fishbone76} provide a solution for a fluid torus around a black hole. Many numerical simulations of black hole accretion are initialized with this, or similar, analytic models.

\subsection{Microphysical viscosity in accretion disks}
To accrete, gas in a Keplerian disk configuration that satisfies hydrostatic and rotational equilibrium has to lose angular momentum. This can happen in two ways. Gas in the disk can {\em lose} angular momentum, for example when a magnetic field threading the disk exerts a torque on the disk surface. Alternatively or additionally, angular momentum can be {\em redistributed} within the disk, such that the inner part of the disk loses angular momentum and accretes while the outer part expands to conserve angular momentum.

A rotating fluid redistributes angular momentum due to (microscopic) viscosity, but this process is too slow to be astrophysically relevant in essentially all disks. For an ionized gas of cosmic composition the kinematic viscosity at temperature $T$ and density $\rho$ is \citep{spitzer62},
\begin{equation}
    \nu = 1.6 \times 10^{-15} \frac{T^{5/2}}{\rho \ln \Lambda} \ {\rm cm^2 \ s^{-1}}.
\end{equation}
Here, $\Lambda$ is the Coulomb logarithm for proton-proton scattering. (See \citet{balbus08} for a discussion and partial derivation of this result.) It is fiendishly hard to measure the density and temperature of most astrophysical disks precisely, but there are many observational and theoretical ways to get an estimate which is good enough for our purposes. For dwarf novae (accreting white dwarfs in mass transfer binary systems), for example, we can appeal to state-of-the-art numerical simulations by \citet{hirose14}. At a radius of $r \approx 10^{10} \ {\rm cm}$, they obtain characteristic temperatures $T \approx 3 \times 10^{4} \ {\rm K}$ and densities $\rho \approx 10^{-7} \ {\rm g \ cm^{-3}}$. The corresponding viscosity is $\nu \approx 2500 / \ln \Lambda \ {\rm cm^2 \ s^{-1}}$. Anticipating somewhat results from \S\ref{sec_time_scales} (or on dimensional grounds) we construct a time scale $\tau \sim r^2 / \nu$ assuming that the viscosity acts diffusively. With these numbers, $\tau \sim 10^{9} \ {\rm yr}$. Observationally, however, dwarf nova disks are seen to evolve dramatically on time scales of just days \citep[for an example with nice {\em Kepler} data, see][]{cannizzo12}. Pretty clearly viscosity due to small-scale kinetic processes in the gas is not responsible. Similar arguments apply to protoplanetary disks and disks in AGN.

\subsection{The shearing sheet}
The low level of microphysical viscosity in accretion disks motivates study of macroscopic instabilities that generate turbulence and angular momentum transport. The most important of these is the magnetorotational instability \citep[MRI;][]{balbus91}, which we will discuss in \S\ref{sec_MRI}. Our analysis will be based on the shearing sheet approximation \citep{goldreich65}, which is frequently used in both analytic and numerical studies of disks. It is a fluid cousin of {\em Hill's equations} \citep{hill78}, developed for the study of lunar motion.

To develop the shearing sheet we start with the inviscid momentum equation. In an inertial frame this is just,
\begin{equation}
  \frac{\partial {\bf v}}{\partial t} + {\bf v} \cdot \nabla {\bf v} =  
  - \frac{\nabla P}{\rho} - \nabla \Phi.
\end{equation}
The velocity is ${\bf v}$, $P$ is the pressure, $\rho$ is the density, and $\Phi$ is the gravitational potential. Equivalently, in terms of the Lagrangian or material derivative,
\begin{equation}
    \frac{D {\bf v}}{D t} = - \frac{\nabla P}{\rho} - \nabla \Phi.
\end{equation}
The right hand side expresses the forces acting on a fluid element, so transforming to a frame rotating at constant angular velocity $\Omega$ requires adding the Coriolis and centrifugal forces in the same way as for point mass dynamics,
\begin{equation} 
    \frac{D {\bf v}}{D t} = - \frac{\nabla P}{\rho} - \nabla \Phi - 2 {\bf \Omega} \times {\bf v} + r \Omega^2 {\hat {\bf r}}.
\end{equation}
Here $\hat {\bf r}$ is a unit vector in the radial direction. No approximation has yet been made.

\begin{figure*}[t]
    \centering
    \includegraphics[width=1.8\columnwidth]{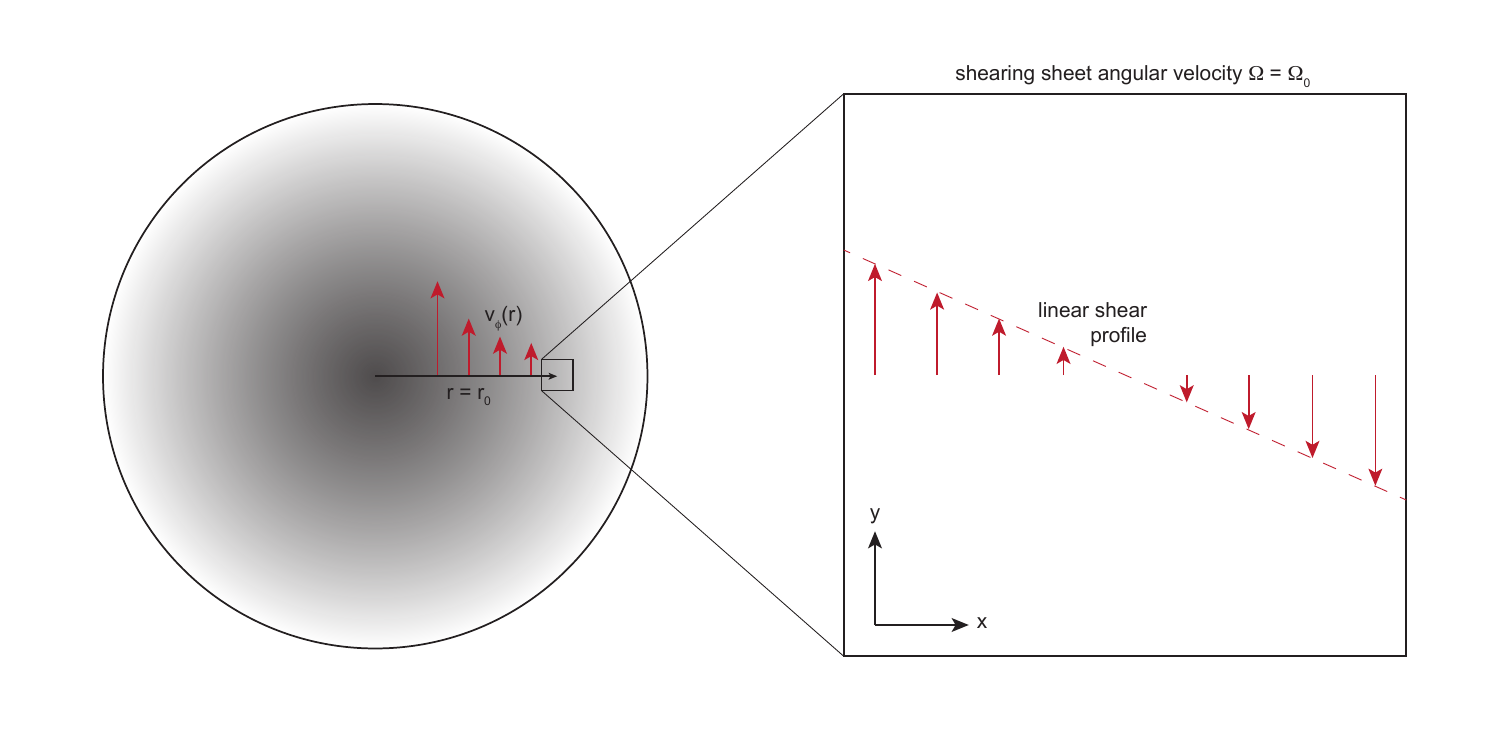}
    \vspace{-0.5cm}
    \caption{Geometry of the {\em local} or {\em shearing sheet} model for accretion disks. The shearing sheet is a Cartesian $(x,y)$ representation of a small patch of the disk, centered at $r=r_0$, that has an angular velocity $\Omega = \Omega_0$. The disk shear is approximated by a linear function across the sheet.}
    \label{fig_shearing_sheet}
\end{figure*}

Figure~\ref{fig_shearing_sheet} shows the geometry of the shearing sheet. The idea is to consider a local patch of the disk, centered at $r_0$, that co-rotates with the fluid at angular velocity $\Omega_0$. The dynamics is modeled in a Cartesian co-ordinate system, within which disk quantities are replaced by constants or via first-order Taylor expansion about $r_0$. Mathematically, assume that the angular velocity can be approximated as a power-law,
\begin{equation}
    \Omega (r) = \Omega_0 \left( \frac{r}{r_0} \right)^{-q}.
\end{equation}
$q=3/2$ corresponds to the Keplerian case (equation~\ref{eq_omegaK}). We define a co-rotating co-ordinate system,
\begin{eqnarray}
 x & = & r - r_0, \\
 y & = & r \phi - \Omega_0 t,
\end{eqnarray}
and sum up the radial component of the gravitational and centrifugal forces, keeping only the first order term in $x/r_0$,
\begin{eqnarray}
 -\nabla \Phi + r \Omega_0^2 {\hat {\bf r}} & = & - r \Omega_0^2 \left( \frac{r}{r_0} \right)^{-q} + r \Omega_0^2, \\
 & \approx & 2q \Omega_0^2 x.
\end{eqnarray}
Adding in the vertical gravitational acceleration, with a value appropriate to the center of the shearing sheet (using equation~\ref{eq_vertical_hydrostatic_eq}), the momentum equation in the shearing sheet approximation is,
\begin{equation}
    \frac{D {\bf v}}{D t} = - \frac{\nabla P}{\rho} - 2 {\bf \Omega}_0 \times {\bf v} + 2 q \Omega_0^2 x {\hat {\bf x}} - \Omega_0^2 z {\hat {\bf z}}.  
\label{eq_momentum_shearing_sheet}    
\end{equation}
$\hat{\bf x}$ and $\hat{\bf z}$ are unit vectors in the $x$ (radial) and $z$ (vertical) directions respectively. This three-dimensional version is called a shearing box.

\begin{figure*}[t]
    \centering
    \includegraphics[width=1.8\columnwidth]{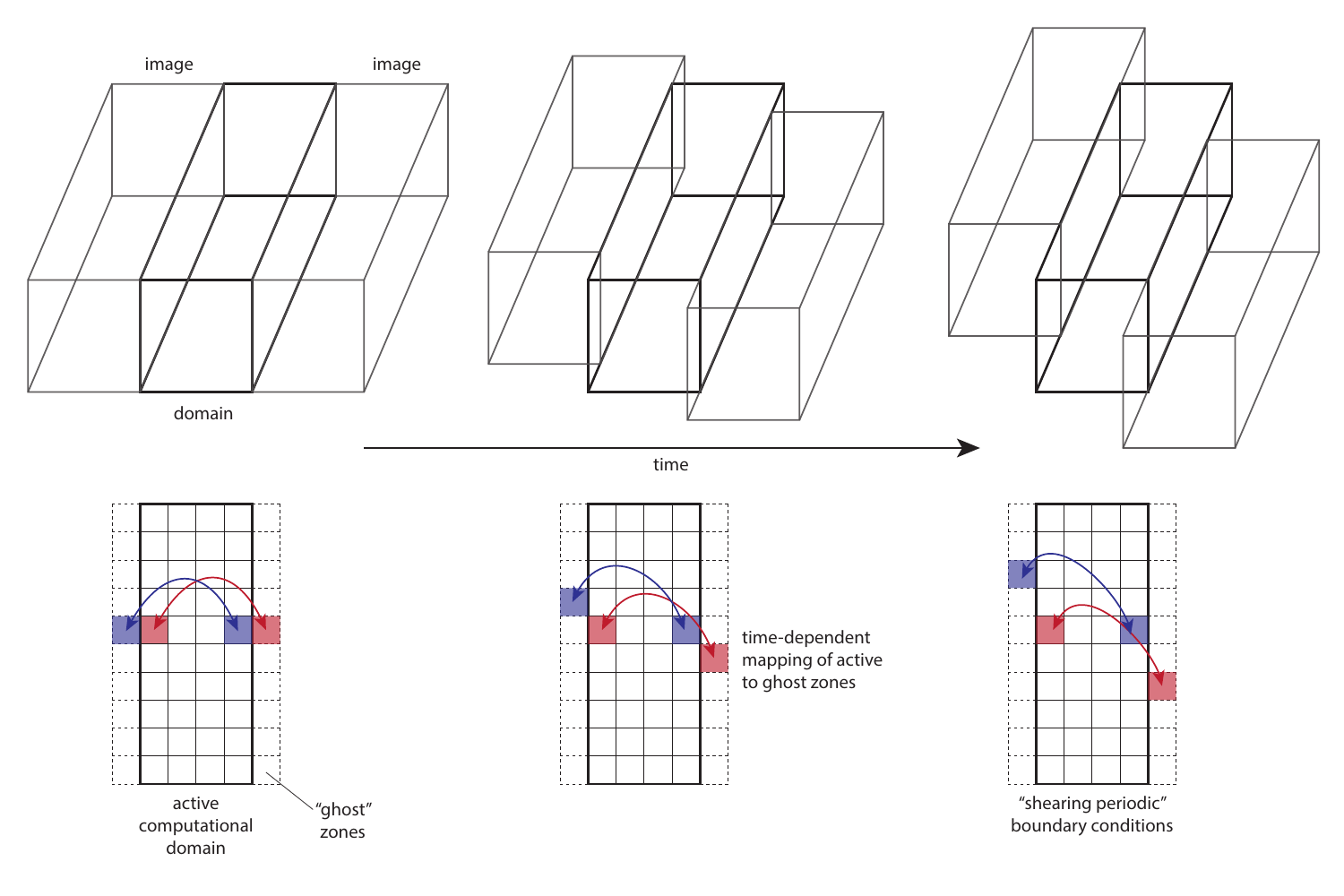}
    \vspace{-0.25cm}
    \caption{Illustration of how radial boundary conditions are imposed in computational studies that use the shearing box approximation \citep{hawley95}. Conceptually, we imagine that the simulation domain is bordered radially by identical copies of itself, which shear azimuthally over time. Practically, these boundary conditions are often implemented using time-dependent copying of the fluid quantities in the active zones into one or more layers of ``ghost" zones, interpolating as necessary when the required azimuthal shift is not an integer number of grid cells.}
    \label{fig_shearing_box}
\end{figure*}

In addition to analytic applications, the shearing box is a useful tool for local numerical simulations of accretion disks. The key point is that because the angular velocity is constant across the box, the dynamical time scale $\Omega^{-1}$ is not a function of the radial co-ordinate $x$. This makes it possible to construct self-consistent {\em shearing periodic} radial boundary conditions. The construction is illustrated in Figure~\ref{fig_shearing_box} \citep{hawley95}. The active computational domain is bordered radially by copies of itself, which shift azimuthally over time at a rate that reflects the shear across the domain. For a domain of radial extent $L_x$, the radial boundary conditions can be written as \citep{hawley95},
\begin{eqnarray}
 f(x,y,z) & = & f(x+L_x, y - q \Omega_0 L_x t, z), \\
 v_y(x,y,z) & = & v_y(x+L_x, y - q \Omega_0 L_x t, z) + q \Omega_0 L_x, 
\end{eqnarray}
where the first expression applies to all flow variables {\em apart} from the azimuthal velocity $v_y$, which requires a boost to account for the shear across the box.

The shearing box is a powerful tool. For analytic calculations, it captures much of the essential small-scale dynamics of disks, while being easier to deal with than global equations. For numerical work, on the other hand, equation~\ref{eq_momentum_shearing_sheet} is not much easier to solve than its global counterpart. The advantage for computation studies derives mainly from the ability to use shearing periodic radial boundary conditions, which are (relatively) non-coercive. If, instead, one attempts to simulate a small {\em cylindrical} patch of disk, it proves hard to devise boundary conditions which do not change the flow dynamics. The shearing box can be readily adapted to include magnetic fields (we will do so shortly), an energy equation, and radiation fields \citep{hirose09,jiang13}. It can also be extended to model disks whose background states are eccentric \citep{ogilvie14} or warped \citep{paris18}.

The shearing box approximation can also set traps for the unwary. At the most basic level, although we defined $x = r-r_0$, going to the shearing sheet introduces a symmetry between $+x$ and $-x$, such that there is no physical sense in which one direction is ``inward" and the other ``outward". This means, for example, that while an MRI simulation may develop sustained turbulence and measurable stresses, the gas does not actually move radially. More perniciously, this symmetry means that a shearing box threaded by a net vertical magnetic field can develop a wind solution in which the gas flows to $+x$ (say) above the mid-plane, and to $-x$ below the mid-plane. This sort of solution is unphysical in a global context, but there is nothing to prevent it developing within a shearing box \citep[e.g.][]{bai13}. Another difficulty comes from attempts to model radial gradients in disk properties, or a vertical gradient in $\Omega$. These gradients are physically important for various disk instabilities, but they are hard to incorporate consistently within the shearing box formalism \citep{mcnally15,latter17}.

\subsection{Magnetorotational instability}
\label{sec_MRI}
The magnetorotational instability (MRI) is the most important process that can generate turbulence and angular momentum transport in accretion disks \citep{balbus91,balbus98}. The MRI is a local, linear instability, that is present in accretion disks provided that,
\begin{itemize}
    \item[(i)] There is differential rotation, with ${\rm d}\Omega / {\rm d}r < 0$. Except in boundary layers, all disks satisfy this condition.
    \item[(ii)] ``Weak" magnetic fields are present. The MRI can be damped by microphysical viscosity for {\em very} weak fields \citep[too weak to be relevant to any disks apart perhaps from those around primordial stars;][]{krolik06}, and stabilized (though not necessarily entitrely shutdown) by magnetic tension effects in disks where the magnetic pressure is larger than the thermal pressure 
    \citep{pessah05,das18}. 
    \item[(iii)] The ionization fraction is high enough to couple the magnetic field to the fluid. This is the most important caveat. Although very low ionization fractions suffice, protoplanetary disks can be so neutral as to call into question the importance of the MRI \citep{gammie96}.
\end{itemize}
These prerequisites are weak, and most astrophysical disks satisfy them. Accordingly, we will skip over the tricky and rich question of whether {\em non-magnetic} non-linear\footnote{Keplerian flows are linearly stable by the Rayleigh criterion, as the specific angular momentum is an increasing function of radius.} fluid instabilities would exist in notional non-magnetic disks, and proceed directly to the MHD case. \citet{lyra19} review potential fluid instabilities in protoplanetary disk systems where the viability of the MRI as a transport mechanism remains doubtful.

\begin{figure*}
    \centering
    \includegraphics[width=1.8\columnwidth]{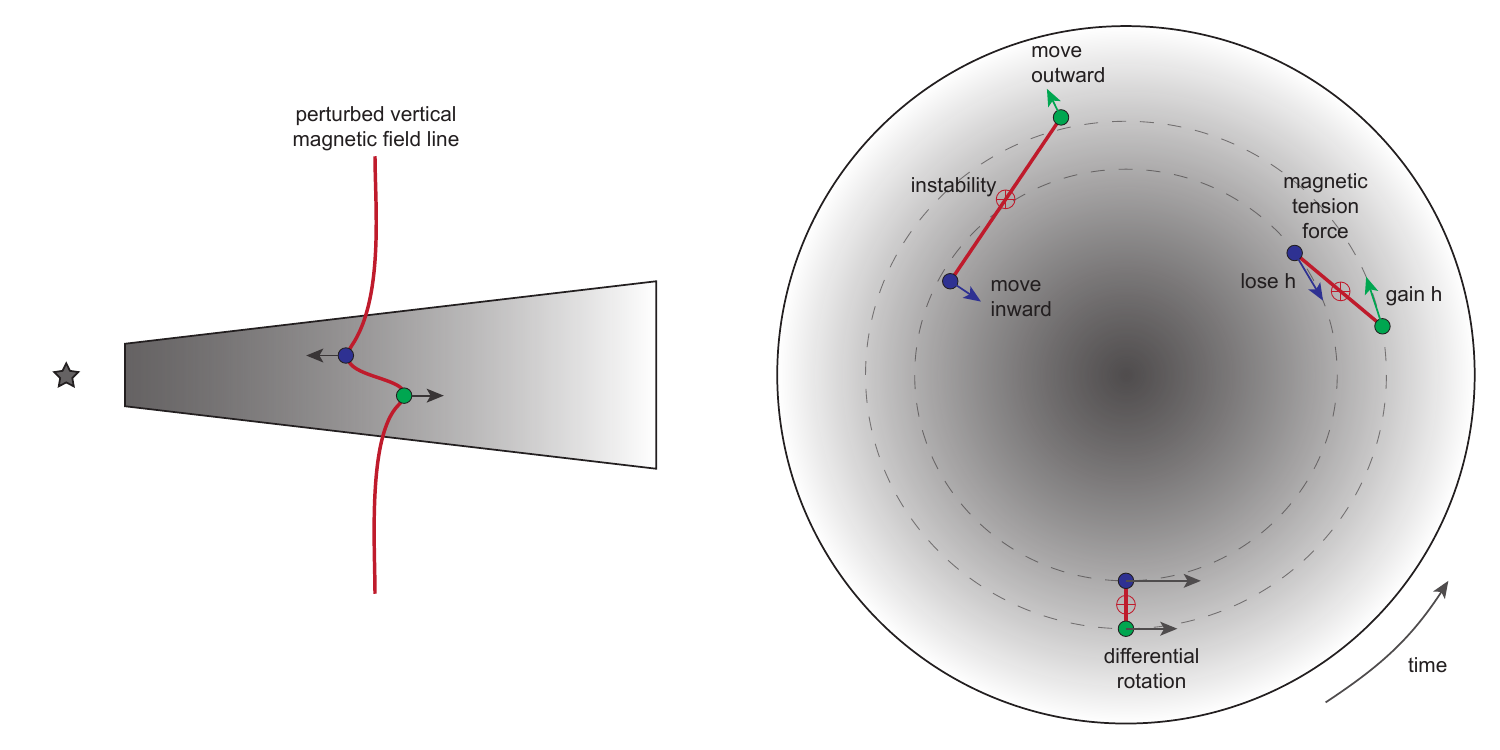}
    \caption{The operation of the magnetorotational instability (MRI) for a perturbed vertical magnetic field in ideal MHD. The MRI is a fluid instability, but the dynamics is closely related (and in some cases mathematically equivalent) to that of two orbiting point masses connected by a spring \citep{balbus92}.}
    \label{fig_MRI_cartoon}
\end{figure*}

The physical origin of the MRI is shown in Figure~\ref{fig_MRI_cartoon}, for the conceptually simplest case of a disk threaded by a weak, initially vertical magnetic field. Consider a radial perturbation to the magnetic field---exaggerated in the cartoon---that results in linkage between an inner (blue) fluid element and an outer (green) one. As the disk rotates, differential rotation causes the fluid elements, initially at the same azimuth, to become azimuthally separated. This separation is opposed by magnetic tension, which leads to a force that acts to {\em decrease} the angular momentum of the inner fluid element and {\em increase} that of the outer element. Because angular momentum is an increasing function of radius in Keplerian disks, this transfer of angular momentum causes the fluid elements to separate further radially, signalling an instability.

The existence of an instability in magnetized rotating flows was demonstrated by \citet{velikhov59} and by \citet{chandrasekhar61}, but the importance of this instability for accretion disks was not recognized. \citet{safronov72} came close, but no cigar. The MRI was rediscovered and applied to accretion disks in breakthrough work by \citet{balbus91}.

\subsubsection{The MRI dispersion relation}
The dispersion relation for the MRI can be derived in several distinct ways. Here, following \citet{fromang13}, we work through a rather explicit calculation of a differentially rotating disk containing a purely vertical magnetic field, whose stability we assess in the shearing sheet approximation. The disk has a power-law angular velocity profile, $\Omega \propto r^{-q}$, 
and a uniform vertical magnetic field $B_0$. Radial and vertical variations in density, and the vertical component of gravity, are ignored. The equation of state is isothermal, $P=\rho c_s^2$, with $c_s$ a constant. As with any linear stability analysis the goal is to find out whether infinitesimal perturbations to the equilibrium are stable, or whether instead they exhibit exponentially growth, 
implying instability.

The relevant equations are the continuity equation, the induction equation in the ideal magnetohydrodynamic (MHD) limit, and the momentum equation (including MHD forces) in the shearing box approximation,
\begin{eqnarray}
 \frac{\partial \rho}{\partial t} + \nabla \cdot (\rho {\bf v}) & = & 0, \\
 \frac{\partial {\bf v}}{\partial t} + ( {\bf v} \cdot \nabla ) {\bf v} & = & 
 -\frac{1}{\rho} \nabla P + 
 \frac{1}{4 \pi \rho} ( \nabla \times {\bf B} ) \times {\bf B} - \nonumber \\
 & & 2 {\Omega}_0 \times {\bf v} + 
 2q \Omega_0^2 x {\hat {\bf x}}, \\
 \frac{\partial {\bf B}}{\partial t} & = & \nabla \times ({\bf v} \times {\bf B}).
\label{eq_sheet_mhd} 
\end{eqnarray} 
The equilibrium state has uniform density, $\rho = \rho_0$, and uniform magnetic field 
${\bf B} = (0,0,B_0)$. In this setup there are no pressure or magnetic forces, and the initial velocity field is set by the balance of the Coriolis and centrifugal terms,
\begin{equation}
 2 \Omega_0 \times {\bf v} = 2q \Omega_0^2 x {\hat{\bf x}}.
\end{equation}
The initial velocity field is then,
\begin{equation}
 {\bf v} = \left( 0, -q \Omega_0 x, 0 \right).
\end{equation}
As expected given that we are working in the shearing sheet, there is a 
linear shear around radius $r_0$.  

To determine the stability of this system, we write the fluid variables as the sum of their equilibrium values plus perturbations. For the MRI, it suffices to consider  
a perturbation which depends only on $z$ and $t$. For the velocity, for example, 
we take,
\begin{eqnarray}
 v_x & = & v_x^\prime (z,t),  \\
 v_y & = & -q \Omega_0 x + v_y^\prime (z,t),  \\
 v_z & = & v_z^\prime (z,t),
\end{eqnarray}
and do similarly for the magnetic field and density. These expressions are substituted into the 
continuity, momentum and induction equations, discarding any terms that are quadratic in 
the primed quantities. This procedure leads to seven equations, but to derive the MRI we need only a subset of them, namely the $x$ and $y$ components of the momentum and induction equations. The four linearized equations are,
\begin{eqnarray}
 \frac{\partial v_x^\prime}{\partial t} & = & \frac{B_0}{4 \pi \rho_0} \frac{\partial B_x^\prime}{\partial z} + 
 2 \Omega_0 v_y^\prime,  \\
 \frac{\partial v_y^\prime}{\partial t} - q \Omega_0 v_x^\prime & = & \frac{B_0}{4 \pi \rho_0} \frac{\partial B_y^\prime}{\partial z} -
 2 \Omega_0 v_x^\prime,  \\ 
 \frac{\partial B_x^\prime}{\partial t} & = & B_0 \frac{\partial v_x^\prime}{\partial z},  \\
 \frac{\partial B_y^\prime}{\partial t} & = & B_0 \frac{\partial v_y^\prime}{\partial z} - q \Omega_0 B_x^\prime.
\end{eqnarray} 
These linearized differential equations are then converted into algebraic equations by taking the perturbations to have the form, 
\begin{equation}
 B_x^\prime = \bar{B}_x^\prime e^{i ( \omega t - k z)},
\end{equation} 
where $\omega$ is the frequency of a perturbation whose vertical wave-number is $k$. The time derivatives give us factors of $i \omega$, while the spatial derivatives give us $i k$. The four algebraic equations we end up with are,
\begin{eqnarray}
 i \omega v_x^\prime & = & -i k \frac{B_0 B_x^\prime}{4 \pi \rho_0} + 2 \Omega_0 v_y^\prime,  \\
 i \omega v_y^\prime & = & -i k \frac{B_0 B_y^\prime}{4 \pi \rho_0} +(q-2) \Omega_0 v_x^\prime,  \\ 
 i \omega B_x^\prime & = & - i k B_0 v_x^\prime,  \\
 i \omega B_y^\prime & = & - i k B_0 v_y^\prime -q \Omega_0 B_x^\prime .
\end{eqnarray}
(Here we've dropped bars on the variables.) The final step is to eliminate the perturbation variables between the equations, leaving us with the {\em dispersion relation} for the MRI. It takes the form,
\begin{eqnarray}
 \omega^4 - \omega^2 \left[ 2 k^2 v_A^2 + 2(2-q) \Omega_0^2 \right] + \nonumber \\ k^2 v_A^2 \left[ k^2 v_A^2 - 2q \Omega_0^2 \right] = 0,
\label{eq_mri_dispersion} 
\end{eqnarray}
where $v_A^2 = B_0^2 / (4 \pi \rho_0)$ is the Alfv\'en speed in the initial state.

\begin{figure}
    \centering
    \includegraphics[width=\columnwidth]{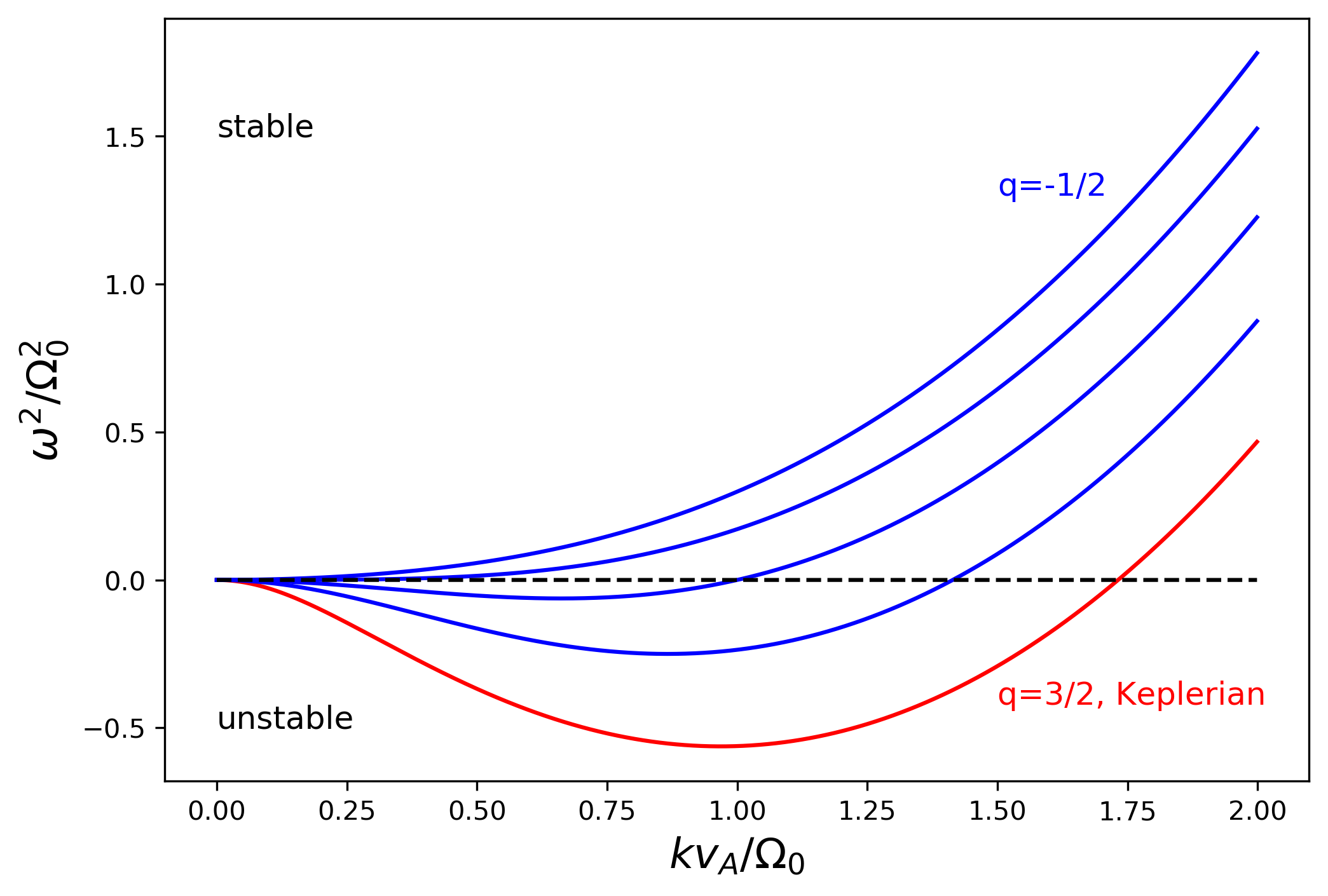}
    \caption{The MRI dispersion relation in ideal MHD is plotted for different rotation laws. The lower red curve shows the $q=3/2$ case appropriate for a Keplerian disk. The blue curves are plotted, from top downwards, for $q=-1/2$, $q=0$, $q=1/2$, and $q=1$. Instability ($\omega^2 / \Omega_0^2 < 0$) is present at sufficiently small $k$ for $q > 0$.}
    \label{fig_MRI_dispersion}
\end{figure}

The MRI dispersion relation is a quadratic in $\omega^2$. It is plotted as a function of $k v_A$, for different values of the shear parameter $q$, in Figure~\ref{fig_MRI_dispersion}. For the $q=3/2$ case that characterizes Keplerian disks $\omega^2 < 0$, implying instability, for all scales larger than some minimum (i.e. for $k v_A$ smaller than some maximum). Graphically, it can be seen that the instability disappears for $q \leq 0$. Requiring that $\omega^2 < 0$ the instability criterion is,
\begin{equation}
    k^2 v_A^2 + \frac{ {\rm d}\Omega^2 }{ {\rm d} \ln r} < 0.
\end{equation}
In the limit where the vertical field $B_0 \rightarrow 0$, the condition for instability is just that the angular velocity decreases outward. This is distinct from the Rayleigh stability criterion for a strictly unmagnetized fluid, which instead requires that the specific angular momentum $r^2 \Omega$ of the fluid decrease outward for an instability 
\citep[see e.g.][]{pringle07}.

A number of other important quantities can be straightforwardly derived from equation~(\ref{eq_mri_dispersion}). Solving for ${\rm d} \omega^2 / {\rm d} (kv_A) =0$ on the unstable branch of the dispersion relation yields the scale with the fastest MRI linear growth rate. For a Keplerian disk the answer is,
\begin{equation}
    (k v_A)_{\rm max} = \frac{\sqrt{15}}{4} \Omega_0.
\end{equation}
The growth rate at this spatial scale is,
\begin{equation}
    \vert \omega_{\rm max} \vert = \frac{3}{4} \Omega_0.
\end{equation}
This is a fast growth rate indeed! The most unstable MRI modes grow by a factor $\exp(3 \pi /2) \sim 10^2$ {\em per orbit}.

Returning to equation~(\ref{eq_mri_dispersion}) we can set $\omega^2 = 0$ and find $k_{\rm crit}$, the largest wavenumber (smallest spatial scale) which is unstable. Specializing again to the Keplerian case we find,
\begin{equation}
    k_{\rm crit} v_A = \sqrt{3} \Omega_0.
\end{equation}
This result can be used to approximately quantify our earlier assertion that the MRI can be stabilized if the vertical field is too strong. For a strong field, the smallest unstable scale $\lambda_{\rm crit}$ may exceed the vertical thickness of the disk, leading to a situation where the MRI is stabilized on account of the disk being too thin to support any unstable modes at all. For an estimate, we can set,
\begin{equation}
    \lambda_{\rm crit} = \frac{2 \pi}{k_{\rm crit}} = 2h, 
\end{equation}
which leads to an estimate of the stabilizing field as,
\begin{equation}
    B_{0,{\rm max}}^2 = \frac{12}{\pi} \rho c_s^2. 
\label{eq_B0_max}    
\end{equation}
Defining the plasma $\beta$ parameter to be the ratio of the gas pressure to the magnetic pressure,
\begin{equation}
    \beta = \frac{8 \pi P}{B_0^2},
\end{equation}
and making use of $h = c_s / \Omega_0$, we can re-express equation~(\ref{eq_B0_max}) as,
\begin{equation}
    \beta = \frac{2 \pi^2}{3}.
\end{equation}
Having ignored vertical density gradients---which by definition are significant on scales of the disk scale height $h$---this can only be an estimate. With that caveat, the conclusion is that disks are unstable to the MRI provided that the vertical magnetic field is significantly sub-thermal, in the sense of the magnetic pressure being smaller than the thermal pressure.

Somewhat surprisingly, relaxing the severe simplifications that we have made in the above analysis---a local shearing sheet, with a simple field geometry, simple perturbations, and no consideration of the flow energetics---does not materially change the answer. The review by \citet{balbus98}, and more recent work by \citet{latter15}, are good places to start for more complete analyses of the MRI.

\begin{figure}
    \centering
    \includegraphics[width=\columnwidth]{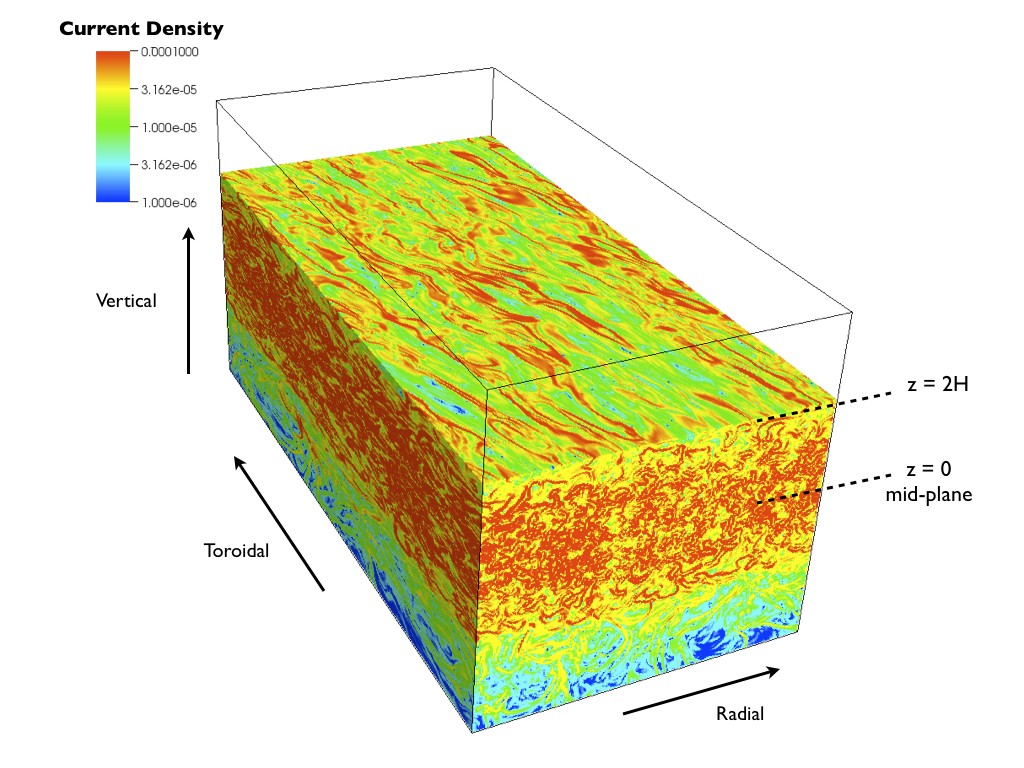}
    \caption{Rendering of the current density from a local shearing box simulation of MRI disk turbulence. Based on simulations from \citet{simon12}.}
    \label{fig_MRI_turbulence}
\end{figure}

The linear analysis says nothing about whether the MRI leads to physically significant levels of turbulence and angular momentum transport in accretion disks. For this, we must turn to simulations, in either shearing box \citep{hawley95,brandenburg95} or global \citep{armitage98} geometries. A local simulation of the MRI for a vertically stratified isothermal disk \citep{simon12} is shown in Figure~\ref{fig_MRI_turbulence}. The most basic simulation results, known now for more than twenty years, include the fact that the MRI can act as a {\em dynamo}, sustaining turbulence and magnetic fields against dissipation in a domain with no external currents. Angular momentum is transported primarily due to the action of MHD (Maxwell) stresses. Stronger levels of turbulence and transport occur if the disk is threaded by a net vertical magnetic field. There remain open questions, of uncertain physical significance, regarding the convergence of the MRI in local domains with simple physics \citep[at the least, very high resolution is needed;][]{ryan17}, but the bulk of recent work has turned to models with more complete treatments of thermal, radiative, relativistic, and plasma physics. \citet{jiang19}, for example, use radiation magnetohydrodynamic simulations to study the global evolution of disks around supermassive black holes.

\subsection{Self-gravity}
\label{sec_selfgravity}
In addition to the MRI, there are other linear instabilities of accretion disk flow. Most of these either occur under less general circumstances than the MRI, or have much lower growth rates, making them less important. Self-gravity is a partial exception. The dynamics of a fluid disk that is massive and / or cold is strongly modified by self-gravity if the Toomre $Q$ parameter,
\begin{equation}
    Q = \frac{c_s \kappa}{\pi G \Sigma} < 1.
\end{equation}
Here $\kappa$ is the epicyclic frequency, $\kappa^2 \equiv 4 \Omega^2 + 2 r \Omega {{\rm d}\Omega}/{{\rm d}r}$. A similar result \citep[and the eponymous work;][]{toomre64} applies to particle disks, for which the sound speed in the equation should be replaced by the velocity dispersion. Conditions where $Q \sim 1$ are likely to occur at sufficiently large radii in Active Galactic Nuclei (AGN) disks \citep{shlosman90}, and during star formation. Gravitational instability can in turn lead to angular momentum transport, or fragmentation \citep[for a review, see][]{kratter16}.

We can gain intuition into where $Q$ comes from using a simple time scale argument. 
Pressure will prevent the collapse of a patch of the disk, with scale $\Delta r$, when the sound-crossing time $\Delta r / c_s$ is 
small compared to the free-fall time $\sqrt{\Delta r^3 / G \Delta r^2 \Sigma}$. (Ignoring factors of the order of unity.) Equating these time scales 
the minimum scale of collapse can be estimated as $\Delta r \sim c_s^2 / G \Sigma$. 
On larger scales, collapse can be stopped by shear if the free-fall time is long compared to the 
time scale for radial shear to separate neighboring fluid elements. 
In a Keplerian disk the shear time scale is $\sim \Omega^{-1}$. For a disk that is marginally unstable the minimum scale set by pressure  
equals the maximum scale set by shear. This condition implies that marginal stability occurs when $c_s \Omega / G \Sigma \sim 1$, as quoted above.

\subsubsection{Dispersion relation}
Proceeding more formally \citep{pringle07} we consider a razor-thin circular gas disk with uniform surface density $\Sigma_0$ 
and sound speed $c_s$ in the $z=0$ plane. In cylindrical polar 
coordinates $(r,\phi,z)$ the density of the disk is given by,
\begin{equation} 
 \rho_0(r,\phi,z)=\Sigma_0 \delta (z),
\label{eq_C4_eq1} 
\end{equation}
where $\delta(z)$ is a Dirac delta-function. The velocity field is,
\begin{equation}
 {\bf v}_0 (r,\phi,z) = (0, r \Omega, 0).
\label{eq_C4_eq2} 
\end{equation}
The angular velocity $\Omega(r)$ is not required to be Keplerian, but for circular orbits the centrifugal force must balance gravity. If the gravitational potential is $\Phi_0$,
\begin{equation}
 \Omega^2 r = - \frac{ {\rm d}\Phi_0}{ {\rm d}r}.
\label{eq_C4_eq3}
\end{equation}
The assumptions of initially constant density and sound speed mean that pressure gradient forces do not enter into the calculation of the equilibrium state.

The stability analysis proceeds in a similar way as for the MRI, except that we now ignore magnetic fields while including both the gravity of the central object and the self-gravity of the disk itself. The equations are the continuity and momentum equations, together with Poisson's 
equation for the gravitational field,
\begin{eqnarray}
 \frac{\partial \Sigma}{\partial t} + \nabla \cdot (\Sigma {\bf v} ) & = & 0, \\
 \frac{\partial {\bf v}}{\partial t} + ( {\bf v} \cdot \nabla ) {\bf v} & = & 
 -\frac{\nabla p}{\Sigma} - \nabla \Phi, \\
 \nabla^2 \Phi & = & 4 \pi G \rho .
\end{eqnarray} 
We will also make use of a two dimensional sound speed, 
\begin{equation}
 c_{\rm s}^2 \equiv \frac{ {\rm d}p}{ {\rm d}\Sigma},
\label{eq_C4_eos} 
\end{equation}
defined in terms of the pressure $p$ and surface density $\Sigma$ in the usual way. 

We consider infinitesimal axisymmetric perturbations to the equilibrium 
state,
\begin{eqnarray}
 \Sigma &=& \Sigma_0 + \Sigma_1 (r,t), \\
 p &=& p_0 + p_1 (r,t), \\
 \Phi & = & \Phi_0 + \Phi_1 (r,t), \\
 {\bf v} &=& {\bf v}_0 + \left[ v_r (r,t), \delta v_\phi (r,t), 0 \right]\!, 
\end{eqnarray} 
that have a spatial and temporal dependence given by (using the surface 
density as an example),
\begin{equation}
 \Sigma_1 (r,t) \propto \exp [ {\rm i}(k r - \omega t) ].
\end{equation}
Here $k$ is the spatial wavenumber of the perturbation (related to the 
wavelength via $\lambda = 2 \pi / k$) and $\omega$ is the temporal 
frequency. Making one further approximation, we assume that for the perturbations of interest, 
\begin{equation}
 kr \gg 1.
\end{equation}
This amounts to considering disturbances that are small compared to the 
radial extent of the disk.  

We now substitute the expressions for the surface density, pressure, gravitational 
potential and velocity into the fluid equations, discarding any terms we 
encounter that are quadratic in the perturbed quantities. For the continuity equation this yields, 
\begin{equation}
 -\!{\rm i} \omega \Sigma_1 + v_r \Sigma_0 \left( \frac{1}{r} + {\rm i} k \right) = 0, 
\end{equation}
which simplifies further in the local limit ($kr \gg 1$) to
\begin{equation}
 - \!\omega \Sigma_1 + k v_r \Sigma_0 = 0. 
\label{eq_C4_elim1} 
\end{equation} 
Deriving the analogous algebraic equations from the momentum equation requires 
us to express the convective operator $( {\bf v} \cdot \nabla ) {\bf v}$ in 
cylindrical coordinates. This takes the form,
\begin{eqnarray}
&&\hspace*{-14pt} ( {\bf v} \cdot \nabla ) {\bf v} = 
 \left[ v_r \frac{\partial v_r}{\partial r} + \frac{v_\phi}{r} \frac{\partial v_r}{\partial \phi}
  + v_z \frac{\partial v_r}{\partial z} - \frac{v_\phi^2}{r}, \right. \nonumber \\
 &&\hspace*{-14pt}\quad v_r \frac{\partial v_\phi}{\partial r} + \frac{v_\phi}{r} \frac{\partial v_\phi}{\partial \phi} + 
 v_z \frac{\partial v_\phi}{\partial z} + \frac{v_r v_\phi}{r}, \nonumber \\
  &&\hspace*{-14pt}\quad v_r \frac{\partial v_z}{\partial r} + \frac{v_\phi}{r} \frac{\partial v_z}{\partial \phi} + 
 v_z \frac{\partial v_z}{\partial z} \Bigg] .
\label{eq_convective_cylindrical} 
\end{eqnarray} 
With this in hand, the momentum equation reduces to, 
\begin{eqnarray}
&\displaystyle -{\rm i} \omega v_r - 2 \Omega \delta v_\phi = - \frac{1}{\Sigma_0} 
 \frac{{\rm d} p_1}{{\rm d} r} - \frac{{\rm d} \Phi_1}{{\rm d} r}
\label{eq_C4_momr},& \\
& \displaystyle -\,{\rm i} \omega \delta v_\phi + v_r \left[ \Omega + \frac{\rm d}{{\rm d} r} 
 \left( r \Omega \right) \right] = 0,&
\label{eq_C4_elim2} 
\end{eqnarray}
where the two equations come from the radial and azimuthal components respectively. 

The next step is to relate the perturbations in pressure and gravitational potential 
expressed on the right-hand-side of equation~(\ref{eq_C4_momr}) to perturbations in the 
surface density. For the pressure term this is straightforward. Equation~(\ref{eq_C4_eos}) 
implies that,
\begin{equation}
 \frac{1}{\Sigma_0} \frac{{\rm d}p_1}{{\rm d}r} = 
 \frac{1}{\Sigma_0} c_{\rm s}^2 {\rm i} k \Sigma_1.
\end{equation}
Dealing with the potential perturbations requires more work. Starting from the 
linearized Poisson equation,
\begin{equation}
 \nabla^2 \Phi_1 = 4 \pi G \Sigma_1 \delta (z),
\end{equation}
we write out the Laplacian explicitly and simplify making use of the fact that 
for short wavelength perturbations $k r \gg 1$. This yields a relation between 
the density and potential fluctuations,
\begin{equation}
 \frac{{\rm d}^2 \Phi_1}{{\rm d}z^2} = k^2 \Phi_1 + 4 \pi G \Sigma_1 \delta (z).
\end{equation}
For $z \neq 0$ the only solution to this equation that remains finite for 
large $|z|$ has the form,
\begin{equation}
 \Phi_1 = C \exp [ - \vert k z \vert ],
\end{equation}
where $C$ remains to be determined. To do so we integrate the Poisson equation 
vertically between $z = - \epsilon$ and $z = + \epsilon$,
\begin{equation}
 \int_{-\epsilon}^{+\epsilon} \nabla^2 \Phi_1 {\rm d}z = 
 \int_{-\epsilon}^{+\epsilon} 4 \pi G \Sigma_1 \delta (z) {\rm d}z.
\end{equation}
Noting that both $\partial^2 \Phi_1 / \partial x^2$ and $\partial^2 \Phi_1 / \partial y^2$ 
are continuous at $z=0$, whereas $\partial^2 \Phi_1 / \partial z^2$ is not, we 
obtain, 
\begin{equation}
 \left. \frac{{\rm d}\Phi_1}{{\rm d}z} \right|_{-\epsilon}^{+\epsilon} = 4 \pi G \Sigma_1.
\end{equation}
Taking the limit $\epsilon \rightarrow 0$ we find that $C = -2 \pi G \Sigma_1 / |k|$, 
and hence that the general relation between potential and surface density fluctuations on the 
$z=0$ plane is,
\begin{equation}
 \Phi_1 = - \frac{2 \pi G \Sigma_1}{|k|}.
\end{equation}  
Taking the radial derivative,
\begin{equation} 
 \frac{{\rm d}\Phi_1}{{\rm d}r} = - \frac{2 \pi {\rm i} k G \Sigma_1}{|k|},
\end{equation}
which allows us to eliminate the potential from the right-hand-side of 
equation~(\ref{eq_C4_momr}) in favor of the surface density. The result is,
\begin{equation}
 -\!{\rm i} \omega v_r -2 \Omega \delta v_\phi = 
 -\frac{1}{\Sigma_0} c_{\rm s}^2 {\rm i} k \Sigma_1 + \frac{2 \pi {\rm i} k G \Sigma_1}{|k|}.
\label{eq_C4_elim3} 
\end{equation}
Finally, we are ready to derive the functional relationship between $\omega$ 
and $k$ (the dispersion relation). Eliminating $v_r$ and $\delta v_\phi$ 
between equations~(\ref{eq_C4_elim1}), (\ref{eq_C4_elim2}), and (\ref{eq_C4_elim3}) 
we find that,
\begin{equation}
 \omega^2 = \kappa^2 +c_{\rm s}^2 k^2 - 2 \pi G \Sigma_0 |k|,
\label{eq_C4_dispersionrelation} 
\end{equation} 
where the {\em epicyclic frequency} $\kappa$ is defined as,
\begin{equation}
 \kappa^2 \equiv 4 \Omega^2 + 2 r \Omega \frac{{\rm d}\Omega}{{\rm d}r}.
\end{equation}
In a Keplerian potential $\kappa^2 = \Omega^2$.  

\subsubsection{Toomre $Q$}

\begin{figure}
    \centering
    \includegraphics[width=\columnwidth]{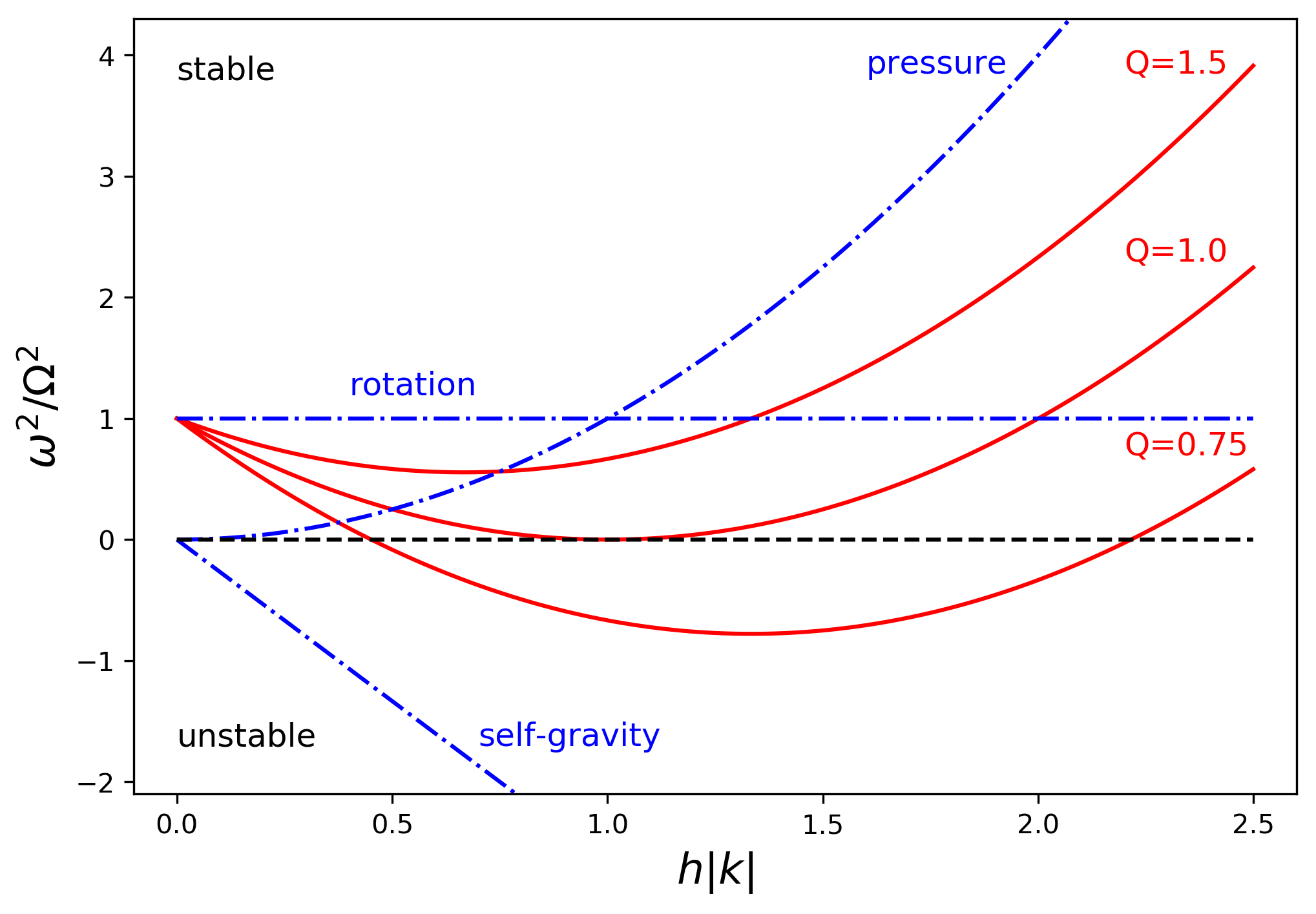}
    \caption{The dispersion relation for a self-gravitating accretion disk in the Keplerian limit is plotted for different values of the Toomre $Q$ parameter, $Q=0.75$, $Q=1$ and $Q=1.5$. Instability is present at some scales for $Q < 1$. The blue dot-dashed lines show the contributions from pressure (dominant at large $|k|$), rotation (dominant as $|k| \rightarrow 0$, and self-gravity (destabilizing).}
    \label{fig_SG_dispersion}
\end{figure}

Let us consider, for simplicity, a Keplerian disk. Noting that $h = c_s / \Omega$, we can write the dispersion relation in the form,
\begin{equation}
    \frac{\omega^2}{\Omega^2} = 1 - \frac{2 h |k|}{Q} + h^2 k^2,
\end{equation}
where $Q=c_s \Omega / (\pi G \Sigma)$ is the dimensionless Toomre $Q$ parameter that we previously deduced using a time scale argument. It is plotted in Figure~\ref{fig_SG_dispersion} for different values of $Q$. Each of the terms on the right-hand-side has a simple physical interpretation. The constant term describes the effect of rotation, which stabilizes all scales but which is dominant at small $h|k|$ (i.e. at large spatial scales). The quadratic term describes the effect of pressure, which has the opposite tendency and preferentially stabilizes small spatial scales. The linear term, which describes self-gravity, is negative and thus destabilizing. The strength of self-gravity is fully parameterized by the value $Q$, with small enough $Q$ leading to $\omega^2 < 0$ and unstable modes. It is easy to verify that the criterion for instability corresponds to $Q < 1$. 

\subsubsection{Angular momentum transport or fragmentation}
The onset of self-gravity in an accretion disk can lead to angular momentum transport or to fragmentation. Loosely speaking, fragmentation is the outcome for disks that either cool too quickly \citep{gammie01,rice05}, or that are fed mass from infall at too high a rate \citep{kratter10}. \citet{kratter16} discuss the quantitative thresholds, determined from simulations, for these outcomes.

\subsection{MHD disk winds}
Several physical processes can lead to mass loss in a disk wind from geometrically thin accretion disks, including,
\begin{itemize}
    \item A vertical gradient of thermal pressure. A sufficiently strong thermal pressure can develop, even in a thin disk, if high energy radiation heats the disk's surfaces to a temperature where $c_s \sim v_{\rm K}$. The hot gas then has positive total energy and can escape to infinity. Thermal or {\em photoevaporative} winds may be important in AGN \citep{begelman83} and in protoplanetary disks \citep{bally82}, where they contribute to the dispersal of the disk at the end of the protoplanetary disk phase \citep{alexander14,ercolano17}.
    \item Radiation pressure acting on spectral lines. The physics of line-driven disk winds is an extension of the accepted theory for mass loss from massive stars \citep{castor75}. For accretion disks, line-driving is efficient for AGN disks that are strong emitters of ultraviolet radiation \citep{proga00}. (Radiation absorbed by the continuum can also be important, but typically only under conditions where the disk is geometrically thick.)
    \item MHD acceleration, which in general involves contributions from {\em centrifugal} acceleration \citep{blandford82} along poloidal magnetic field lines (i.e. in the $(r,\theta)$ plane of spherical polar co-ordinates), and from a gradient of toroidal magnetic pressure \citep{lynden-bell96}.
\end{itemize}
In addition to these processes, which in principle can drive a wind from a broad range of disk radii, there are others whose applicability is limited to the innermost disk region. Disk magnetic fields interacting with a spinning black hole can extract spin energy via the Blandford-Znajek effect \citep{blandford77}, while for accreting objects with a material surface the interaction between the inner disk and a magnetosphere can generate outflow \citep{shu94}. Inner disk processes are implicated in the formation of well-collimated jets in accreting systems.

Of these mass loss processes, magneto-centrifugal disk winds are particularly interesting because they can extract angular momentum as well as mass from the underlying disk flow. The basic idea is that gas, accelerating outwards along a poloidal field line, rotates with the angular velocity of the gas at the field line's foot point in the disk. The departing gas thus gains angular momentum, which is removed from the disk by a magnetic torque that can be considered to act on the disk's surface. The angular momentum loss leads to inflow of gas in the disk, independent of  internal angular momentum transport processes such as the MRI or self-gravity.

The clearest description of the physics of MHD disk winds that I'm aware of can be found in \citet{spruit96}. Here, we content ourselves with a derivation of one of the basic properties of a \citet{blandford82} disk wind---the existence of a critical inclination angle for a poloidal disk-threading field to launch a cold wind. 

\subsubsection{Blandford-Payne disk winds}

\begin{figure*}
    \centering
    \includegraphics[width=1.8\columnwidth]{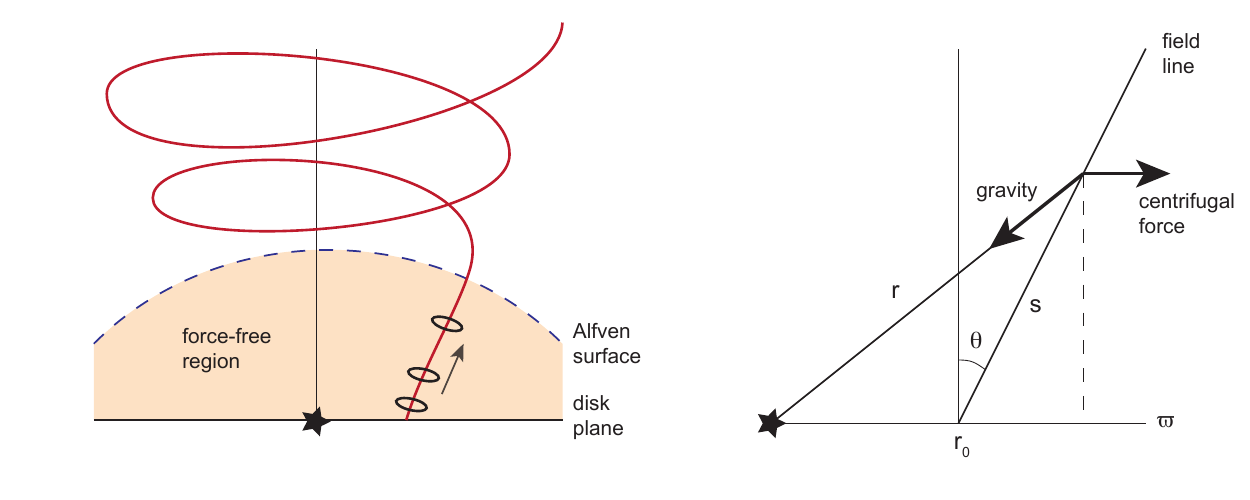}
    \vspace{-0.25cm}
    \caption{Illustration \citep[adapted from the pedagogical review by][]{spruit96} of a magnetized disk wind driven by centrifugal force. Left panel: the region above the disk surface is force-free. If magnetic field lines threading the disk are inclined by a large enough angle with respect to the vertical, centrifugal force can accelerate gas along the field lines even in the absence of a pressure gradient. The dynamics is equivalent to a mechanical system of a bead (the gas) on a rigid wire (the magnetic field line) that rotates with the angular velocity of the disk at the foot point. Right panel: the geometry for calculating the critical inclination angle for cold magneto-centrifugal wind launching.}
    \label{fig_BP_geometry}
\end{figure*}

The geometry of a Blandford-Payne wind \citep{blandford82} wind is illustrated in Figure~\ref{fig_BP_geometry}. We envisage a 
Keplerian disk threaded by a large-scale poloidal magnetic field, in the limit of ideal MHD. 
Within the disk the energy density in the magnetic field, $B^2 / 8 \pi$, 
is usually smaller than $\rho c_s^2$, the thermal energy. Due to flux conservation, 
however, the energy in the vertical field component, $B_z^2 / 8 \pi$, is roughly 
constant with height for $z < r$, while the gas pressure decreases rapidly (for a thin isothermal disk, as a gaussian with a scale height $h \ll r$). This leads to a region above the 
disk surface where magnetic forces dominate. The magnetic force per unit 
volume can be written as the sum of a magnetic pressure 
gradient and a force due to magnetic tension, 
\begin{equation}
 \frac{ {\bf J} \times {\bf B} }{c} = - \nabla \left( \frac{B^2}{8 \pi} \right) + 
 \frac{ {\bf B} \cdot \nabla {\bf B} }{4 \pi},
\end{equation} 
where the current,
\begin{equation}
 {\bf J} = \frac{c}{4 \pi} \nabla \times {\bf B}.
\end{equation} 
In the disk wind region where magnetic 
forces dominate, the requirement that they exert a finite acceleration on the low density 
gas can only be satisfied if the force approximately vanishes, i.e. that,
\begin{equation}
 {\bf J} \times {\bf B} \approx 0.
\end{equation}
The structure of the magnetic field in the magnetically dominated region is then described as 
being ``force-free", and in the disk wind case (where $B$ changes slowly with $z$) the field 
lines must be approximately straight to ensure that the magnetic tension term is  
small. If the field lines support a wind, the force-free structure persists up to where the 
kinetic energy density in the wind, $\rho v^2$, first exceeds the magnetic energy density. This 
criterion defines the {\em Alfv\'en surface}. Beyond the Alfv\'en surface, the inertia of the gas in the wind 
is sufficient to bend the field lines, which wrap up into a spiral structure as the disk 
below them rotates.

Magneto-centrifugal driving can launch a wind from the surface of a cold gas disk if the 
magnetic field lines are sufficiently inclined to the disk normal. The critical inclination angle 
in ideal MHD can be derived via an exact mechanical analogy. To proceed, we 
note that in the force-free region the magnetic field lines are (i) basically straight lines, and 
(ii) enforce rigid rotation out to the Alfv\'en surface at an angular velocity equal to that 
of the disk at the field line's footpoint. The geometry is shown in Figure~\ref{fig_BP_geometry}. We consider a field line that intersects the disk 
at radius $r_0$, where the angular velocity is $\Omega_0 = \sqrt{GM/r_0^3}$, and that 
makes an angle $\theta$ to the disk normal. We define the spherical polar radius $r$, 
the cylindrical polar radius $\varpi$, and measure the distance along the field line from its intersection with the disk at $z=0$ as $s$. In the frame co-rotating with $\Omega_0$ 
there are no magnetic forces along the field line to affect the acceleration of a wind; 
the sole role of the magnetic field is to constrain the gas to move along a straight line 
at constant angular velocity. Following this line of argument, the condition for the acceleration of the 
wind can be described in terms of an effective potential,
\begin{equation}
 \Phi_{\rm eff} (s) = -\frac{GM}{r(s)} -\frac{1}{2} \Omega_0^2 \varpi^2 (s), 
\end{equation}
that is the sum of the gravitational potential and the 
centrifugal potential in the rotating frame. 

\begin{figure}
    \centering
    \includegraphics[width=\columnwidth]{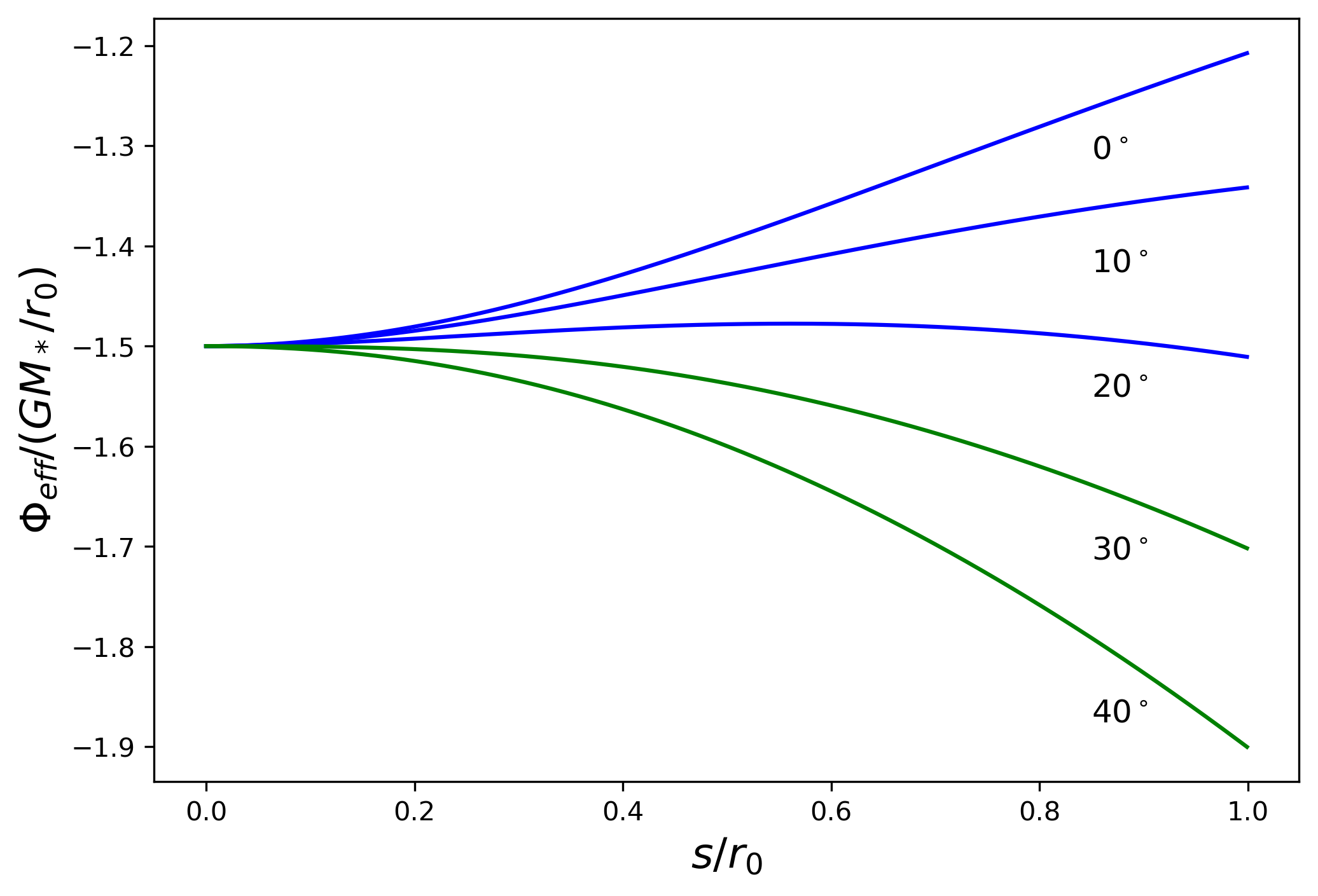}
    \caption{The effective potential for a cold magnetocentrifugally driven disk wind. Acceleration occurs for field lines that are inclined by $30^\circ$ or more away from the vertical.}
    \label{fig_BP_potential}
\end{figure}

Written out explicitly the effective potential is,
\begin{eqnarray}
 \Phi_{\rm eff} (s) = -\frac{GM}{ (s^2 + 2 s r_0 \sin \theta + r_0^2 )^{1/2}} \nonumber \\
 - \frac{1}{2} \Omega_0^2 \left( r_0 + s \sin \theta \right)^2.
\label{eq_eff_potential} 
\end{eqnarray} 
This function is plotted in Figure~\ref{fig_BP_potential} for various values of the angle $\theta$. 
If we consider first a vertical field line ($\theta = 0$) the effective potential is a monotonically increasing function 
of distance $s$. For modest values of $\theta$ there is a potential barrier defined by a maximum 
at some $s = s_{\rm max}$, while for large enough $\theta$ the potential decreases monotonically from $s=0$. 
In this last case purely magneto-centrifugal forces suffice to accelerate a wind off the disk 
surface, even in the absence of any thermal effects. We compute the critical inclination angle  $\theta_{\rm crit}$, 
defined as the minimum angle that allows magneto-centrifugal wind driving, via the condition,
\begin{equation}
 \left. \frac{\partial^2 \Phi_{\rm eff}}{\partial s^2} \right\vert_{s=0} = 0.
\end{equation} 
Evaluating this condition, we find,
\begin{eqnarray} 
 1 - 4 \sin^2 \theta_{\rm crit} & = & 0 \nonumber \\
 \Rightarrow \theta_{\rm crit} & = & 30^\circ,
\end{eqnarray}
as the minimum inclination angle from the vertical needed for unimpeded wind launching 
in ideal MHD. 

The rigid rotation of the field lines interior to the Alfv\'en surface means that gas being 
accelerated along them increases its specific angular momentum. The magnetic field, 
in turn, applies a torque to the disk that removes a corresponding amount of angular 
momentum. If a field line, anchored to the disk at radius $r_0$, crosses the Alfv\'en 
surface at (cylindrical) radius $r_A$, it follows that the angular momentum flux is,
\begin{equation}
 \dot{L}_w = \dot{M}_w \Omega_0 r_A^2,
\end{equation}
where $\dot{M}_w$ is the mass loss rate in the wind. Removing angular momentum 
at this rate from the disk results in a local accretion rate $\dot{M} = 2 \dot{L}_w / \Omega_0 r_0^2$. 
The ratio of the disk accretion rate to the wind loss rate is,
\begin{equation}
 \frac{\dot{M}}{\dot{M}_w} = 2 \left( \frac{r_A}{r_0} \right)^2.
\end{equation} 
If $r_A$ substantially exceeds $r_0$ (by a factor of a few, which is reasonable for detailed
disk wind solutions) a relatively weak wind can carry away enough angular momentum 
to support a much larger accretion rate.

\section{Effective viscous theory of accretion disks}
The MRI, self-gravity, and disk winds are physical processes that lead to angular momentum transport and evolution of accretion disks. In many cases, it is not possible to simulate these processes with sufficient fidelity, or over enough radial range and for long enough, to compare against observations. It can therefore be necessary to turn to viscous disk theory, historically developed much earlier, which can be used to model long-term disk evolution \citep{shakura73,lynden-bell74}.

The basic assumption of an effective viscous disk theory is that angular momentum transport within the disk can be represented approximately as a {\bf fluid viscosity}, using some parameterization that does not explicitly involve {\bf B}. The continuity and momentum equations are then,
\begin{eqnarray}
  \frac{\partial \rho}{\partial t} + \nabla \cdot \left( \rho {\bf v} \right) & = & 0, \\
  \frac{\partial {\bf v}}{\partial t} + {\bf v} \cdot \nabla {\bf v} & = & 
  - \frac{\nabla P}{\rho} - \nabla \Phi + \frac{1}{\rho} \nabla \cdot {\bf T}.
\label{eq_fluid_equations}  
\end{eqnarray}
Here $\rho$ is density, ${\bf v}$ is velocity, $P$ is pressure, $\Phi$ is gravitational potential, 
and ${\bf T}$ is stress (represented by a tensor). We further assume that we are dealing with a geometrically thin disk for which the pressure gradient term is small. We can then make progress using simplified versions of the equations for mass and momentum conservation.

\subsection{One dimensional time-dependent disk evolution}
The fluid equations (\ref{eq_fluid_equations}) apply generally. We first specialize to the case of a geometrically thin, circular, and planar disk, and derive a one-dimensional (in radius $r$) evolution equation. We then look for time-independent and time-dependent solutions.

\subsubsection{1D evolution equation}
In cylindrical polar co-ordinates $(r,\phi,z)$ the continuity equation is,
\begin{equation}
    \frac{\partial \rho}{\partial t} + 
      \frac{1}{r} \frac{\partial}{\partial r} \left( r \rho v_r \right) + 
      \frac{1}{r} \frac{\partial}{\partial \phi} \left( \rho v_\phi \right) + 
      \frac{\partial}{\partial z} \left( \rho v_z \right) = 0.
\end{equation}
Integrating over $\phi$ $[0, 2 \pi]$ and over $z$ $[-\infty, \infty]$ the first term becomes,
\begin{equation}
    \frac{\partial}{\partial t} \int_0^{2 \pi} \int_{-\infty}^{\infty} \rho dz d\phi = 
    \frac{\partial}{\partial t} \left( 2 \pi \Sigma \right),
\end{equation}
where $\Sigma (r,t)$ is the azimuthally averaged surface density. We are working toward an evolution 
equation for this quantity. Integrating the second term,
\begin{equation}
    \frac{1}{r} \frac{\partial}{\partial r} \int_0^{2 \pi} \int_{-\infty}^{\infty} r \rho v_r dz d \phi= 
    \frac{1}{r} \frac{\partial {\cal{F}}}{\partial r},
\end{equation}
where $\cal{F}$, the radial mass flux,
\begin{equation}
    {\cal{F}} = \int_0^{2 \pi} \int_{-\infty}^{\infty} r \rho v_r dz d \phi = 2 \pi r \Sigma \bar{v}_r,
\label{def_F}    
\end{equation}
can be written in terms of the surface density and the density-weighted radial velocity $\bar{v}_r$. On integration, the third and fourth terms vanish (the latter assuming that there is no mass loss from the surfaces of the disk) leaving,
\begin{equation} 
 \frac{\partial \Sigma}{\partial t} + \frac{1}{r} \frac{\partial}{\partial r} \left( r \Sigma \bar{v}_r \right) = 0,
\label{eq_disk_continuity} 
\end{equation}
as the one-dimensional version of the continuity equation.

Dealing with the momentum equation in the same way, we can simplify the algebra by assuming at the outset that the disk is axisymmetric such that the orbital velocity,
\begin{equation}
    v_\phi = r \Omega,
\end{equation}
depends only on the angular velocity $\Omega = \Omega(r)$ of circular orbits in a fixed gravitational potential $\Phi$. The only non-zero terms in the $\phi$ component of the momentum equation then come from ${\bf v} \cdot \nabla {\bf v}$ and $\nabla {\bf \cdot T}$. Looking up the forms for these in cylindrical polar co-ordinates the surviving terms are,
\begin{eqnarray}
 \left. {\bf v} \cdot {\nabla {\bf v}} \right|_\phi & = & v_r \frac{\rm d v_\phi}{\rm d r} + \frac{v_\phi v_r}{r}, \\
 \left. \frac{1}{\rho} \nabla \cdot {\bf T} \right|_\phi & = & \frac{1}{r^2 \rho} \frac{\partial}{\partial r} \left( r^2 T_{r \phi} \right) + \frac{1}{\rho} \frac{\partial T_{\phi z}}{\partial z}.
\end{eqnarray}
Multiplying these expressions through by $r \rho$ the azimuthal component of the momentum equation takes the form,
\begin{equation}
    \rho v_r \frac{{\rm d}h}{{\rm d}r} = \frac{1}{r} \frac{\partial}{\partial r} \left( r^2 T_{r \phi} \right) + r \frac{\partial T_{\phi z}}{\partial z},
\label{eq_aximuthal_mom_partial}    
\end{equation}
where the specific angular momentum $h$ is defined as,
\begin{equation}
    h \equiv r^2 \Omega.
\end{equation}
Finally we multiply equation~(\ref{eq_aximuthal_mom_partial}) through again by $r$, and integrate over $\phi$ and $z$. If $T_{\phi z}$ vanishes as $z \rightarrow \pm \infty$ the result is,
\begin{equation}
    {\cal{F}} \frac{{\rm d}h}{{\rm d}r} = - \frac{\partial {\cal G}}{\partial r},
\label{eq_FhG}    
\end{equation}
where ${\cal F}$ is given by equation~(\ref{def_F}) and,
\begin{equation}
    {\cal{G}} = - \int_0^{2 \pi} \int_{-\infty}^{\infty} r^2 T_{r \phi} dz d\phi,
\end{equation}
is the viscous torque.

Getting to this point from the conservation laws expressed in equation~(\ref{eq_fluid_equations}) requires only some rather transparent assumptions: axisymmetry, a time-independent potential, and no mass or angular momentum loss from the disk surfaces. One could stop there, and consider $T_{r \phi}$ to be the key quantity whose dependence on disk conditions needs to be determined. Conventionally, however, we instead write the torque in terms of an effective fluid viscosity that follows a Navier-Stokes form. For a fluid with viscosity $\mu$ and bulk viscosity $\mu_{\rm b}$ we have,
\begin{equation}
    {\bf T} = \mu \left[ \nabla {\bf v} + \left( \nabla {\bf v} \right)^{T} \right] + 
    \left( \mu_{\rm b} - \frac{2}{3} \mu \right) \left( \nabla \cdot {\bf v} \right) {\bf I}.
\end{equation}
To leading order the divergence of a thin disk velocity field vanishes, so we don't have to worry about bulk viscosity at all. The $r \phi$ component of the stress is,
\begin{equation}
    T_{r \phi} = \mu r \frac{{\rm d}\Omega}{{\rm d}r}.
\label{eq_Trphi}    
\end{equation}
Defining the kinematic viscosity (later just ``the viscosity") $\nu$ as,
\begin{equation}
    \nu = \frac{1}{2 \pi \Sigma} \int_0^{2 \pi} \int_{-\infty}^{\infty} \mu dz d \phi,
\end{equation}
the viscous torque has a fairly intuitive form that is the product of the circumference, the viscous force per unit length, and the lever arm,
\begin{equation}
    {\cal G} = - 2 \pi r \cdot \nu \Sigma r \frac{{\rm d}\Omega}{{\rm d}r} \cdot r.
\end{equation}
Equation~(\ref{eq_FhG}) is then,
\begin{equation}
    \Sigma \bar{v}_r \frac{{\rm d}h}{{\rm d}r} = \frac{1}{r} \frac{\rm d}{{\rm d} r} \left( \nu \Sigma r^3 \frac{{\rm d}\Omega}{{\rm d}r} \right).
\label{eq_disk_momentum}    
\end{equation}
Given the aforementioned assumptions, this equation expresses angular momentum conservation for a viscous fluid in a disk geometry.

Eliminating the radial velocity $\bar{v}_r$ between equation~(\ref{eq_disk_continuity}) and equation~(\ref{eq_disk_momentum}) we obtain,
\begin{equation}
    \frac{\partial \Sigma}{\partial t} = -\frac{1}{r} 
    \frac{\partial}{\partial r} \left[
    \left( \frac{{\rm d}h}{{\rm d}r} \right)^{-1} 
    \frac{\partial}{\partial r} \left( 
    \nu \Sigma r^3 \frac{{\rm d}\Omega}{{\rm d}r} \right) \right].
\end{equation}
This form is valid for an arbitrary (fixed) profile of angular velocity and angular momentum in the disk. Very often we are interested in the case of a disk that orbits a Newtonian point mass $M$. In that limit,
\begin{eqnarray}
 \Omega &=& \Omega_{\rm K} = \sqrt{GM/r^3}, \\
 h &=& \sqrt{GMr}.
\end{eqnarray}
The radial velocity is given by equation~(\ref{eq_disk_momentum}) as, 
\begin{equation}
    \bar{v}_r = - \frac{3}{\Sigma r^{1/2}} 
      \frac{\rm d}{{\rm d}r} \left(  \nu \Sigma r^{1/2} \right),
\end{equation}
and the evolution equation has the form,
\begin{equation}
    \frac{\partial \Sigma}{\partial t} = \frac{3}{r}
    \frac{\partial}{\partial r} \left[ 
    r^{1/2} \frac{\partial}{\partial r} \left( \nu \Sigma r^{1/2} \right) \right].
\label{eq_1D_disk_evolution}    
\end{equation}
The surface density thus evolves according to a diffusive partial differential equation, whose precise character depends upon the nature of the viscosity. The equation is linear if $\nu \neq f(\Sigma)$, though there is no general reason for this to be the case.
 
\subsubsection{Steady solutions}
Steady solutions to equation~(\ref{eq_1D_disk_evolution}) are easily derived. Setting $\partial \Sigma / \partial t = 0$ and integrating we find that,
\begin{equation}
    \nu \Sigma = c_1 + c_2 r^{-1/2}.
\end{equation}
Determining the constants of integration takes a little more work, and the right answer depends on the physics of the disk being modeled. It's easiest to start from equation~(\ref{eq_disk_momentum}). Noting that the accretion rate $\dot{M}$ is,
\begin{equation}
    \dot{M} = -2 \pi r \Sigma \bar{v}_r,
\end{equation}
and that $\dot{M}$ must be constant for a steady solution, we integrate equation~(\ref{eq_disk_momentum}). The result is,
\begin{equation}
    -\frac{\dot{M}}{2 \pi} h = \nu \Sigma r^3 \frac{{\rm d}\Omega}{{\rm d}r} + {\rm const}.
\label{eq_disk_mom_integrated}    
\end{equation}
For Keplerian (point mass) forms for $h$ and $\Omega$, the constant term is negligible at large $r$. We have, as $r \rightarrow \infty$,
\begin{equation}
    \nu \Sigma \simeq \frac{\dot{M}}{3 \pi}.
\end{equation}
The surface density of the disk is inversely proportional to the viscosity.

To get at the second constant of integration, we note that the constant appearing in equation~(\ref{eq_disk_mom_integrated}) is proportional to an angular momentum flux $\dot{M} h$. To obtain the standard form of the steady disk solution we assume that at some radius $r = \tilde{r}$ the first term on the right-hand-side of equation~(\ref{eq_disk_mom_integrated}), which is proportional to the viscous torque ${\cal G}$, vanishes. If $h$ and $\Omega$ are given by Keplerian expressions, we then find,
\begin{equation}
    \nu \Sigma = \frac{\dot{M}}{3 \pi} \left[ 1 - \sqrt{\frac{\tilde{r}}{r}} \right].
\label{eq_disk_nuSigma}    
\end{equation}
This is the solution for a steady-state disk subject to a zero-torque boundary condition at $r=\tilde{r}$. Classically, this boundary condition can be physically justified for a disk around a slowly rotating, non-magnetized star, with $\tilde{r} \simeq r_*$, the stellar radius \citep{pringle77}, and for a disk around a black hole, in which case 
$\tilde{r}$ can be identified with the innermost stable circular orbit \citep{bardeen70,shakura73,page74}. However, in neither situation is the justification watertight \citep[for the black hole case see, e.g.;][]{gammie99,agol00}.

The heating rate per unit volume in the disk is given by,
\begin{equation}
    q_+ = T_{r \phi} r \frac{{\rm d}\Omega}{{\rm d}r} = \mu \left( r \frac{{\rm d}\Omega}{{\rm d}r} \right)^2 = \frac{9}{4} \mu \Omega_{\rm K}^2,
\label{eq_qplus}    
\end{equation}
where the last equality applies only for a point-mass Keplerian potential. Integrating over $z$, the heating rate per unit surface area in the disk plane is,
\begin{equation}
    Q_+ = \int_{-\infty}^{\infty} \frac{9}{4} \mu \Omega_{\rm K}^2 dz = \frac{9}{4} \nu \Sigma \Omega_{\rm K}^2.
\label{eq_qplus2}    
\end{equation}
This heat may result in an increase in the temperature of the disk, and it may be transported radially by the disk flow. If these effects are negligible and the energy is radiated locally, the disk effective temperature is,
\begin{equation}
    2 \sigma T_{\rm eff}^4 = \frac{9}{4} \nu \Sigma \Omega_{\rm K}^2,
\end{equation}
where $\sigma$ is the Stefan-Boltzmann constant and the factor of two comes from the fact that the disk radiates from both its upper and lower surfaces. This equation does not require that the disk be in a steady state. If the disk {\em is} in a steady state, however, with the profile given by equation~(\ref{eq_disk_nuSigma}), the temperature distribution is,
\begin{equation}
    T_{\rm eff}^4 = \frac{3 G M \dot{M}}{8 \pi \sigma r^3} \left[ 1 - \sqrt{\frac{\tilde{r}}{r}} \right].
\label{eq_disk_Teff_r}    
\end{equation}
The steady state temperature profile does not depend on the viscosity.

The temperature profile for a steady viscous disk is not what you get from a fully local toy model in which available gravitational potential energy is lost as radiation at every radius. To see this, suppose that mass $\Delta m$ at radius $r$ moves to $r - \Delta r$ while remaining on a near circular orbit. Half of the liberated potential energy goes into increased kinetic energy, so the energy available to heat up the gas is $GM \Delta m \Delta r / 2 r^2$. If the time scale for mass to move inward is $\Delta t$, and the heat is radiated uniformly from the annulus with total surface area $4 \pi r \Delta r$, the expected temperature profile would be $T_{\rm eff}^4 = G M \dot{M} / (8 \pi \sigma r^3)$. At large $r$, this differs by a factor of three from the disk profile given by equation~(\ref{eq_disk_Teff_r}). Viscous torques cause a significant radial redistribution of energy in accretion disks.

\subsubsection{Irradiated disks}

\begin{figure}
    \centering
    \includegraphics[width=\columnwidth]{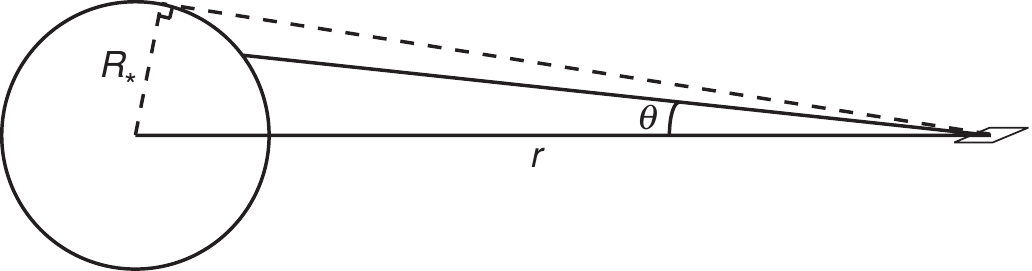}
    \caption{Geometry for computing the radial temperature profile of a disk primarily heated by {\em irradiation} from a central source, rather than by internal dissipation.}
    \label{fig_irradiation}
\end{figure}

The temperature profile given by equation~(\ref{eq_disk_Teff_r}) applies if the dominant source of disk heating is internal dissipation. It is also possible for the dominant source to be external irradiation, for example if the accreting object is a star. The temperature profile in this limit depends upon the {\em shape} of the disk, which determines the fraction of stellar photons that are absorbed at each radius. The simplest case is a flat disk in the equatorial plane, that absorbs all the incident stellar photons and re-emits the energy locally as a single temperature blackbody. 

To compute the resulting $T_{\rm eff} (r)$, consider a surface in the plane of the disk at distance $r$ from a star of radius $R_*$. The star is assumed to be a sphere of constant brightness $I_*$. Setting up spherical polar coordinates such that the axis of the coordinate system points to the center of the star, as shown in Figure~\ref{fig_irradiation}, the stellar flux passing through this surface is
\begin{equation} 
 F = \int I_* \sin \theta \cos \phi {\rm d} \Omega,
\label{eq_C2_fluxintegral} 
\end{equation}
where ${\rm d} \Omega$ represents the element of solid angle. We count the flux coming from the top half of the star only (and equate that to radiation from only the top surface of the disk), so the limits on the integral are,
\begin{eqnarray}
 -\pi / 2 < & \phi & \leq \pi / 2 \nonumber \\
 0 < & \theta & < \sin^{-1} \left( \frac{R_*}{r} \right).
\end{eqnarray} 
Substituting ${\rm d}\Omega = \sin \theta {\rm d}\theta {\rm d}\phi$, the integral for the flux is,
\begin{equation}
 F = I_* \int_{-\pi/2}^{\pi/2} \cos \phi {\rm d}\phi 
 \int_0^{\sin^{-1}(R_*/r)} \sin^2 \theta {\rm d}\theta,
\end{equation} 
which evaluates to,
\begin{eqnarray}
 F = I_* \left[ \sin^{-1} \left( \frac{R_*}{r} \right) - 
 \left( \frac{R_*}{r} \right) \sqrt{1 - \left( \frac{R_*}{r} \right)^2} \right].
\end{eqnarray} 
For a star with effective temperature $T_*,$ the brightness $I_* = (1 / \pi) \sigma T_*^4$, with $\sigma$ the Stefan--Boltzmann constant \citep[e.g.][]{rybicki79}.   Equating $F$ to the one-sided disk emission $\sigma T_{\rm eff}^4$ we obtain a radial temperature profile,
\begin{eqnarray}
  \frac{T_{\rm eff}^4}{T_*^4}  = \frac{1}{\pi} 
 \left[ \sin^{-1} \left( \frac{R_*}{r} \right) - 
 \left( \frac{R_*}{r} \right) \sqrt{1-\left(\frac{R_*}{r}\right)^2} \right].
\label{equation_C2_tpassive} 
\end{eqnarray}
Integrating over radii, we obtain the total disk luminosity,
\begin{eqnarray}
 L_{\rm disk} & = & 2 \times \int_{R_*}^\infty 2 \pi r \sigma T_{\rm
 eff}^4 {\rm d}r \nonumber \\
              & = & \frac{1}{4} L_*.
\end{eqnarray}
A flat disk that extends all the way to the stellar equator intercepts a quarter of the stellar flux. 

The temperature profile given by equation~(\ref{equation_C2_tpassive}) is approximately a power-law at large radii. Expanding the right-hand-side in a Taylor series in the limit that $(R_* / r) \ll 1$ (i.e. far from the stellar surface), we obtain,
\begin{equation}
 T_{\rm eff} \propto r^{-3/4},
\label{eq_tpassive2} 
\end{equation} 
as the limiting temperature profile of a thin, flat, passive disk. For fixed molecular weight $\mu$ this in turn implies a sound speed profile
\begin{equation}
 c_{\rm s} \propto r^{-3/8},
\end{equation} 
and a scaling of the geometric thickness with radius,
\begin{equation}
 \frac{h}{r} \propto r^{1/8}.
\end{equation} 
An irradiated disk therefore {\em flares} (i.e. has a concave shape) toward larger radii. If the disk does flare then the outer regions intercept a larger fraction of stellar photons, leading to a higher temperature. As a consequence, a temperature profile $T_{\rm eff} \propto r^{-3/4}$ is the steepest profile we would expect to obtain for a passive disk.

Irradiation is frequently important for protoplanetary disks, with standard models (that include self-consistent treatments of disk flaring) having effective temperature profiles close to $T_{\rm eff} \propto r^{-1/2}$ \citep{kenyon87,chiang97}. It can also be important in high energy accretion environments, for example in X-ray binaries where irradiation of the outer disk by X-rays from the inner disk can dominate the local thermal balance \citep{dubus99}. 

\subsubsection{Green's function solution}
Assume for simplicity that the viscosity $\nu(\Sigma,r,\ldots)$ is a constant. The surface density of the 
disk $\Sigma(r,t)$ then obeys the equation,
\begin{equation}
    \frac{\partial \Sigma}{\partial t} = 
    \frac{3 \nu}{r} \frac{\partial}{\partial r} \left[
    r^{1/2} \frac{\partial}{\partial r} \left(
    \Sigma r^{1/2}\right)\right].
\end{equation}
To solve this equation we first manipulate it into the standard form of a Bessel's equation\footnote{We're working toward the famous solution found by \citet{lynden-bell74}, but here following Gordon Ogilvie's notes on ``Accretion Disks" from Part III of the Cambridge Mathematical Tripos. The solution strategy is still not all that obvious, though you might note that we have a diffusion equation in cylindrical co-ordinates, which is analogous to a classical example of Bessel's equation---heat diffusion in a cylinder.}. We look for a solution in which the variables are separated, and modes have a decaying time dependence,
\begin{equation}
    \Sigma(r,t) = r^\beta \sigma(r) \exp[-\lambda t].
\end{equation}
Here $\lambda > 0$ and by writing the spatial dependence as $r^\beta \sigma(r)$ we have given ourselves a free parameter in $\beta$. Substituting,
\begin{equation}
    -\lambda r^\beta \sigma = \frac{3 \nu}{r} 
    \frac{\rm d}{{\rm d}r} \left[ r^{1/2} 
    \frac{\rm d}{{\rm d}r} \left( \sigma r^{\beta+1/2} \right)\right].
\end{equation}
After evaluating the derivatives and dividing through by $r^{\beta-2}$ we have,
\begin{equation}
    r^2 \frac{{\rm d}^2 \sigma}{{\rm d}r^2} + 
    \left( 2 \beta + \frac{3}{2} \right) r \frac{{\rm d}\sigma}{{\rm d}r} + 
    \beta \left( \beta + \frac{1}{2} \right) \sigma + 
    \frac{\lambda}{3 \nu} r^2 \sigma = 0.
\end{equation}
Defining $k^2 \equiv \lambda / (3 \nu)$ and using the freedom to choose $\beta=-1/4$, 
\begin{equation}
    r^2 \frac{{\rm d}^2 \sigma}{{\rm d}r^2} + 
    r \frac{{\rm d}\sigma}{{\rm d}r} +
    \left( k^2 r^2 - \frac{1}{16} \right) \sigma = 0.
\end{equation}
This is in the form of Bessel's equation, which has a general solution,
\begin{equation}
    \sigma = c_1 J_{1/4} (kr) + c_2 Y_{1/4} (kr),
\end{equation}    
where $c_1$ and $c_2$ are constants and $J_{1/4}$ and $Y_{1/4}$ are Bessel functions of the first and second kinds respectively. The term involving $Y_{1/4} (kr)$ implies a non-zero torque as $r \rightarrow 0$, so in the case of a point mass that does not spin up the disk material $c_2 =0$. The solution is therefore,
\begin{equation}
    \Sigma \propto r^{-1/4} J_{1/4} (kr) \exp[-3 \nu k^2 t].
\end{equation}
The properties of Bessel functions allow us to write a general initial condition for the surface density in the form,
\begin{equation}
    \Sigma(r,0) = \int_0^\infty g(k) r^{-1/4} J_{1/4} (kr) dk.
\label{eq_bessel_decomposition}    
\end{equation}
The time-dependent solution will then be,
\begin{equation}
    \Sigma(r,t) = \int_0^\infty g(k) r^{-1/4} J_{1/4} (kr) \exp[-3 \nu k^2 t] dk.
\label{eq_disk_general_solution}    
\end{equation}
The problem is thus solved provided that we can determine the decomposition of the initial surface density into Bessel functions, given by $g(k)$.

\begin{figure}
    \centering
    \includegraphics[width=\columnwidth]{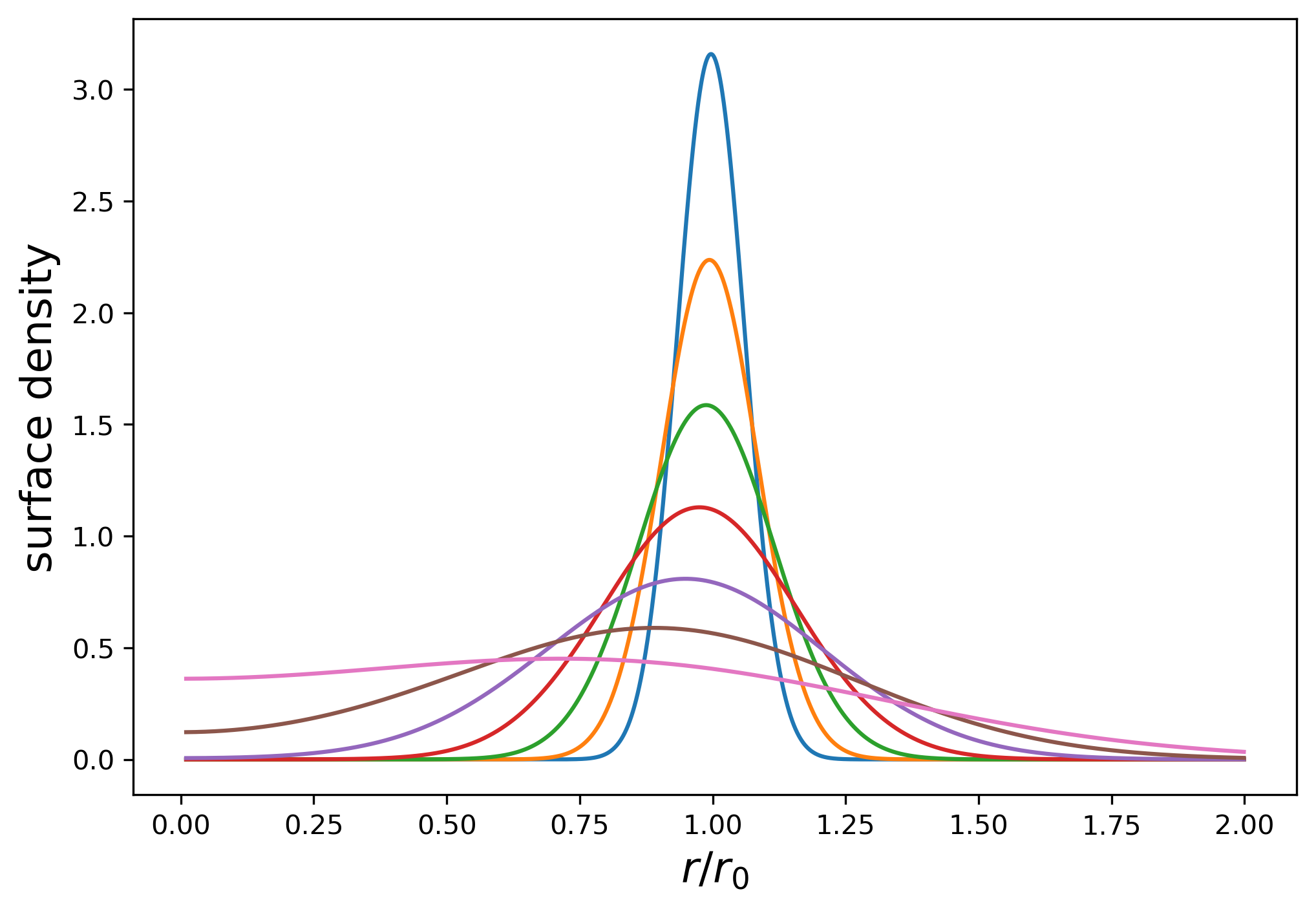}
    \caption{The evolution of a thin ring of gas, initially at $r=r_0$, under the action of a constant viscosity (equation~\ref{eq_ring_solution}). The curves are plotted at dimensionless times that are multiples of two, starting at $\tau=0.008$ and going up to $\tau=0.512$.}
    \label{fig_disk_sigma_evolve}
\end{figure}

To determine $g(k)$ we make use of the Fourier-Bessel (or Hankel) transform pair. The textbook definition of this pair is,
\begin{eqnarray}
 g(k) & = & \int_0^\infty f(r) J_m (kr) r dr, \\
 f(r) & = & \int_0^\infty g(k) J_m (kr) k dk.
\end{eqnarray}
Writing equation~(\ref{eq_bessel_decomposition}) in this form,
\begin{equation}
    r^{1/4} \Sigma(r,0) = \int_0^\infty k^{-1} g(k) J_{1/4} (kr) k dk,
\end{equation}
the inverse transform is,
\begin{equation}
    k^{-1} g(k) = \int_0^\infty s^{1/4} \Sigma(s,0) J_{1/4} (ks) s ds.
\end{equation}
Substituting in equation~(\ref{eq_disk_general_solution}) the general solution is,
\begin{eqnarray}
    \Sigma(r,t) =  r^{-1/4} \int_0^\infty \int_0^\infty \Sigma(s,0) J_{1/4} (ks)
    J_{1/4} (kr) \nonumber \\ \times \exp[-3 \nu k^2 t] s^{5/4} k ds dk.
\end{eqnarray}
We express this in the form,
\begin{equation}
    \Sigma(r,t) = \int_0^\infty G(r,s,t) \Sigma(s,0) ds,
\label{eq_greens_fn_def}    
\end{equation}
where,
\begin{eqnarray}
    G(r,s,t) = s^{5/4} r^{-1/4} \int_0^\infty J_{1/4} (ks) J_{1/4} (kr) \nonumber \\ 
    k \exp[-3 \nu k^2 t] dk,
\label{eq_greens_fn}    
\end{eqnarray}
is the Green's function. This integral evaluates to,
\begin{equation}
    G(r,s,t) = \frac{r^{-1/4} s^{5/4}}{6 \nu t} I_{1/4} \left( \frac{rs}{6 \nu t} \right)
    \exp \left[ -\frac{(r^2 + s^2)}{12 \nu t}\right],
\end{equation}
with $I_{1/4}$ being a modified Bessel function. It is illustrative to consider the solution for an initial ring of gas orbiting at $s = r_0$. Taking the initial condition as,
\begin{equation}
    \Sigma(s,0) = \frac{m}{2 \pi r_0} \delta (s - r_0),
\end{equation}
with $\delta$ being a Dirac delta function, the solution follows immediately from equations~(\ref{eq_greens_fn_def}) and (\ref{eq_greens_fn}). It can be written compactly in terms of dimensionless variables $x$ and $\tau$,
\begin{eqnarray}
 x & = & \frac{r}{r_0}, \\
 \tau & = & \frac{12 \nu}{r_0^2} t. 
\end{eqnarray}
In terms of these variables,
\begin{equation}
    \Sigma(x,\tau) = \frac{m}{\pi r_0^2} \frac{x^{-1/4}}{\tau} 
    I_{1/4} \left( \frac{2x}{\tau} \right) \exp \left[ - \frac{(1+x^2)}{\tau} \right]. 
\label{eq_ring_solution}    
\end{equation}
The solution described by this equation is shown in Figure~\ref{fig_disk_sigma_evolve}. It displays asymmetric diffusion, with mass flowing toward $r=0$ while the conserved angular momentum is carried by a vanishing fraction of the mass toward $r = \infty$. Although derived under the restrictive and normally unrealistic assumption that $\nu$ is constant, these properties are qualitative features of viscous disk evolution in the case where there is zero-torque at the inner edge of the disk.

Although we will not discuss the details here, time-dependent solutions can also be derived that dispense with the zero-torque assumption \citep{rafikov16,nixon20}. Particularly simple {\em decretion} disk solutions \citep[aspects of which were already discussed in \S2.5 of][]{lynden-bell74} exist if one assumes that an external torque maintains a $v_r = 0$ boundary condition at a finite radius $r_{\rm in}$ \citep{pringle91}. It is also possible to derive a relativistic version of the disk evolution equation \citep{balbus17}.

\subsubsection{Self-similar solution}
\citet{lynden-bell74} also derived a self-similar solution to the disk evolution equation (equation~\ref{eq_1D_disk_evolution}), for the case where the viscosity is a power-law function of radius,
\begin{equation} 
 \nu \propto r^\gamma.
\end{equation}
If a disk with characteristic size $r_1$ at $t=0$ has a surface density profile of the form,
\begin{equation} 
 \Sigma (t=0) = \frac{C}{3 \pi \nu_1 \tilde{r}^\gamma} 
 \exp \left[ {-\tilde{r}^{(2-\gamma)}} \right],
\end{equation}
where $C$ is a constant, $\tilde{r} \equiv r / r_1$, 
and $\nu_1 \equiv \nu(r_1)$, then the time-dependent solution is,
\begin{eqnarray}
 \Sigma(\tilde{r},T) = \frac{C}{3 \pi \nu_1 \tilde{r}^\gamma} 
 T^{-(5/2-\gamma)/(2-\gamma)} 
 \exp \left[ -\frac{\tilde{r}^{(2-\gamma)}}{T} \right],
\label{eq_C2_selfsimilar} 
\end{eqnarray}
 where,
\begin{eqnarray}
 T & \equiv & \frac{t}{t_{\rm s}} + 1,  \\
 t_{\rm s} & \equiv & \frac{1}{3(2-\gamma)^2} \frac{r_1^2}{\nu_1}.
\label{eq_selfsimilar} 
\end{eqnarray}  
This solution has proved to be quite useful for comparing theoretical models of viscous disk evolution to data \citep[e.g.][]{hartmann98}. It can be generalized to the case where disk evolution is driven by a combination of viscous transport and MHD winds \citep{tabone21}.

\subsection{The $\alpha$-prescription}
\label{sec_alpha}
A predictive model for disk evolution follows from equation~(\ref{eq_1D_disk_evolution}) if we can write down how the stress, or equivalently the viscosity, depends on properties of the disk. \citet{shakura73} advanced physical arguments in favor of the form,
\begin{equation}
    T_{r \phi} = - \alpha P,
\label{eq_SS_original}    
\end{equation}
where $P$ is the pressure and $\alpha$ is a dimensionless parameter\footnote{\citet{shakura73} first introduce $\alpha$ with an expression involving the magnetic field ($H$ in their notation), similar but not identical to our equation~(\ref{eq_def_alpha}). The famous version is their equation~(1.2), $T_{r \phi} = -\alpha \rho c_s^2$, where the sound speed includes contributions from both gas and radiation pressure.}. This ansatz is known as the ``$\alpha$ prescription". Using equation~(\ref{eq_Trphi}) and writing $\nu = \mu / \rho$, for a Keplerian disk an equivalent form is,
\begin{equation}
    \nu = \frac{2}{3} \alpha c_s h \simeq \alpha c_s h.
\label{eq_alpha_prescription}    
\end{equation}
In keeping with the approximate nature of this exercise, the factor of two-thirds is typically ignored and the $\alpha$-prescription written as just $\nu = \alpha c_s h$.

To order of magnitude, the microphysical viscosity of a fluid can be written in terms of the thermal velocity of the molecules $v_{\rm th}$ and the mean-free-path $l$ as,
\begin{equation}
    \nu \sim v_{\rm th} l.
\end{equation}
By analogy, one can view equation~(\ref{eq_alpha_prescription}) as describing an effective viscosity due to turbulent eddies whose speed scales with the sound speed and whose size scales with the disk thickness. Since turbulent velocities exceeding the sound speed would cause shocks and rapid dissipation, and isotropic eddies could not significantly exceed $h$, this argument bounds $\alpha < 1$. This argument is not terribly useful, as physical mechanisms for disk turbulence do not yield turbulent structures that look much like eddies on scales of the order of $h$. It is more instructive to follow Shakura and Sunyaev's original train of thought, and express $\alpha$ in terms of fluid (Reynolds) and magnetic (Maxwell) stresses in a turbulent fluid \citep{balbus98},
\begin{equation}
 \alpha = \left\langle \frac{ \delta v_r \delta v_\phi }{c_s^2} - \frac{ B_r B_\phi }{4 \pi \rho c_s^2} \right\rangle_{\rho},
\label{eq_def_alpha} 
\end{equation} 
where the angle brackets indicate a density weighted average over space and time. 
The first term is the Reynolds stress from correlated fluctuations in the radial velocity and 
perturbed azimuthal velocity, the second is the Maxwell stress from MHD turbulence. 

\subsection{Time scales}
\label{sec_time_scales}

For a thin disk we can express several relevant time scales as simple functions of the disk properties. The {\em dynamical time scale},
\begin{equation}
    t_{\rm dyn} = \frac{1}{\Omega},
\end{equation}
is evidently $1/2\pi$ of the orbital period. The time scale for establishing vertical hydrostatic equilibrium is the sound crossing time across a scale height, $t_{\rm hydro} \sim h / c_s$ Using $h = c_s / \Omega$ we have,
\begin{equation}
    t_{\rm hydro} \sim \frac{h}{c_s} \sim \frac{1}{\Omega} \sim t_{\rm dyn}.
\label{eq_hydrostatic_equilibrium}    
\end{equation}
Vertical hydrostatic equilibrium is thus established in a circular disk on the shortest possible time scale. In an eccentric disk, however, where the gravitational potential experienced by a fluid element varies around the orbit, the approximate equality between $t_{\rm hydro}$ and $t_{\rm dyn}$ means that hydrostatic equilibrium is not established and interesting coupled dynamics between the radial and vertical structure is possible \citep{ogilvie14}.

The {\em thermal time scale} is the time scale on which the disk would cool if heating processes were instantaneously cut off. The thermal energy per unit surface area of the disk is $U \sim \Sigma c_s^2$. Using equation~(\ref{eq_qplus2}) and equation~(\ref{eq_alpha_prescription}),
\begin{equation}
    t_{\rm th} = \frac{U}{Q_+} \sim \frac{1}{\alpha \Omega}.
\end{equation}
The thermal time scale is the shortest time scale on which we expect the emission from an annulus of the disk, heated by viscous-like dissipation, to change.

The {\em viscous time scale} is the time scale on which redistribution of angular momentum leads to gas inflow. If the surface density is not in a steady-state, it is also the time scale over which the surface density evolves. Starting from equation~(\ref{eq_1D_disk_evolution}), we can estimate the viscous time scale by writing the evolution equation in a form that more closely resembles a prototypical one-dimensional diffusion equation. Defining,
\begin{eqnarray}
 X &\equiv& 2 r^{1/2},  \\
 f &\equiv& \frac{3}{2} \Sigma X, 
\end{eqnarray}
and assuming that $\nu$ is a constant, the evolution equation is,
\begin{equation}
  \frac{\partial f}{\partial t} = D \frac{ \partial^2 f }{\partial X^2},
\label{eq_diffusion_prototype}  
\end{equation}
with diffusion coefficient $D$ given by,
\begin{equation}
 D = \frac{12 \nu}{X^2}.
\end{equation} 
The diffusion time scale across scale $\Delta X$ implied by equation~(\ref{eq_diffusion_prototype}) 
is $(\Delta X)^2 / D$. Returning to the original variables, the 
time scale over which viscosity smooths out surface density gradients on radial 
scale $\Delta r$ is,
\begin{equation}
 \tau_{\rm visc} \sim \frac{(\Delta r)^2}{\rm \nu}.
\label{eq_viscous_timescale} 
\end{equation}
If the disk has size $r$, the surface density can  
evolve on a time scale,
\begin{equation}
 \tau_{\rm visc} \approx \frac{r^2}{\nu}.
\end{equation}
Using the $\alpha$-prescription, we obtain,
\begin{equation}
    t_{\rm visc} \sim \frac{1}{\alpha \Omega} \left( \frac{h}{r} \right)^{-2}.
\end{equation}
For a thin disk the viscous time scale is substantially longer than the thermal time scale, and absent special circumstances thermal equilibrium is maintained in the vertical direction while the surface density evolves due to accretion.

\begin{figure}
    \centering
    \includegraphics[width=\columnwidth]{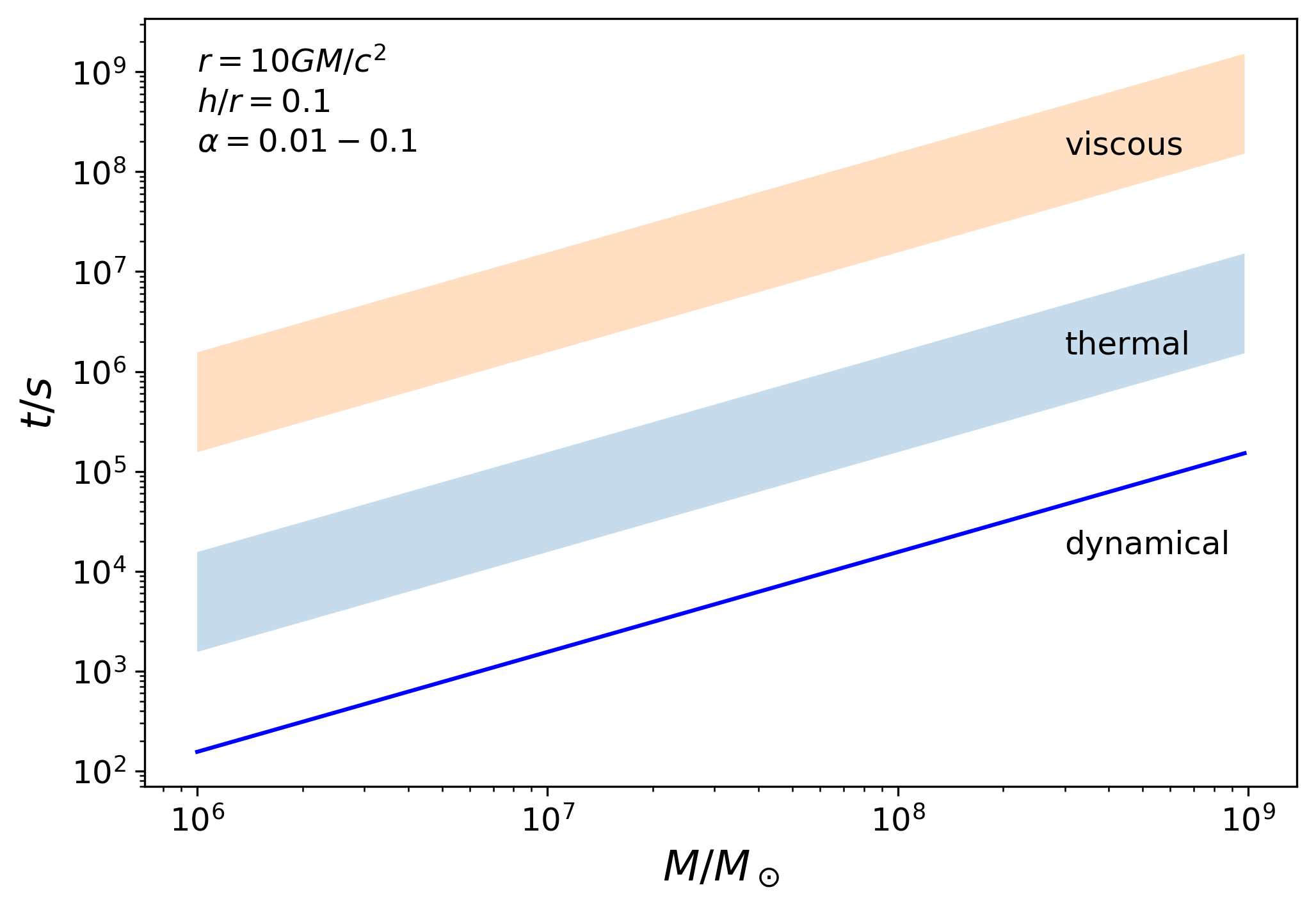}
    \caption{Estimates of the dynamical, thermal and viscous time scales at $r=10 GM/c^2$ for disks around supermassive black holes of different masses. The disk aspect ratio is taken to be $h/r=0.1$. The colored bands for the thermal and viscous time scales show the range obtained assuming $10^{-2} \leq \alpha \leq 0.1$.}
    \label{fig_timescales}
\end{figure}

The time scale hierarchy,
\begin{equation}
    t_{\rm dyn} \sim t_{\rm hydro} \ll t_{\rm th} \ll t_{\rm visc},
\end{equation}
is a generic property of geometrically thin disks, and is one of the main reasons why thin disk theory is internally consistent and useful \citep[e.g.][]{pringle81}. Figure~\ref{fig_timescales} gives a concrete example of these time scales at ten gravitational radii around supermassive black holes of various masses, for $h/r = 0.1$ and different assumed values for $\alpha$. For a $10^9 \ M_\odot$ black hole, for example, the dynamical time scale in the inner disk is of the order of a day, the thermal time scale is around a month, and the viscous time scale is around ten years.
  
\subsection{$\alpha$-model disks}
Adopting the $\alpha$-prescription (\S\ref{sec_alpha}) the dependence of $\nu$ on the local disk conditions and on $\alpha$, $\nu (\alpha, \Sigma, \Omega_{\rm K})$, can be determined. With this function in hand, equation~(\ref{eq_1D_disk_evolution}) can be solved (usually numerically) for the time-dependent evolution of an arbitrary initial surface density profile. Steady-state solutions (usually analytic) for the surface density profile $\Sigma(r,\dot{M})$ can also be found. These are ``$\alpha$-model" or ``Shakura-Sunyaev" disk solutions. Recall that none of this effort is necessary if our only concern is the profile of the disk effective temperature in steady state, as that is given by equation~(\ref{eq_disk_Teff_r}) independent of the form of the viscosity.

A toy example shows how $\alpha$-model disks are constructed. Assume, for no reason other than simplicity, that the vertical structure of the disk is {\em isothermal}. The effective temperature $T_{\rm eff}$ is then the only temperature characterizing the disk at some radius, and the viscosity can be derived from a triplet of already-introduced equations,
\begin{eqnarray}
 \nu & = & \alpha c_s h, \label{eq_alpha_disks_start} \\
 h & = & \frac{c_s}{\Omega_{\rm K}}, \\
 2 \sigma T_{\rm eff}^4 & = & \frac{9}{4} \nu \Sigma \Omega_{\rm K}^2.
\label{eq_alpha_disks_end}
\end{eqnarray}
The sound speed is related to the temperature through,
\begin{equation}
    c_s^2 = \frac{ k_{\rm B} T_{\rm eff} }{\mu m_p},
\end{equation}
where $k_{\rm B}$ is Boltzmann's constant, $m_p$ is the mass of the proton, and $\mu$ is the mean molecular weight in units of $m_p$. Using this, we eliminate $T_{\rm eff}$ from equations~(\ref{eq_alpha_disks_start})-(\ref{eq_alpha_disks_end}) and obtain an expression for the viscosity,
\begin{equation}
    \nu = \frac{9^{1/3}}{2 \sigma^{1/3}}
    \left( \frac{k_{\rm B}}{\mu m_p} \right)^{4/3} \alpha^{4/3} \Omega_{\rm K}^{-2/3} \Sigma^{1/3}.
\end{equation}
For fixed central mass $M$, the predicted viscosity scales as $\nu \propto \alpha^{4/3} r \Sigma^{1/3}$. The equation for the evolution of the disk surface density, equation~(\ref{eq_1D_disk_evolution}), would be non-linear with this viscosity. In steady-state, $\nu \Sigma \propto \dot{M}$ (equation~\ref{eq_disk_nuSigma}), so away from the inner boundary the disk surface density would scale as $\Sigma \propto \dot{M}^{3/4} r^{-3/4}$. We have not specified $\alpha$, but as long as this can be taken to be fixed (determined, perhaps, from simulations of physical angular momentum transport mechanisms or inferred from observations of time-dependent disk systems) we have a full solution for the evolution of geometrically thin disks.

The toy model given above captures the spirit of $\alpha$-model disks, but it's not quite the full story. Real disks will not be vertically isothermal. This extra complexity can be captured at different levels of approximation,
\begin{itemize}
    \item[(i)] At each radius, characterize the disk's vertical structure in terms of a {\em central temperature} $T_c$ as well as an effective temperature $T_{\rm eff}$. We can derive a relation between $T_c$ and $T_{\rm eff}$ by considering how energy is transported vertically within the disk. Then, assuming that the sound speed that enters into the expression for the viscosity ($\nu = \alpha c_s^2 / \Omega_{\rm K}$) corresponds to the {\em central} temperature, we can proceed as before and derive the functional form of the viscosity. This is described as a ``one-zone" model for the disk vertical structure.
    \item[(ii)] Alternatively, we can write down and solve (numerically) differential equations for the vertical disk structure, $\rho(z)$, $T(z)$, in a manner directly analogous to the equations of stellar structure. This approach requires a point-by-point specification of the stress, for which we could adopt the original \citet{shakura73} form, $T_{r \phi} = -\alpha P$, or something else perhaps derived from simulations. Because of the separation between the thermal and viscous time scales in a geometrically thin disk, it is normally consistent to solve for the vertical structure separately from the radial structure. This is sometimes called a ``1+1D" disk model.
\end{itemize}
Both of these approaches are well-defined. Whether the additional complexity of a vertical integration leads to a physically more realistic model, however, is an open question. Unlike in the case of stellar structure---where the energy transport processes and rate of nuclear energy generation are quite well-known---the physical processes entering into calculations of disk vertical structure are uncertain. I wouldn't ascribe much physical reality to modest differences between disk models with differing formal degrees of approximation.

To write down a version of the one zone equations, suitable for deducing the properties of steady-state $\alpha$-disks, we need only a relation between the central temperature $T_c$ and the effective temperature $T_{\rm eff}$. As in stellar structure, energy can be transported from the hot interior to the cooler photosphere by radiative diffusion or by convection\footnote{Turbulent transport or transport by waves are also in principle possible. In fact, they may well be important, and we ignore them here only because they are harder to capture in simple analytic formulae.}. Consider the limit of an optically thick disk with radiative transport. The vertical energy flux is \citep[e.g.][]{rybicki79},
\begin{equation}
    F(z) = - \frac{16 \sigma T^3}{3 \kappa_{\rm R} \rho} 
    \frac{{\rm d}T}{{\rm d}T} = \sigma T_{\rm eff}^4,
\end{equation}
where $\kappa_{\rm R}$ is the Rosseland mean opacity (with units of ${\rm cm}^2 \ {\rm g}^{-1}$). In equating the flux to a constant we have assumed that energy dissipation is strongly concentrated at $z=0$. Noting that the increment of optical depth ${\rm d}\tau = \kappa_{\rm R} {\rm d}z$, we integrate from the mid-plane to the photosphere,
\begin{equation}
    -\frac{16}{3} \int_{T_c}^{T_{\rm eff}} T^3 {\rm d}T = 
    T_{\rm eff}^4 \int_0^{z_{\rm ph}} {\rm d}\tau.
\end{equation}
If the disk is sufficiently optically thick that $T_c \gg T_{\rm eff}$ then we obtain,
\begin{equation}
    \left( \frac{T_c}{T_{\rm eff}} \right)^4 \simeq \frac{3}{4} \tau,
\label{eq_Tc_Teff}    
\end{equation}
as the relation between the central and photospheric disk conditions.

With this expression in hand, we can write down a set of equations that determine the steady-state radial structure of $\alpha$-model disks in the one zone approximation. The disk is specified by the central mass $M$, accretion rate $\dot{M}$, and innermost radius $\tilde{r}$, where a zero-torque inner boundary condition is imposed. The variables to be determined are the mid-plane density $\rho_0$, pressure $P$, temperature $T_c$, sound speed $c_s$, surface density $\Sigma$, scale height $h$, optical depth to the mid-plane $\tau$, Rosseland mean opacity $\kappa_{\rm R}$, and viscosity $\nu$. We assume an $\alpha$-prescription, with $\alpha$ a constant, and approximate the opacity as a power-law function of the central density and temperature.
Collecting together previous results (equations~\ref{eq_rho0}, \ref{eq_thickness}, \ref{eq_EOS}, \ref{eq_disk_Teff_r}, \ref{eq_alpha_prescription}, and \ref{eq_disk_nuSigma}), and adding in an equation of state together with some straightforward definitions, the set is,
\begin{eqnarray}
 \rho_0 & = & \frac{1}{\sqrt{2 \pi}} \frac{\Sigma}{h}, \label{eq_alpha_start} \\
 h & = & \frac{c_s}{\Omega_{\rm K}}, \\
 c_s^2 & = & \frac{P}{\rho_0}, \\
 P & = & \frac{k_{\rm B}}{\mu m_p} \rho_0 T_c + \frac{4 \sigma}{3 c} T_c^4, \\
 T_c^4 & = & \frac{9 \dot{M} \Omega_{\rm K}^2}{32 \pi \sigma} \left[ 1 - \sqrt{\frac{\tilde{r}}{r}} \right] \tau, \\
 \nu \Sigma & = & \frac{\dot{M}}{3 \pi} \left[ 1 - \sqrt{\frac{\tilde{r}}{r}} \right], \\
 \nu & = & \alpha c_s h, \\
 \tau & = & \frac{1}{2} \kappa_{\rm R} \Sigma, \\
 \kappa_{\rm R} & = & \kappa_0 \rho_0^a T_c^b \label{eq_alpha_end}.
\end{eqnarray}
Up to a few not-so-important numerical factors, these are the standard equations used to determine thin disk structure in the Newtonian limit. Additional discussion of them can be found in \citet{frank02}.

As written, the mid-plane pressure is the sum of a gas pressure component and one due to radiation pressure. Usually one or the other of these pressure sources is much larger than the other, with as a rule of thumb radiation pressure dominating in black hole disks close to the innermost stable circular orbit, and gas pressure dominating otherwise\footnote{Under some circumstances, specifically when the disk is threaded by a net magnetic field, magnetic pressure, $P_B = B^2 / 8 \pi$, may also contribute to vertical support against gravity \citep{bai13,salvesen16,zhu18,mishra20}. This possibility is not normally considered in classical disk models.}. Dropping either gas or radiation pressure, we can solve the set of equations for the steady-state disk structure, verifying after the fact that we dropped the right one. Away from the inner boundary, the solutions take the form of power-laws, e.g. $\Sigma \propto r^w M^x \dot{M}^y \alpha^z$, with power-law indices that depend upon the source of pressure and upon the functional form of the opacity. Because radius and central mass only enter the equations combined in the form of the Keplerian angular velocity, the solutions only depend on $\Omega_{\rm K}$.

\subsection{Self-gravitating $\alpha$-disks}
\label{sec_SG_alpha}
In most cases, and in particular when the source of angular momentum transport is the MRI, the $\alpha$-disk equations in no way {\bf determine} $\alpha$ (though the formalism would be inconsistent for $\alpha$ values large enough to induce supersonic inflow). Self-gravitating disks can be an exception. Their stability is a function of the Toomre $Q = c_s \Omega_{\rm K}/(\pi G \Sigma)$ (\S\ref{sec_selfgravity}, specializing to a Keplerian disk). Suppose, somewhat reasonably, that the angular momentum transport rate is a function of the local disk conditions and increases rapidly as $Q$ drops below some critical value $Q_0 \sim 1$. Under these assumptions, the combination of self-gravitating transport and local thermal equilibrium can establish a stable feedback loop that maintains $Q \approx Q_0$. If $Q > Q_0$, reduced transport leads to reduced heating, lowering $c_s$ to re-establish $Q=Q_0$. The reverse happens for $Q < Q_0$. Self-gravitating disks are then expected to be everywhere marginally stable, with $Q \approx Q_0$. This type of model was introduced by \citet{paczynski78}.

Imposing $Q=Q_0$ in addition to the usual set of $\alpha$-disk equations has an important consequence: $\alpha$ is no longer a free parameter but rather a specified function of the local disk conditions \cite{gammie01,levin07}. The evolution of the disk in this limit is then fully determined. \citet{rafikov15}, and references therein, detail such ``gravito-turbulent" disk models. They are useful provided that non-local angular momentum transport and mass infall (a complication that often accompanies self-gravity in astrophysically relevant settings) can be consistently ignored.

\subsection{The values of $\alpha$}
Although a great deal of effort has been expended over the years in trying to determine ``the" value of $\alpha$, it should be clear from the discussion so far that this is an illusory quest. Even to the extent that $\alpha$ provides a good parameterization of the strength of accretion disk turbulence, its value ought to depend upon the MHD properties of the disk, on the strength of self-gravity, and so on. The consensus theoretical expectation is that for disks that are well-described by ideal MHD, turbulence in the absence of net magnetic flux yields \citep{davis10,simon12},
\begin{equation}
    \alpha_{\rm ZNF} \simeq 0.01-0.02.
\end{equation}
There remains some uncertainty about how well converged this computational result is \citep{ryan17}. In the presence of a net vertical magnetic field $B_z$ transport is stronger, with local simulation results indicating that \citep{hawley95,bai13,salvesen16}, 
\begin{equation}
    \alpha_{\rm VF} \approx \beta_z^{-0.5},
\end{equation}
where $\beta_z$ is defined as the ratio of the gas pressure to the magnetic pressure in the net field at the disk mid-plane,
\begin{equation}
    \beta_z \equiv \frac{P_{\rm gas}}{P_{\rm B}}.
\end{equation}
This scaling implies that disks with moderately strong vertical fields, $\beta_z \lesssim 10^2$, are strongly turbulent and generate mid-plane toroidal fields with magnetic pressure comparable to the gas pressure. Such magnetically dominated or ``magnetically elevated" disks have stability properties that differ in interesting ways from standard Shakura-Sunyaev disks \citep{begelman07}.

For self-gravity, local gravito-turbulent models predict that $\alpha$ scales with the local cooling time as \citep{gammie01},
\begin{equation}
    \alpha_{\rm SG} \propto \frac{1}{\Omega_{\rm K} t_{\rm cool}},
\end{equation}
with an upper limit set by fragmentation at $\alpha_{\rm SG} \sim 0.1$ \citep{gammie01,rice05}. As with the MRI, pinning down these numbers precisely from numerical simulations is none too easy a task.

Observationally, $\alpha$ can be estimated in systems where variability exposes the viscous time scale (equation~\ref{eq_viscous_timescale}), with the most important example being  dwarf novae, whose disks show limit cycle behavior due to thermal instability (see \S\ref{sec_thermal_instability}). Dwarf nova outbursts can be well-described as time-dependent $\alpha$-disks \citep{meyer81,bath82,mineshige83,smak84}. Under well-ionized conditions, modeling of these systems suggests $\alpha \approx 0.1$ \citep{king07,hameury20}. The inferred larger value of $\alpha$, as compared to predictions from simplified MRI simulations, may be due to convection in dwarf nova disks \citep{hirose14}. The strength of turbulence in protoplanetary disks can be constrained more directly from observations of molecular line broadening \citep{hughes11}. Such analyses suggest that much lower levels of turbulence (in some cases only upper limits are obtained) occur in very weakly ionized disks \citep{flaherty17,flaherty20}.

\subsection{The Shakura-Sunyaev solution}
Thin disk solutions depend upon whether the main source of pressure is gas or radiation, and upon the opacity under conditions encountered near the disk mid-plane. For disks around compact objects (black holes, neutron stars, and white dwarfs) two opacity regimes cover most conditions of interest. At high temperatures, {\em electron scattering} dominates. For plasma with a typical astrophysical distribution of elements, the opacity is,
\begin{equation}
    \kappa_{\rm es} = 0.34 \ {\rm cm^2 \ g^{-1}}.
\end{equation}
At lower temperatures, {\em free-free} opacity, which can be approximated using Kramers' law, applies,
\begin{equation}
    \kappa_{\rm ff} = 6.4 \times 10^{22} \rho T^{-7/2} \ {\rm cm^2 \ g^{-1}}.
\end{equation}
(Here, $\rho$ and $T$ are understood to be expressed in c.g.s. units.)
Kramers' law remains valid down to the temperature where hydrogen recombines, at $T \approx 1-2 \times 10^4 \ {\rm K}$. At lower temperatures, which can be encountered at large radii in AGN disks and which are typical of protoplanetary disks, molecules, dust and ice grains provide the opacity.

The properties of Shakura-Sunyaev disk solutions are not terribly intuitive, but one important result---the disk thickness in the innermost radiation pressure dominated region---is readily derived. The vertical flux of momentum carried by radiation, $F_z/c$, is equal to (using equation~\ref{eq_disk_Teff_r}),
\begin{equation}
 \frac{F_z}{c} = \frac{\sigma T_{\rm eff}^4}{c} = \frac{3}{8 \pi} \Omega_{\rm K}^2 \dot{M}
 \left[ 1 - \sqrt{\frac{\tilde r}{r}} \right].
\end{equation}
At high enough temperatures the force that the radiation exerts per unit mass on the gas is $F_z \sigma_{\rm T} / c$, where $\sigma_{\rm T} = 6.7 \times 10^{-25} \ {\rm cm^2}$ is the Thomson cross-section appropriate for electron scattering. Setting this equal to the vertical acceleration due to the gravity of the central object, which in the Newtonian limit is just $\Omega_{\rm K}^2 z$, the scale height in the radiation pressure / electron scattering dominated regime is,
\begin{equation}
    h = \frac{3 \sigma_{\rm T}}{8 \pi c} \left[ 1 - \sqrt{\frac{\tilde r}{r}} \right] \dot{M}.
\end{equation}
Away from the inner boundary, the scale height ($h$ itself, {\em not} $h/r$) is predicted to be  constant with radius, with a value that is proportional to the accretion rate.

Without further ado, we quote without derivation the \citet{shakura73} disk solution, which takes the form of piece-wise power-laws corresponding to three regimes,
\begin{itemize}
    \item 
    An inner disk, dominated by radiation pressure and electron scattering opacity.
    \item
    A middle disk, dominated by gas pressure and electron scattering opacity.
    \item
    An outer disk, dominated by gas pressure and free-free opacity.
\end{itemize}
The extent of the outer disk is limited by the validity of the free-free opacity formula---which as noted already fails at low temperature---and / or by the onset of other physical processes such as gravitational instability. The model is conceptually just the solution of equations~(\ref{eq_alpha_start}-\ref{eq_alpha_disks_end}), though there is some added subtlety that arises from the fact that electron scattering is a true scattering process that does not alter the energies of either photons or electrons. This means that even at high temperatures, sub-dominant absorption opacity plays a critical role in thermalizing the emergent radiation. There are also order unity numerical factors that differ between our equation set and that of the original \citet{shakura73} paper.

The \citet{shakura73} solutions can usefully be expressed in terms of dimensionless variables for mass, accretion rate, and radius,
\begin{eqnarray}
  m & \equiv & \frac{M}{M_\odot}, \\
  \dot{m} & \equiv & \frac{\dot{M}}{3 \times 10^{-8} (M / M_\odot) \ M_\odot \ {\rm yr^{-1}}}, \\
  r^\prime & \equiv & \frac{r}{3 r_{\rm S}}, 
\end{eqnarray}
where $r_{\rm S}$, the Schwarzschild radius, is given by,
\begin{equation}
    r_{\rm S} = \frac{2 GM}{c^2}.
\end{equation}
These scaling are evidently intended for black hole accretion problems. The Schwarzschild radius is the radius of the event horizon for a non-rotating black hole, which has an innermost stable orbit at $r = 3 r_{\rm S}$. The accretion rate scaling corresponds roughly to the Eddington limit, which is also most directly relevant to black hole and other energetic accretion environments (see~\ref{sec_Eddington}). Nonetheless, these are fundamentally Newtonian solutions, which can be rewritten with scalings more appropriate to, e.g., white dwarfs, without difficulty. Denoting the inner disk with a subscript $i$, the scale height, surface density, and central temperature are,
\begin{eqnarray}
 h_i & = & 3.2 \times 10^6 \dot{m} m f \ {\rm cm}, \nonumber \\
 \Sigma_i & = & 4.6 \alpha^{-1} \dot{m}^{-1} {r^\prime}^{3/2} f^{-1} \ {\rm g \ cm^{-2}}, \nonumber \\
 T_{ci} & = & 2.3 \times 10^7 \alpha^{-1/4} m^{-1/4} {r^\prime}^{-3/4} \ {\rm K}.
\end{eqnarray}
In these expressions, $f \equiv \left( 1 - {r^\prime}^{-1/2} \right)$. 
For the middle disk, denoted with a subscript $m$,
\begin{eqnarray}
 h_m & = & 1.2 \times 10^4 \alpha^{-1/10} \dot{m}^{1/5} m^{9/10} {r^\prime}^{21/20} f^{1/5} \ {\rm cm}, \nonumber \\
 \Sigma_m & = & 1.7 \times 10^5 \alpha^{-4/5} \dot{m}^{3/5} m^{1/5} {r^\prime}^{-3/5} f^{3/5} \ {\rm g \ cm^{-2}}, \nonumber \\
 T_{ci} & = & 3.1 \times 10^8 \alpha^{-1/5} \dot{m}^{2/5} m^{-1/5} {r^\prime}^{-9/10} f^{2/5} \ {\rm K}.
\end{eqnarray}
For the outer disk, denoted with a subscript $o$,
\begin{eqnarray}
 h_o & = & 6.1 \times 10^3 \alpha^{-1/10} \dot{m}^{3/20} m^{9/10} {r^\prime}^{9/8} f^{3/20} \ {\rm cm}, \nonumber \\
 \Sigma_o & = & 6.1 \times 10^5 \alpha^{-4/5} \dot{m}^{7/10} m^{1/5} {r^\prime}^{-3/4} f^{7/10} \ {\rm g \ cm^{-2}}, \nonumber \\
 T_{co} & = & 8.6 \times 10^7 \alpha^{-1/5} \dot{m}^{3/10} m^{-1/5} {r^\prime}^{-3/4} f^{3/10} \ {\rm K}.
\end{eqnarray}
The transition radii between the regimes ($r_{mi}^\prime$ for the inner to middle disk, $r^\prime_{om}$ for the middle to outer disk) are given implicitly by solving,
\begin{eqnarray}
 r^\prime_{mi} & = & 150 \left( \alpha m \right)^{2/21} \dot{m}^{16/21} \left( 1 - {r^\prime}_{mi}^{-1/2} \right)^{16/21}, \\
 r^\prime_{om} & = & 6.3 \times 10^3 \dot{m}^{2/3} \left( 1 - {r^\prime}_{om}^{-1/2} \right)^{2/3}.
\end{eqnarray}
The dependence on $\alpha$ is weak, so these radii are mostly dependent on the accretion rate and black hole mass (bearing in mind that the latter can vary across many orders of magnitude).

\begin{figure}
    \centering
    \includegraphics[width=\columnwidth]{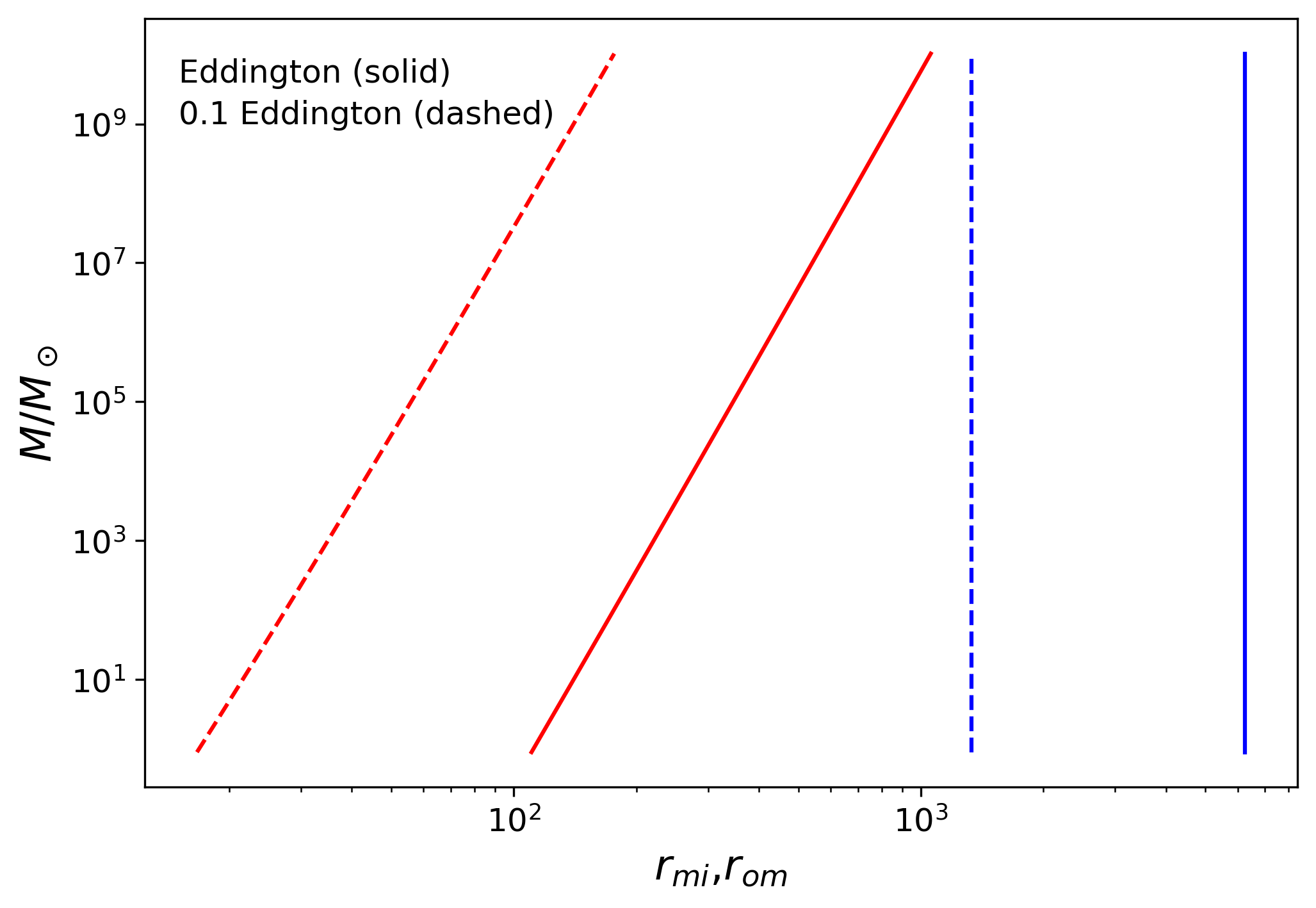}
    \caption{Transition radii $r^\prime_{mi}$ (red: between the inner radiation pressure dominated disk and the middle gas pressure / electron scattering dominated disk) and $r^\prime_{om}$ (blue: between the middle gas pressure / electron scattering dominated disk and the outer gas pressure / free-free opacity dominated disk), plotted for different black hole masses in the Shakura-Sunyaev solution. The solid lines assumes $\dot{m} =1$, the dashed lines $\dot{m}=0.1$, in both cases for $\alpha=0.1$.}
    \label{fig_SS_regions}
\end{figure}

Figure~\ref{fig_SS_regions} shows the dependence of $r^\prime_{mi}$ and $r^\prime_{om}$ as a function of black hole mass. For a stellar mass black hole with $M = 10 \ M_\odot$, and $\dot{m} = 1$, $r^\prime_{mi} \approx 140$ (i.e. $840 \ GM/c^2$) and $r^\prime_{om} \approx 6250$. For a supermassive mass black hole with $M = 10^7 \ M_\odot$, again assuming $\dot{m} = 1$, $r^\prime_{mi} \approx 540$ and $r^\prime_{om} \approx 6250$. In dimensional units, in the supermassive case the transition from radiation to gas pressure occurs at about $5 \times 10^{15} \ {\rm cm}$, while the transition from electron scattering to free-free opacity is at about $6 \times 10^{16} \ {\rm cm}$ (0.02~pc). A lower accretion rate of $\dot{m} = 0.1$ moves both radii in by a factor of 5-6. Qualitatively we observe that (1) all three Shakura-Sunyaev regions are predicted to be present for reasonable parameter choices, and that (2) radiation pressure is relatively more important for disks around supermassive black holes as compared to stellar mass examples.

\begin{figure}
    \centering
    \includegraphics[width=\columnwidth]{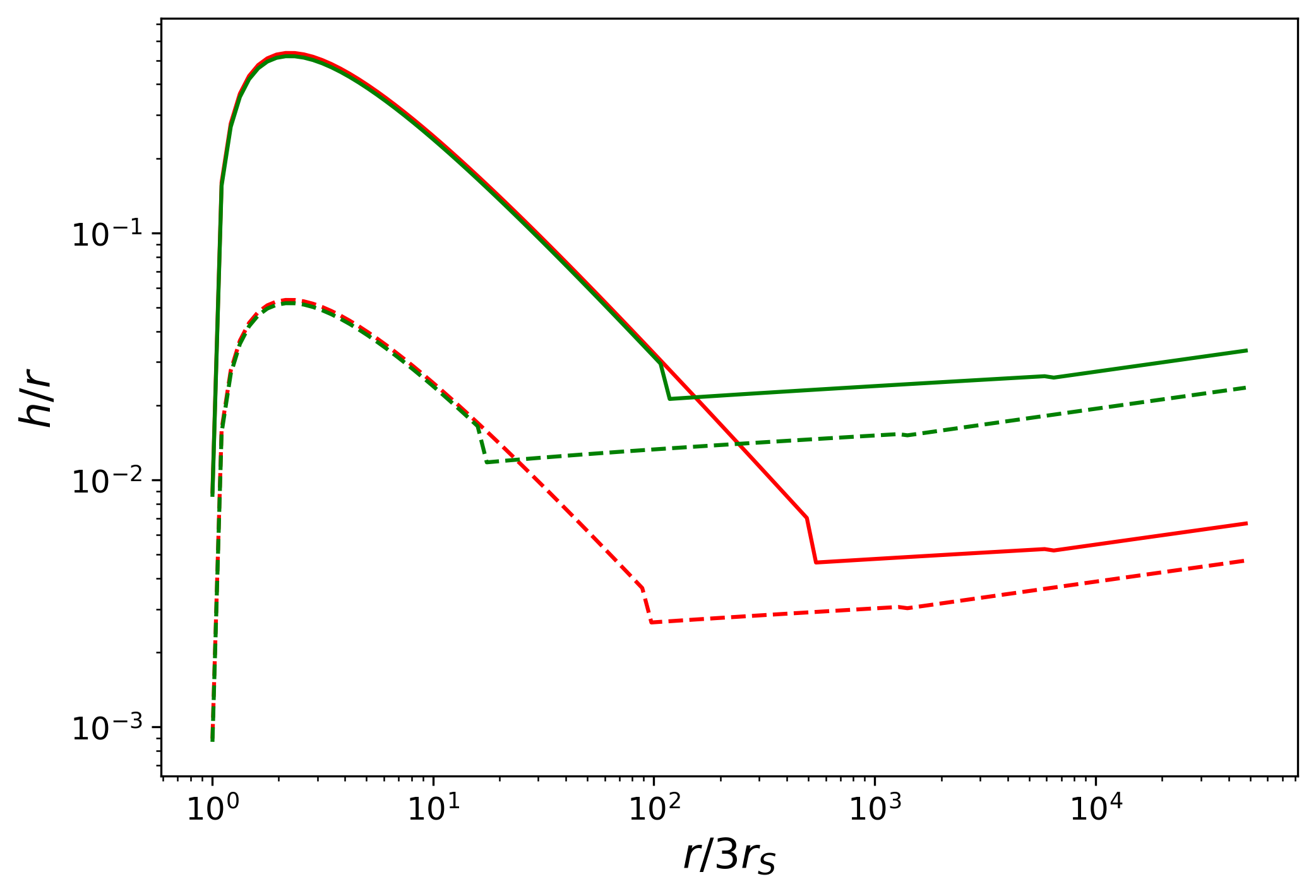}
    \caption{The geometric thickness of Shakura-Sunyaev disks as a function of dimensionless disk radius $r^\prime = r/ 3 r_{\rm S}$ is shown for supermassive ($m=10^7$, red) and stellar mass ($m=10$, green) black holes. The solid lines show high accretion rate solutions ($\dot{m}=1$), the dashed lines a lower accretion rate ($\dot{m}=0.1$). In all cases $\alpha=0.1$. The geometric thickness of the inner radiation pressure dominated region depends strongly on the accretion rate, and can be quite large for high accretion rates. The middle and outer disk regions are much thinner, especially for disks around supermassive black holes, where values of $h/r \sim {\rm few} \times 10^{-3}$ are characteristic. (The small discontinuities in the plotted curves are due to approximations and do not have any physical significance.)}
    \label{fig_SS_hr}
\end{figure}

Figure~\ref{fig_SS_hr} shows the radial dependence of the predicted geometric thickness of some selected Shakura-Sunyaev disk solutions. Although these are ``thin" disk solutions, the region where radiation pressure dominates (and $h$ is constant) is not actually thin at all for $\dot{m} \sim 1$. Values of $h/r$ significantly in excess of 0.1 are predicted at $r \sim 10-20 \ GM/c^2$. The resulting violation of the assumptions underlying thin disk theory is remedied in {\em slim accretion disk} models \citep{abramowicz88}, which should really be used to give a consistent treatment of this region. Conversely, the gas pressure dominated middle and outer regions of the Shakura-Sunyaev solution are quite thin, especially in the case of supermassive black holes. For $M \gtrsim 10^7 \ M_\odot$ we expect $10^{-3} < h/r < 10^{-2}$ in these regions. As a consequence, disk self-gravity (\S\ref{sec_selfgravity}, \ref{sec_SG_alpha}) is predicted to become important for disk masses that are much smaller than the mass of the black hole. The onset of self-gravity and the likelihood of ensuing fragmentation, in turn, has far-reaching consequences for the radial extent of AGN disks, for the formation of stars and compact objects {\em within} the accretion disk, and for how supermassive black holes are fuelled and grow \citep{shlosman90,goodman03,king06,levin07}.

\citet{novikov73} generalized the Newtonian Shakura-Sunyaev thin disk solution to include relativity. The Novikov-Thorne solution does not introduce any novelties in its treatment of the disk physics, but the proper inclusion of all the relativistic effects is just as tricky as you might expect. There are plenty of opportunities for making mistakes (even the original authors, no slouches when it comes to relativity, didn't get it quite right). I recommend \citet{abramowicz13} as a source for the explicit solution, and as a starting point for reading the literature on Novikov-Thorne disks.

\section{Energetics of disk accretion}
The thin disk solutions are predicated on two assumptions: that the energy released by accretion can be radiated on a time scale that is short compared to the local inflow time scale, and that the outgoing radiation has a negligible impact on the flow dynamics. We now turn to what happens when these assumptions fail. Radiation becomes dynamically important when the luminosity of the disk (or that of the central object) reaches the {\em Eddington limit}. Radiative cooling ceases to be efficient both when the accretion rate is very low, due to plasma physics effects (the regime of {\em radiatively inefficient accretion}), and when it is very high, due to photon trapping (the regime of {\em hyperaccretion}).

\subsection{Eddington limit}
\label{sec_Eddington}
The Eddington limit is the luminosity at which radiation pressure from a central point source balances gravity, curtailing spherically symmetric accretion. Noting that photons of energy $E$ carry momentum $p=E/c$, the momentum flux at distance $r$ from an isotropic source with luminosity $L$ is $L / (4 \pi c r^2$). If the opacity of the gas is $\kappa$, the outward radiative force per unit mass of gas is,
\begin{equation}
    f_{\rm rad} = \frac{\kappa L}{4 \pi c r^2}.
\end{equation}
Equating to the inward force per unit mass of gravity,
\begin{equation}
    f_{\rm grav} = \frac{GM}{r^2},
\end{equation}
the Eddington limiting luminosity is,
\begin{equation}
    L_{\rm Edd} = \frac{4 \pi c G M}{\kappa}.
\end{equation}
At sufficiently high temperatures the opacity is due to Thomson (electron) scattering, and $\kappa = \sigma_{\rm T} / m_{\rm H}$, where $\sigma_{\rm T} = 6.7 \times 10^{-25} \ {\rm cm^2}$ is the Thomson cross-section and $m_{\rm H} = 1.66 \times 10^{-24} \ {\rm g}$ is the mass of a hydrogen atom. (The radiative force acts on the electrons, but these are tightly coupled to the protons electrostatically.) Under these conditions, numerically,
\begin{eqnarray}
    L_{\rm Edd} & = & \frac{4 \pi c m_{\rm H} GM}{\sigma_{\rm T}}, \\
        & \simeq & 1.2 \times 10^{38} \left( \frac{M}{M_\odot} \right) \ {\rm erg \ s^{-1}}, \\
        & \simeq & 3.2 \times 10^4 \left( \frac{M}{M_\odot} \right) \ L_\odot.
\label{eq_Ledd}        
\end{eqnarray}
A modest correction is needed for Thomson scattering on gas that is not pure hydrogen, with a larger one being in order if the surrounding gas is cool and the opacity is due to dust. Dust opacity can be relevant to accretion during massive star formation, and to the dynamics of gas at relatively large distances from supermassive black holes in AGN.

It is hopefully obvious that the Eddington limit, derived assuming spherical symmetry, is best-regarded as a characteristic luminosity above which radiative forces are guaranteed to matter for surrounding accretion flows. As with the Pirate's Code in {\em Pirates of the Caribbean}, the Eddington limit is more what you call a guideline than an actual rule.

\subsection{Radiative efficiency of thin disks}
The Newtonian estimate for the luminosity of accretion at rate $\dot{M}$ onto a star with mass $M$ and radius $R_*$ is just,
\begin{equation}
    L = \frac{GM\dot{M}}{R_*}.
\end{equation}
For a black hole the presence of an event horizon means that a fraction of the potential energy liberated by accretion can, in principle, be ``lost" across the horizon, augmenting the mass of the hole as measured at large distance. The luminosity associated with black hole accretion then depends on the mode of accretion, and on some uncertain MHD physics close to the hole. The standard estimate is based upon the assumption that the black hole accretes from a disk that is geometrically thin, with a zero-torque boundary condition at an innermost radius that is close to the hole.

\begin{figure}
    \centering
    \includegraphics[width=\columnwidth]{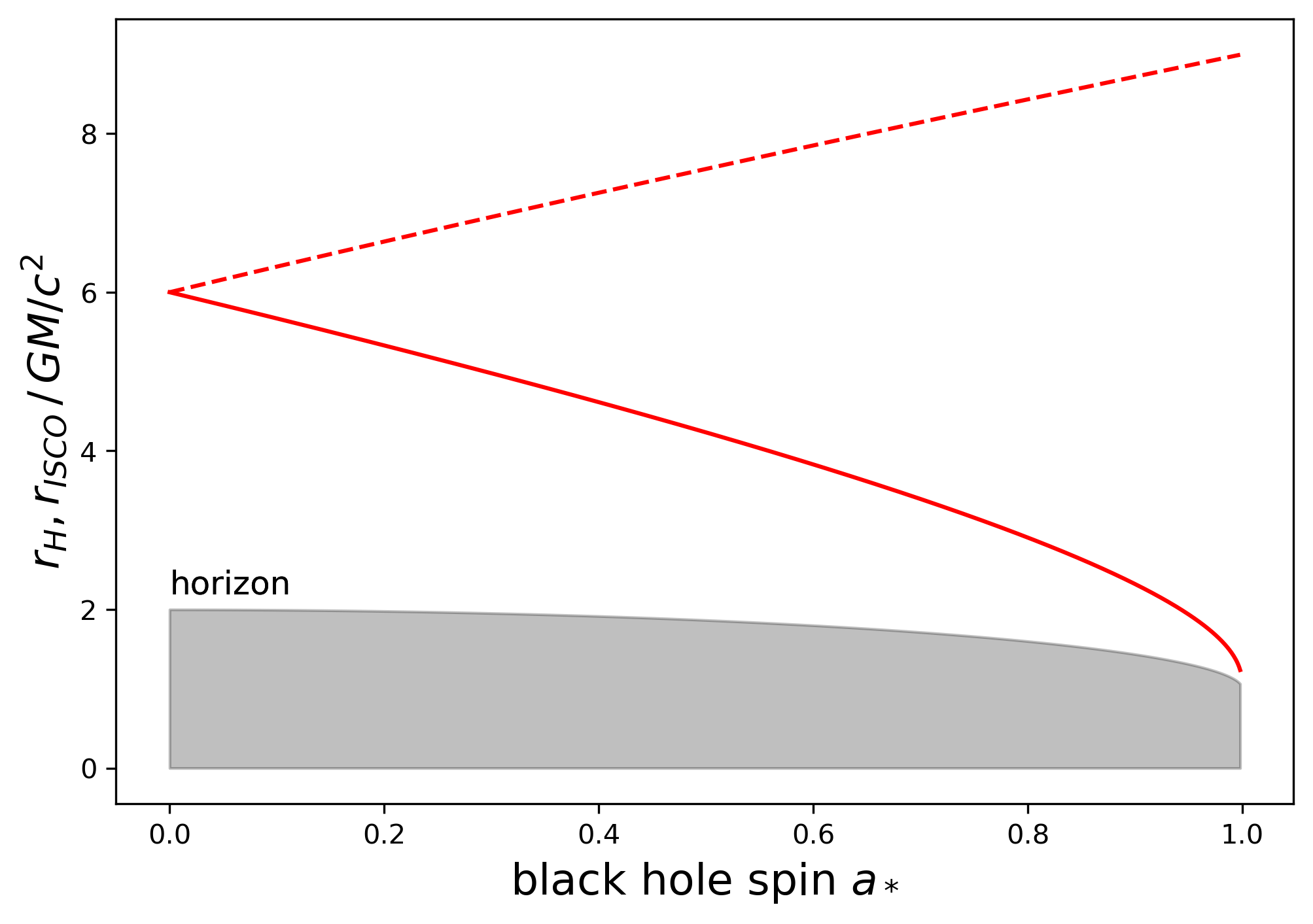}
    \caption{The horizon radius, and that of the innermost stable circular orbit for prograde (solid) and retrograde (dashed) orbits, is plotted as a function of the dimensionless black hole spin parameter $a_*$.}
    \label{fig_ISCO}
\end{figure}

The fiducial estimate for the radiative efficiency of thin disks relies on various properties of Kerr black holes \citep{kerr63}. For a rotating, uncharged, black hole, we define the dimensionless spin parameter $a_*$ in terms of the mass $M$ and angular momentum $J$ via,
\begin{equation}
    a_* \equiv \frac{cJ}{GM^2}.
\end{equation}
The spin is limited to $0 \leq a_* < 1$ (or $-1 < a_* < 1$ using negative values to denote orbits that are counter-rotating with respect to the spin). Amongst the many important properties of the Kerr metric are the radius of the event horizon,
\begin{equation}
    r_{\rm H} = \left( 1 + \sqrt{1-a_*^2} \right) \frac{GM}{c^2},
\end{equation}
and the radii of the innermost stable circular orbits, which differ depending on whether the orbits are co-rotating or counter-rotating in the equatorial plane of the hole \citep{bardeen72,shapiro83},
\begin{eqnarray}
    r_{\rm ISCO} &=& \left[ 3 + Z_2 \mp \sqrt{ (3-Z_1)(3+Z_1+2Z_2)} \right] \frac{GM}{c^2}, \nonumber \\
    Z_1 &\equiv& 1 + \left( 1-a_*^2 \right)^{1/3} 
    \left[ \left( 1 + a_* \right)^{1/3} + \left( 1 - a_* \right)^{1/3} \right], \nonumber \\
    Z_2 &\equiv& \sqrt{ 3 a_*^2 + Z_1^2}.
\end{eqnarray}
These quantities are plotted in Figure~\ref{fig_ISCO}. The innermost stable circular orbit lies at $6 GM/c^2$ for a non-rotating (Schwarzschild) black hole, and tends towards $GM/c^2$ and $9 GM / c^2$ for co-rotating and counter-rotating orbits respectively as $a_* \rightarrow 1$. The variation of $r_{\rm ISCO}$ with $a_*$ provides the physical underpinning for various black hole spin estimates derived from electromagnetic observables \citep{reynolds20}.

\begin{figure}
    \centering
    \includegraphics[width=\columnwidth]{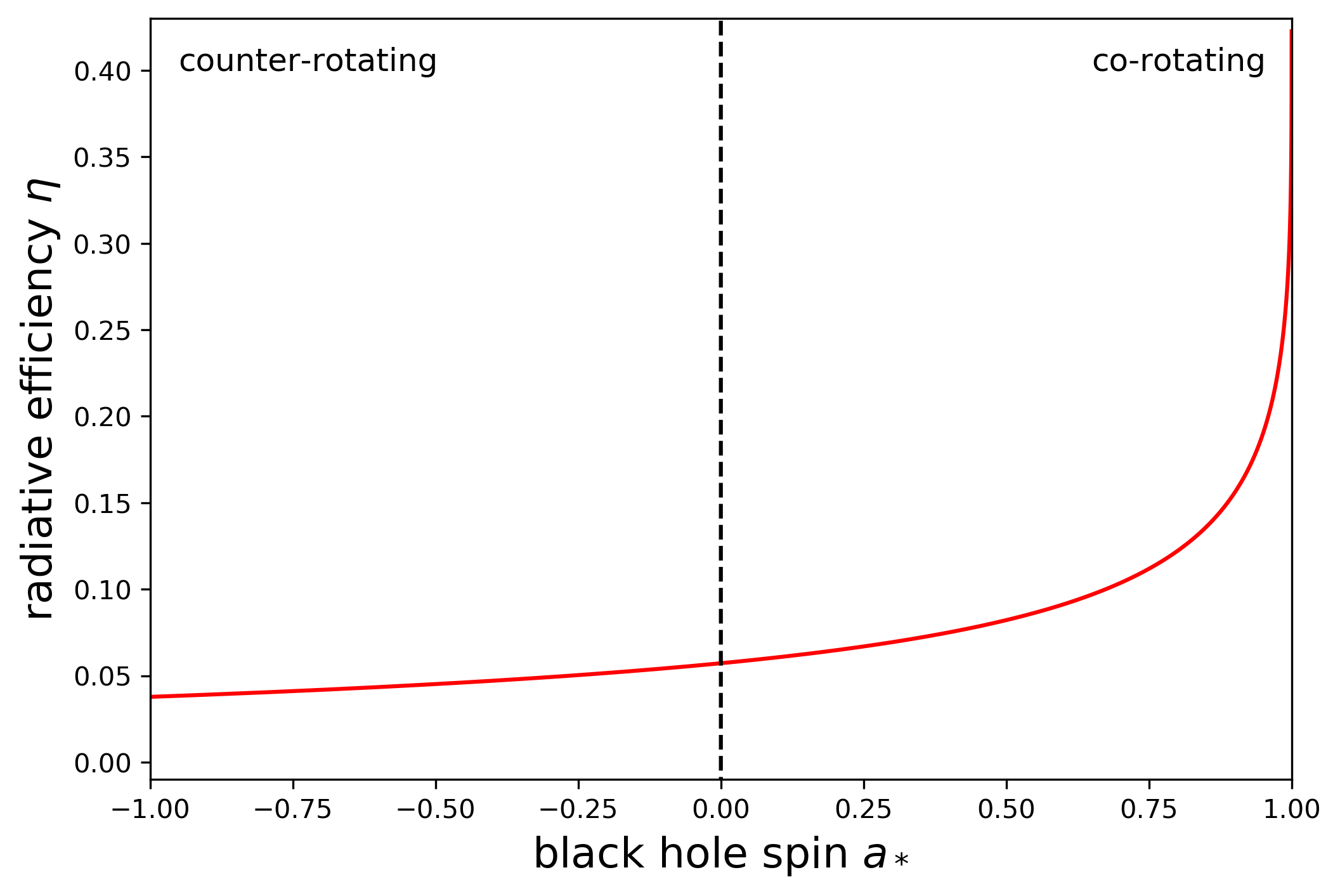}
    \caption{The nominal radiative efficiency of thin disk accretion is plotted as a function of the dimensionless black hole spin parameter $a_*$. Negative values of $a_*$ denote accretion from equatorial counter-rotating orbits, positive values accretion from equatorial co-rotating orbits. For $a_* = 0$ the efficiency $\eta = 1 - \sqrt{8/9} = 0.0572$, for $a_* = -1$ we have $\eta = 1 - \sqrt{25/27} = 0.0377$, while for $a_* = 1$ (this value of the spin is not quite physically realizable) $\eta = 1 - \sqrt{1/3} = 0.423$.}
    \label{fig_efficiency}
\end{figure}

Armed with these results, we can define the fiducial estimate of the radiative efficiency properly. Assume that a thin disk extends from large radius, where the mass accretion rate is $\dot{M}$, down to the innermost stable circular orbit, where a zero-torque boundary condition is applied. Gas interior to the disk---between $r_{\rm ISCO}$ and $r_{\rm H}$---is assumed to neither radiate nor to exert any feedback effects on the disk flow. The radiative efficiency $\eta$, defined via,
\begin{equation}
    L = \eta \dot{M} c^2,
\end{equation}
is then fully specified by the value of the binding energy of particle orbits at $r_{\rm ISCO}$. Defining the ancillary variable $E=1-\eta$ the efficiency is found by solving the equation \citep{shapiro83},
\begin{equation}
    a_* = \mp \frac{ 4 \sqrt{2} \left( 1-E^2 \right)^{1/2} - 2E }{3 \sqrt{3} \left( 1 - E^2 \right)}.
\end{equation}
The resulting function $\eta(a_*)$ is shown in Figure~\ref{fig_efficiency}. For $a_* = 0$ the efficiency is 5.7\% (i.e. that fraction of the rest mass of the accreting gas is radiated from the disk), while as $a_* \rightarrow 1$ the limiting efficiency is 42\%.\footnote{Although mechanically it is possible to spin up a Kerr black hole arbitrarily close to $a_* = 1$, a black hole spun up by disk accretion necessarily consumes disk photons on orbits that counteract the spin up. \citet{thorne74} found that this effect limits the maximum spin to $a_* \simeq 0.998$, with a radiative efficiency $\eta \simeq 0.3$.}

\subsubsection{Salpeter time}
The results for the Eddington limit and the radiative efficiency of disk accretion can be combined to give an estimate of how rapidly black holes can grow from thin disk accretion. To do so, assume that the accretion rate is limited to the value that would yield an Eddington-limited luminosity. For a black hole of mass $M_{\rm BH}$,
\begin{equation}
    \eta \dot{M} c^2 = \frac{4 \pi c m_{\rm H} G}{\sigma_{\rm T}} M_{\rm BH}.
\end{equation}
Subtracting the rest-mass equivalent of the energy lost via radiation, accretion at rate $\dot{M}$ increases the mass of the black hole according to,
\begin{equation}
    \dot{M}_{\rm BH} = \left( 1 - \eta \right) \dot{M}.
\end{equation}
A black hole that always grows as fast as it can---at the Eddington limit---then obeys,
\begin{equation}
    \frac{{\rm d}M_{\rm BH}}{{\rm d}t} = \frac{4 \pi \left(1-\eta\right)}{\eta} \frac{m_{\rm H}G}{c \sigma_{\rm T}} M_{\rm BH}.
\end{equation}
At fixed radiative efficiency (i.e. at fixed spin) the result is exponential growth,
\begin{equation}
    M_{\rm BH} = M_0 \exp \left[ \frac{t}{t_{\rm S}} \right],
\end{equation}
where $M_0$ is some initial mass and $t_{\rm S}$, the {\em Salpeter time} \citep{salpeter64}, is a characteristic time scale,
\begin{equation}
    t_{\rm S} = \frac{\eta}{4 \pi \left( 1 - \eta \right)} \frac{c \sigma_{\rm T}}{m_{\rm H} G}.
\end{equation}
For $\eta = 0.1$, the Salpeter time is about 50~Myr.

The Salpeter time is a useful characteristic time scale for how fast black holes accreting from geometrically thin disks grow. You will often see it referenced as part of an argument about supermassive black hole formation, which goes as follows. We observe luminous quasars, likely hosting black holes with masses $M_{\rm BH} \sim 10^9 \ {M_\odot}$, at redshifts exceeding $z=7$ \citep{mortlock11,wang21}. If we assume (say) that the seed for such a quasar formed at $z=20$, the time available between $z=20$ and $z=7$ for it to grow is only approximately $5.9 \times 10^{8} \ {\rm yr}$. This is 11.6 Salpeter times (for $\eta=0.1$), while 18 e-foldings are needed to grow from a $10 \ M_\odot$ stellar mass black hole to a $10^9 \ M_\odot$ supermassive black hole. The observed early growth of massive black holes in the Universe appears to pose problems, or at least to provide strong constraints on the masses of black hole seeds \citep[which may not be stellar mass black holes at all;][]{begelman06}.

This argument, in my opinion, is rather tired from overuse, and in its strong form requires undue faith in the Eddington limit being a strict limit. However, it does justify the weaker statement that the existence of high redshift quasars requires a high duty cycle of prior accretion, at a rate high enough to imply that radiative forces are important.

A separate argument, credited to \citet{soltan82}, takes off from the observation that the mass accumulated in supermassive black holes, and the total amount of energy radiated from accretion during their growth, are two sides of the same coin. If supermassive black holes grow primarily via thin disk accretion, the total mass in black holes per comoving Mpc$^3$ at $z=0$ is related to the integral of the AGN luminosity function over redshift, and an appropriate comparison between the two constrains the radiative efficiency \citep{yu02}.

\subsection{Electron-ion coupling in low density plasmas}
The validity of thin disk solutions for black hole accretion is bounded above at $\dot{m} \sim 1$, both by the onset of radiatively driven outflows as the Eddington limit is exceeded and by the importance of radial advection of energy as $h/r$ becomes larger. In the other direction, thin disk solutions remain internally consistent for $\dot{m} \ll 1$, but they are not unique. A hot, geometrically thick disk solution also exists below a critical accretion rate. The physical origin of the hot solution is tied to an asymmetry in the microphysics of electron-ion plasmas. Heating, due to small-scale dissipation of turbulent energy, gives energy predominantly to the {\em ions}, while cooling, due to processes such as free-free emission and synchrotron radiation, is much more efficient for {\em electrons}. At low density, the time scale for Coulomb collisions to transfer energy from the ions to the electrons becomes long---in some cases longer than the time scale for the gas to accrete on to the black hole. Unable to cool efficiently, the accretion flow becomes (or remains) geometrically thick \citep{shapiro76,ichimaru77,rees82}. It resembles a torus or a doughnut more than a disk.

\begin{figure}
    \centering
    \includegraphics[width=\columnwidth]{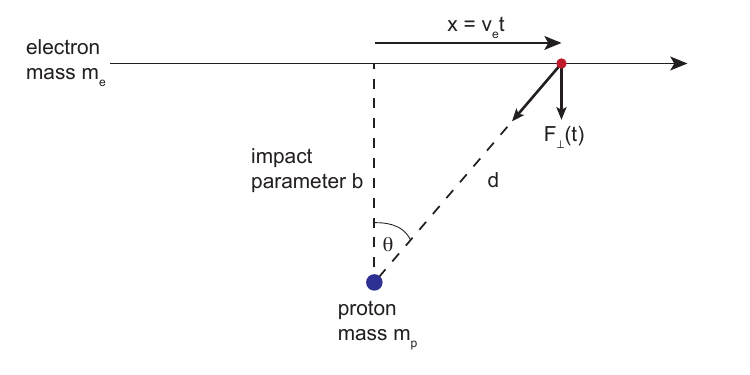}
    \caption{Setup for computing the transverse momentum imparted by an electron-proton encounter in the impulse approximation.}
    \label{fig_impulse}
\end{figure}

The time scale for electron-proton thermal equilibration can be computed to better accuracy than we need via elementary methods. Step one is to calculate the transverse momentum that is imparted when an electron, with mass $m_e$, flies by a proton, with mass $m_p$, at velocity $v_e$ and impact parameter $b$. The setup is shown in Figure~\ref{fig_impulse}. We work in the {\em impulse approximation}, and compute the momentum change along the unperturbed straight-line trajectory of the electron. The force perpendicular to the direction of motion is,
\begin{equation}
    F_\perp = \frac{e^2}{d^2} \cos \theta = \frac{e^2 b}{\left( b^2 + x^2 \right)^{3/2}}.
\end{equation}
The total momentum imparted in the perpendicular direction as a result of the encounter is just the integral of the force over time,
\begin{equation}
    \Delta p_\perp = \int_{-\infty}^{\infty} F_\perp (t) {\rm d}t = 
    \frac{e^2}{b^2 v_e} \int_{-\infty}^{\infty} \left( 1 + x^2 / b^2 \right)^{-3/2} {\rm d}x.
\end{equation}
A quick trignometric substitution, or an even quicker trip to {\em Wolfram Alpha}, and we have the answer,
\begin{equation}
    \Delta p_\perp = \frac{2 e^2}{b v_e}.
\end{equation}
With this result in hand we proceed to step two, which is to sum up the effect of many encounters occurring with a range of impact parameters $b_{\rm min} < b < b_{\rm max}$. The mean-square momentum change is the lowest order quantity that does not vanish. If the number density of particles is $n$,
\begin{eqnarray}
    \left\langle \frac{\rm d}{{\rm d}t} \left( \Delta p_\perp \right)^2 \right\rangle 
    & = & \int_{b_{\rm min}}^{b_{\rm max}} 2 \pi b n v_e \left( \frac{2 e^2}{b v_e} \right)^2 {\rm d}b \nonumber \\
    & = & \frac{8 \pi n e^4}{v_e} \int_{b_{\rm min}}^{b_{\rm max}} \frac{{\rm d}b}{b}.
\end{eqnarray}
The integral gives us $\ln (b_{\rm max} / b_{\rm min})$, which goes by the name of the {\em Coulomb logarithm} and is written as $\ln \Lambda$. The same quantity crops up, for the same reasons, in discussions of relaxation in gravitating point-mass systems such as globular clusters. In a plasma, $b_{\rm max}$ should be set to be the {\em Debye length} \citep[e.g.][]{thorne17},
\begin{equation}
    \lambda_D = \left( \frac{kT}{4 \pi n e^2} \right)^{1/2},
\end{equation}
because this is the spatial scale on which charges are effectively screened. The lower limit, $b_{\rm min}$, is either the impact parameter where a single deflection $\Delta \theta \sim 1$~radian (as in the gravitational case), or the particle's de Broglie wavelength, whichever is the larger \citep{spitzer62}. All this can be calculated properly as a function of density and temperature, but for our purposes the details are neither very interesting nor very important. The answer depends on $b_{\rm max} / b_{\rm min}$ only logarithmically, and taking $\ln \Lambda \approx 20$ is a good enough approximation.

For the final step, we switch from thinking about momentum to thinking about energy. The electron, with kinetic energy $(1/2)m_e v_e^2$, changes the energy of the proton by an amount $(\Delta p_\perp)^2 / 2 m_p$ in a single encounter. Considering all encounters, the equilibration time is,
\begin{equation}
    t_{ep} = \frac{m_e v_e^2}{\langle {\rm d}/{\rm d}t ( \Delta p_\perp )^2 / m_p \rangle} = \frac{m_e m_p v_e^3}{8 \pi n e^4 \ln \Lambda}.
\label{eq_tep_1}    
\end{equation}
Let us assume that the typical electron velocity is,
\begin{equation}
    v_e = \left( \frac{3 k_{\rm B}T_e}{m_e} \right)^{1/2}.
\end{equation}
Then, noting that the Thomson cross-section \citep{rybicki79},
\begin{equation}
    \sigma_{\rm T} = \frac{8 \pi}{3} \frac{e^4}{m_e^2 c^4},
\end{equation}
and defining a parameter expressing how relativistic the electrons are,
\begin{equation}
    \theta_e \equiv \frac{k_{\rm B}T_e}{m_e c^2},
\end{equation}
equation~(\ref{eq_tep_1}) becomes,
\begin{equation}
    t_{ep} = \frac{\sqrt{3}}{\sigma_{\rm T}c n \ln \Lambda} \left( \frac{m_p}{m_e} \right) \theta_e^{3/2}.
\end{equation}
The rather sloppy methods we have employed here get us surprisingly close to the answer obtained by more diligent investigators. A two-temperature electron-proton plasma, where the two species have Maxwellian velocity distributions, comes to a common temperature on a time scale\footnote{In the plasma context this result is usually cited as \citet{spitzer62}, but Spitzer's textbook passes the buck to his paper on stellar dynamics \citep{spitzer40} for the details of the calculation. That's fine because, up to the question of the correct value of $\ln \Lambda$, the gravitational and plasma calculations are the same.},
\begin{equation}
    t_{ep} = \frac{\sqrt{2 \pi}}{2 \sigma_{\rm T}c n \ln \Lambda} \left( \frac{m_p}{m_e} \right) \left( \theta_e + \theta_p \right)^{3/2},
\label{eq_ep_equilibration}    
\end{equation}
where $\theta_p \equiv k_{\rm B}T_p / m_p c^2$. When $\theta_e \gg \theta_p$ this differs from our rough and ready result by only a modest numerical pre-factor.

\subsection{Radiatively inefficient accretion models}
A geometrically thick disk solution is possible if the density is low enough that the time scale for electron-ion thermal equilibration exceeds the time scale for gas to flow into the black hole. This can be written as an upper limit on the accretion rate, in units of the accretion rate that would generate an Eddington limiting luminosity, that depends only on $\alpha$. To obtain this result, we consider a quasi-spherical but rotationally supported accretion flow, such that $n=n(r)$ only but $v_r$ is set by the local rate of angular momentum transport. The continuity equation is,
\begin{equation}
    \dot{M} = - 4 \pi r^2 v_r \rho.
\end{equation}
Substituting the order of magnitude estimates, $v_r = - r / t_{\rm visc}$ and $t_{\rm visc} \sim 1 / (\alpha \Omega_{\rm K})$ (remember that $h/r \sim 1$, by assumption), and using $\rho = n m_{\rm H}$, gives,
\begin{equation}
    n(r) \sim \frac{\dot{M}}{4 \pi m_{\rm H} \alpha \Omega_{\rm K} r^3}.
\end{equation}
We now substitute this expression for $n$ into equation~(\ref{eq_ep_equilibration}) and impose the physical condition that,
\begin{equation}
    t_{ep} \gtrsim t_{\rm visc}.
\end{equation}
Although the electrons cool to lower temperatures than the protons, they're also {\em much} less massive, so we can assume that $\theta_e > \theta_p$ and drop the latter. The inequality gives,
\begin{equation}
    \dot{M} \lesssim \frac{\sqrt{2 \pi}}{2 \ln \Lambda} 
    \frac{4 \pi m_{\rm H} GM \alpha^2}{\sigma_{\rm T} c} \left( \frac{m_p}{m_e} \right) 
    \theta_e^{3/2},
\end{equation}
which is approximately as clear as mud. The expression simplifies when we note that many of the quantities on the right-hand-side also appear in the formula for the Eddington limit (equation~\ref{eq_Ledd}). Let us define an Eddington {\em mass} accretion rate via, 
\begin{equation}
    L_{\rm Edd} = \eta \dot{M}_{\rm Edd} c^2,
\end{equation}
such that,
\begin{equation}
    \dot{M}_{\rm Edd} = \frac{4 \pi m_{\rm H} GM}{\eta c \sigma_{\rm T}}.
\label{eq_mdot_Edd}    
\end{equation}
(Note that some authors define $\dot{M}_{\rm Edd}$ without the factor of $\eta$, or fixing $\eta=0.1$.) The inequality then simplifies to,
\begin{equation}
    \frac{\dot{M}}{\dot{M}_{\rm Edd}} \lesssim 
        \sqrt{\frac{\pi}{2}} \frac{\eta}{\ln \Lambda} \left( \frac{m_p}{m_e} \right) 
        \theta_e^{3/2} \alpha^2.
\end{equation}
It's not easy to give an elementary estimate for $\theta_e$, but adopting reasonable values for the parameters, $\theta_e = 0.1$, $\eta=0.1$, and $\ln \Lambda = 20$, the result is,
\begin{equation}
    \frac{\dot{M}}{\dot{M}_{\rm Edd}} \lesssim 0.4 \alpha^2.
\end{equation}
Our argument has been a simplified version of that given in \citet{mahadevan97}, but there are multiple routes to this result. If $\alpha \sim 0.1$, the conclusion is that a geometrically thick, two-temperature accretion flow, is a consistent possibility provided that the accretion rate, measured in Eddington units, is below $10^{-3}$--$10^{-2}$.

The way in which plasma microphysics and flow macrophysics combine to give a simple result for the critical accretion rate is quite appealing. It should be obvious that our derivation is only good to an order of magnitude, but there are also other caveats. The fact that a hot two-temperature flow {\em can} exist below a critical accretion rate does not mean that it {\em must}---in principle a denser thin disk could be a valid solution below the critical rate. There has also been sporadic discussion over many years as to whether the plasma physics fundamentals underpinning the argument---that dissipation primarily heats the ions \citep{quataert98,gruzinov98}, and that Coulomb coupling is the sole channel for ion-electron energy transfer \citep{begelman88}---are secure. The review by \citet{yuan14} is a good place to start for a discussion of these questions, and the extent to which the uncertain answers impact astrophysically interesting conclusions about hot accretion flows. A promising development is that the basic plasma physics involved is increasingly amenable to direct simulation using particle-in-cell techniques \citep{zhdankin19,schekochihin19,sironi15}.

\subsubsection{The limiting ADAF solution}
The microphysical inability of a low-density plasma to cool efficiently has two important consequences for the structure of low $\dot{m}$ accretion flows. The plasma, or at least the ions within it, becomes hot, and hence geometrically thick. The larger value of $h/r$, in turn, means that the inflow velocity $v_r$ becomes larger than it would be in a thin disk. These two properties mean that radial advection of heat is important, and can be dominant, for the structure of low accretion rate accretion flows. The moniker {\em Advection Dominated Accretion Flow} (ADAF) is sometimes used as generic term for radiatively inefficient accretion flows, though it can refer to the specific disk model introduced by \citet{narayan94,narayan95a,narayan95b}, whose work highlighted the key role of advection and kicked off a resurgence of interest in the properties of low accretion rate disks.

In general, geometrically thick disks require at least a two-dimensional spatial description (axisymmetric seems like it ought to be a reasonable simplification, with flow variables being functions of $r$ and polar angle $\theta$), and of course they may be time-dependent. Radial pressure gradients can not be ignored, the angular velocity need not be Keplerian, and the energy equation must include the advective term. This all adds up to many additional degrees of freedom than we have for thin disks. We may abandon all hope and turn to numerical simulations, but before reaching that point we can make some progress via aggressive simplification of the problem. \citet{narayan94}, in their original paper on ADAFs, assumed that the flow was axisymmetric, steady, and could be described via height-integrated equations rather than a full $(r,\theta)$ treatment (height-integrated here implying something more akin to averaging over spherical rather than cylindrical shells). The governing equations then read,
\begin{eqnarray}
    \frac{{\rm d}}{{\rm d}r} \left( \rho r h v_r \right) &=& 0, \\
    v_r \frac{{\rm d}v_r}{{\rm d}r} - \Omega^2 r &=& -\Omega_{\rm K}^2 r - \frac{1}{\rho} 
        \frac{\rm d}{{\rm d}r} \left( \rho c_s^2 \right), \label{eq_NY2} \\
    v_r \frac{{\rm d} (\Omega r^2)}{{\rm d}r} &=& \frac{1}{\rho r h} 
        \frac{\rm d}{{\rm d}r} \left( \nu \rho r^3 h 
        \frac{{\rm d}\Omega}{{\rm d}r} \right), \\
    2 \rho h v_r T \frac{{\rm d}s}{{\rm d}r} &=& Q^+ - Q^-.   
\end{eqnarray}
The symbols here have the same meanings as in the thin disk equations, but we've added an energy equation involving the temperature $T$, entropy $s$, and the rates of viscous dissipation $Q_+$ and radiative cooling $Q^-$ per unit area. \citet{narayan94} go on to derive a self-similar solution to these equations for a general $\gamma$, in terms of a parameter $f$ that measures the importance of the advective term relative to radiative cooling. 

The essence of the ADAF solution is already present in the limiting case where there is no radiative cooling whatsoever. This limit is amenable to a rather simple analysis. Following \citet{blandford99}, we consider a quasi-spherical accretion flow with angular velocity $\Omega(r)$, density $\rho(r)$, and isothermal sound speed $c_s(r)$. Angular momentum transport is assumed to be inefficient enough that $v_r \ll \Omega r$. In a steady-state flow, with mass accretion rate $\dot{M}$ and inward-directed angular momentum flux $F_l$, we have,
\begin{eqnarray}
    -4 \pi r^2 v_r \rho & = & \dot{M}, \\
    \label{eq_BB1}
    \dot{M} r^2 \Omega - \cal{G} & = & F_l.
    \label{eq_BB2}
\end{eqnarray}
As before, $\cal{G}$ is the torque that the disk at radius $r$ exerts on the disk exterior to that point. Whereas for a thin disk we imagine thin annuli interacting through an effective viscous process, here we imagine coupling between spherical shells.

Another conserved quantity is derived from the energy equation. The total energy density has contributions from potential energy (assumed Newtonian),
kinetic energy (by assumption of slow inflow the relevant velocity is $\Omega r$), and enthalpy. The isothermal sound speed $c_s$ is given in terms of the enthalpy $h$ by,
\begin{equation}
    c_s^2 = \frac{(\gamma-1)}{\gamma} h,
\end{equation}
where $\gamma$ is the effective ratio of specific heats. If non-relativistic ions dominate the energy density, we'd expect $\gamma \simeq 5/3$ as usual for an ideal gas. Radiation pressure would give $\gamma = 4/3$, though this is not a limit that is so relevant to very low $\dot{m}$ flows. Accretion of gas containing energetically dominant small-scale magnetic fields would also give $\gamma \approx 4/3$. Leaving $\gamma$ as a parameter for now, the outward-directed energy flux $F_E$ is,
\begin{equation}
    {\cal{G}} \Omega - 
    \left( \frac{1}{2} r^2 \Omega^2 - \frac{GM}{r} + 
    \frac{\gamma}{\gamma-1} c_s^2 \right) \dot{M} = F_E.
    \label{eq_BB3}
\end{equation}
This expression has a viscous term and an advective term, but omits cooling which is assumed to be neglibly small by comparison.

We now assume that the accretion flow is {\em self-similar}, which means that it has no preferred scale and ``looks the same" when rescaled by a multiplicative factor. A black hole accretion flow has an inner boundary that is set by the radius of the event horizon / innermost stable circular orbit, and it must be fed at an outer scale by some physical process. These realities necessarily break self-similarity. If the inner and outer radii are separated by many orders of magnitude, however, it is reasonable to think that a flow that cannot cool at all ought to be well-approximated as self-similar across much of that radial interval\footnote{The same is obviously not true at all for thin disks. As you can see from Figure~\ref{fig_SS_hr}, radial variations in the dominant sources of pressure and opacity grossly break self-similarity.}. The self-similar assumptions implies that the fluid variables scale as,
\begin{eqnarray}
    \Omega (r)  & \propto & r^{-3/2}, \\
    v_r (r) & \propto & c_s (r) \propto r^{-1/2}.
\end{eqnarray}
The continuity equation then implies that $\rho(r) \propto r^{-3/2}$, and that the pressure $P=\rho c_s^2 \propto r^{-5/2}$. This allows us to simplify the radial momentum equation~(\ref{eq_NY2}). Dropping the $v_r ({\rm d}v_r / {\rm d}r)$ term because $v_r \ll \Omega r$, and inserting the self-similar scaling for the pressure, 
\begin{equation}
    r^2 \Omega^2 - \frac{GM}{r} + \frac{5}{2} c_s^2 = 0.
\label{eq_BB4}    
\end{equation}
With these simplifications the equations left to work with are (\ref{eq_BB1}), (\ref{eq_BB2}), (\ref{eq_BB3}) and (\ref{eq_BB4}).

We next observe that the term $\dot{M} r^2 \Omega$ on the left-hand side of equation~(\ref{eq_BB2}) scales as $r^{1/2}$ (as must $\cal{G}$), and that similarly the terms on the left-hand side of equation~(\ref{eq_BB3}) scale as $r^{-1}$. These equations can only be satisfied for arbitrary ranges of $r$ if the right-hand sides vanish, $F_l = F_E = 0$. Eliminating ${\cal G}$ between equation~(\ref{eq_BB2}) and equation~(\ref{eq_BB3}) we have,
\begin{equation}
    \frac{1}{2} r^2 \Omega^2 + \frac{GM}{r} - \frac{\gamma}{\gamma-1} c_s^2 = 0.
\end{equation}
Using equation~(\ref{eq_BB4}) the solution for the angular velocity and sound speed is,
\begin{eqnarray}
 c_s^2 & = & \frac{6 (\gamma-1)}{(9 \gamma - 5)} \frac{GM}{r}, \\
 \Omega^2 & = & \frac{2(5 - 3 \gamma)}{(9 \gamma - 5)} \Omega_{\rm K}^2.
\end{eqnarray}
The difference between this solution and the thin disk solutions we have discussed previously is clear. The angular velocity of ADAF solutions can be {\em strongly} sub-Keplerian, indeed $\Omega \rightarrow 0$ as $\gamma \rightarrow 5/3$.

To complete the solution $\cal{G}$ has to be specified. This can be done in various ways that are all consistent with the spirit of the Shakura-Sunyaev $\alpha$-prescription. \citet{blandford99}, for example, take ${\cal{G}} = 4 \pi r^3 \alpha P$, which can be seen as {\em defining} $\alpha$ for this type of flow. Other choices are possible too \citep{narayan94}. Once $\cal{G}$ is fixed, the density and radial velocity of the accretion flow are readily derived. The solution generalizes to the case where radiative cooling is a fixed but non-zero fraction of the viscous dissipation rate \citep{narayan94}\footnote{Essentially the same equations admit a {\em time-dependent} self-similar solution for an accretion flow of finite extent \citep{ogilvie99b}. Although this solution is not often discussed, it's a useful way to think about obviously time-dependent problems such as accretion that ensues following tidal disruption events.}. 

\subsubsection{Variations on a theme of radiatively inefficient accretion}

\begin{figure*}
    \centering
    \includegraphics[width=0.9\textwidth]{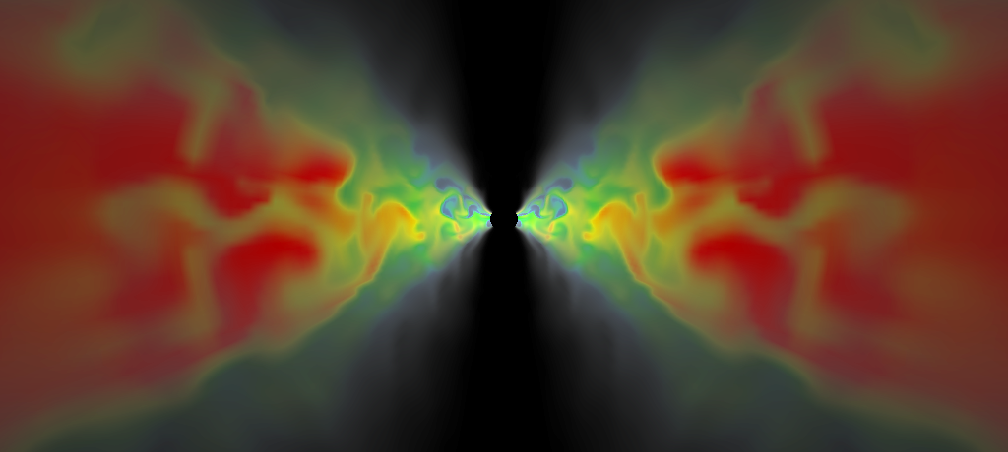}
    \caption{Visualization of a radiatively inefficient accretion flow in the viscous hydrodynamic limit. Such flows are geometrically thick, and develop large-scale flow structures that resemble convection. Outflows are likely. (Unpublished simulations; Armitage and Dullemond, 2000.)}
    \label{fig_hydro_ADAF}
\end{figure*}

The governing equations from which the ADAF solution is derived hard-wire assumptions about the radial constancy of the mass flux, and the nature of radial transport of angular momentum and energy, that can be questioned. It is therefore possible to construct alternate models for radiatively inefficient accretion flows, that are based on different physical principles. One such alternative is motivated by the observation that the ADAF solution is unstable to radial convection (see for example the simulation in Figure~\ref{fig_hydro_ADAF}). Convection in thick disks transports energy outward, but angular momentum {\em inward}, opposing viscous transport. If convection is dominant, these properties suggest that the governing principle that determines the flow structure is {\em marginal stability against convective instability} \citep{narayan00,quataert00}. A {\em Convection Dominated Accretion Flow} (CDAF) has a radial density profile that scales as $\rho(r) \propto r^{-1/2}$, substantially shallower than the ADAF's $\rho(r) \propto r^{-3/2}$. A second alternative is motivated by the fact that the supposedly accreting gas in the ADAF solution is so hot that it is, at best, weakly bound to the central object. The {\em Adiabatic Inflow-Outflow Solution} (ADIOS) model assumes that the disk responds by driving an outflow from its upper and lower surfaces at all radii, that co-exists with inflow near the equator \citep{blandford99}. $\dot{M}$ is then no longer a constant, but an increasing function of radius. In ADIOS models very little of the gas that is supplied to the system at large radii ever reaches the vicinity of the black hole. A third possibility, though one more typically studied with simulations rather than self-similar models, is that the magnetic field is strong enough to directly dictate the flow dynamics, at least close to the black hole. This is the regime of {\em Magnetically Arrested Disks} (MAD) \citep{tchekhovskoy11,igumenshchev08}. Analytic arguments suggest that geometrically thick disks transport magnetic flux inward more easily than thin disks \citep{lubow94}, though the exact conditions that lead to MAD solutions and their governing physical principles remain a topic of investigation \citep{begelman22}.

Which of these physical considerations is most determinative of the structure of radiatively inefficient accretion flows in nature cannot be divined by pure thought. Numerical simulations, and observations of hot flows in systems such as the Galactic Center \citep{marrone07} and M87 \citep{EHT19}, are essential. \citet{yuan14} and \citet{davis20} are good starting points for reviews of the current state of the art.

\subsection{Hyperaccretion}
As we have seen, geometrically thick accretion flows arise at low accretion rates because plasma physics effects prevent low density gas from cooling efficiently. Thick disks can also occur at very high accretion rates, when the gas is so dense that photons cannot diffuse away faster than they are advected into the black hole. This is the regime of {\em hyperaccretion}.

The new aspect of hyperaccreting systems is the existence of a trapping radius $r_{\rm trap}$, interior to which photons cannot escape being dragged into the black hole. \citet{begelman78} derived the trapping radius by considering a generalized version of spherical Bondi accretion \citep{bondi52}, which we will get to in \S\ref{sec_Bondi}. We can obtain the main result, however, without needing to specify many details of the accretion flow.

The trapping radius is defined as the location where the local infall time matches the time for outward photon diffusion. For gas at radius $r$, with inflow speed $v_r$, the infall time is just,
\begin{equation}
    t_{\rm infall} \sim \frac{r}{v_r}.
\end{equation}
To find the diffusion time, we use basic results from the theory of random walks \citep[e.g.][]{rybicki79}. If the photon mean-free-path is $l$, then after $N$ scatterings the photon will have diffused an expected distance, 
\begin{equation}
    d \sim \sqrt{N} l,
\end{equation}
after a time,
\begin{equation}
    \Delta t \sim \frac{Nl}{c}.
\end{equation}
The diffusion time scale from radius $r$ is thus, to order of magnitude,
\begin{equation}
    t_{\rm diffuse} \sim \frac{r^2}{lc}.
\end{equation}
Suppose now that the temperature of the accreting gas is high enough that electron scattering provides the main source of opacity. The mean-free-path is then,
\begin{equation}
    l = \frac{1}{n \sigma_{\rm T}} = \frac{m_{\rm H}}{\rho \sigma_{\rm T}}, 
\end{equation}
where $\rho$ is the density, $m_{\rm H}$ is the mass of a hydrogen atom, and $\sigma_{\rm T}$ is the Thomson cross-section as usual. The density is related to the accretion rate through,
\begin{equation}
    \dot{M} = 4 \pi r^2 v_r \rho.
\end{equation}
Combining equations, $v_r$ cancels out and we find that $t_{\rm infall}$ equals $t_{\rm diffuse}$ at a trapping radius that is given by,
\begin{equation}
    r_{\rm trap} = \frac{\sigma_{\rm T}}{4 \pi c m_{\rm H}} \dot{M}.
\end{equation}
Rewriting this in terms of the Eddington mass accretion rate (equation~\ref{eq_mdot_Edd}),
\begin{equation}
    r_{\rm trap} = \frac{\dot{M}}{\dot{M}_{\rm Edd}} \frac{GM}{\eta c^2},
\end{equation}
we see that photon trapping should become important roughly when the mass accretion rate first exceeds the value where the nominal luminosity is the Eddington limit. At higher $\dot{m}$, the radiative efficiency will necessarily drop as photons produced close to the black hole are trapped and advected across the event horizon without any chance to escape to infinity. Although derived here in spherical symmetry, the same physics applies to geometrically thick disk flows \citep{begelman82,abramowicz88}. This regime can now be simulated using radiation hydrodynamics \citep{jiang19b}.

\begin{figure*}
    \centering
    \includegraphics[width=0.9\textwidth]{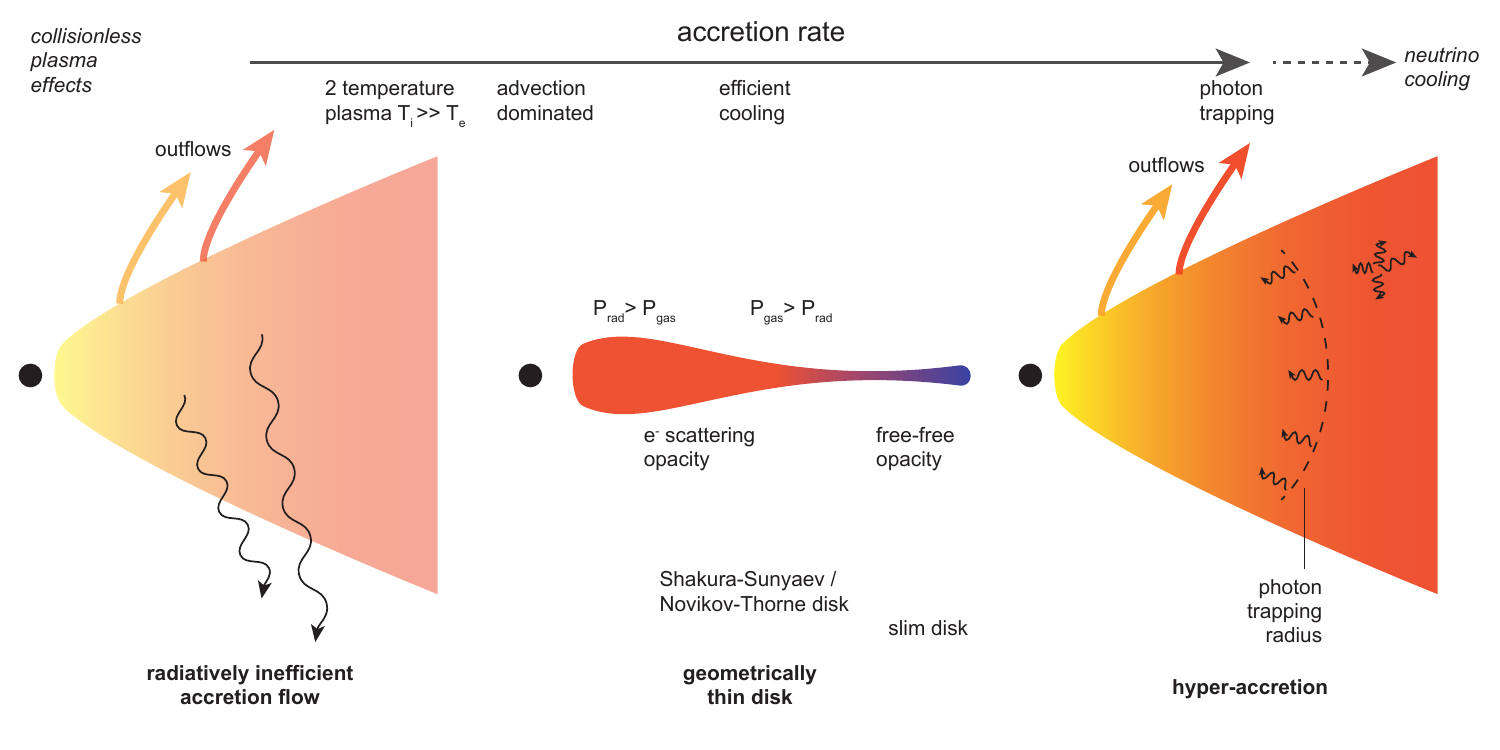}
    \caption{Illustration of the main regimes of black hole accretion as a function of the accretion rate, scaled to the accretion rate that would generate the Eddington limiting luminosity for radiatively efficient accretion. (There is a strong argument that this illustration should have a second dimension, with the extra parameter being the net magnetic flux that threads the disk. The regimes above are appropriate if the net flux is relatively weak.)}
    \label{fig_accretion_regimes}
\end{figure*}

Stepping back from the details of the last few pages, Figure~\ref{fig_accretion_regimes} summarizes what we have deduced about the qualitative structure of black hole accretion flows. The crucial parameter---at least as far as energetics go---is the accretion rate in units of the Eddington mass accretion rate. The thin disk solutions of \citet{shakura73} and \citet{novikov73}, although they are surely wrong in many details due to their {\em ad hoc} treatment of angular momentum transport, probably provide a semi-quantitative description of disk structure across about three orders of magnitude in accretion rate, $10^{-3} \lesssim \dot{M} / \dot{M}_{\rm Edd} \lesssim 1$. At both extremities, the thin disk solutions fail and the expected outcome of accretion with non-zero angular momentum is a geometrically thick disk. The cause is inefficient radiative cooling, at low $\dot{m}$ because the low density keeps the energy locked away in the ions, which cannot radiate, and at high $\dot{m}$ because the high density traps photons within the inflowing gas. Outflows are likely to accompany both geometrically thick disk regimes, which are surprisingly similar despite the different physics at low and high $\dot{m}$. Net magnetic flux is an additional parameter that can qualitatively modify the properties of both thick and thin accretion disk flows.

What happens if the accretion rate is pushed even further to the extremes? At extremely low $\dot{m}$, the plasma is highly collisionless, and the long mean-free-path means that processes such as anisotropic conduction and viscosity are potentially important for the bulk flow \citep{chandra15}. At extremely high $\dot{m}$, cooling due to neutrino emission opens up a second regime of at least moderately thin disk accretion \citep{popham99,dimatteo02,chen07}. High enough accretion rates, exceeding $\dot{M} \sim 10^{-2} \ M_\odot \ {\rm s}^{-1}$ onto a stellar mass black hole, are realized during the collapse of the cores of massive stars to form a black hole and a massive disk \citep{macfadyen99}.

\section{Waves in disks}
\label{sec_waves}
Consider the following situations,
\begin{itemize}
    \item A planet, on a circular orbit, interacts gravitationally with the surrounding gas in the disk.
    \item A binary black hole merges due to the emission of gravitational radiation, which leaves the system at the speed of light. From the point of view of a surrounding circumbinary disk, the merger leads to a near-instantaneous reduction in the central point mass from $M$ to $M(1-\epsilon)$, with $\epsilon$ being of the order of a few percent.
    \item
    A star-star flyby warps the outer part of a disk around one of them into a non-planar shape.
\end{itemize}
Although physically quite different, these are all situations where perturbations excite  {\em waves} within an accretion disk. Disk waves assume a variety of forms, due to the different forces involved (pressure, rotation, gravity, magnetic fields) and because the perturbations of physical interest differ from system to system. A book could be written on the mathematical description of these waves, and in fact at least one book {\em has} been written \citep{kato16}. Here, we briefly describe some simple examples.

\subsection{Modified sound waves}
\label{sec_waves_sound}
The simplest disk wave is an axisymmetric disturbance in an unmagnetized, non-self-gravitating, two-dimensional fluid. We have already done the work needed to understand this situation. Dropping the term that results from disk self-gravity, the dispersion relation (equation~\ref{eq_C4_dispersionrelation}) is,
\begin{equation}
 \omega^2 = \kappa^2 +c_{\rm s}^2 k^2,
\end{equation} 
where $\omega$ is the frequency of the wave, $k$ is the wavenumber, $c_s$ is the sound speed, and $\kappa$ is the epicyclic frequency,
\begin{equation}
 \kappa^2 \equiv 4 \Omega^2 + 2 r \Omega \frac{{\rm d}\Omega}{{\rm d}r}.
\end{equation}
For a Keplerian disk, $\kappa = \Omega_{\rm K}$. This dispersion relation describes sound waves whose properties are modified by the Coriolis force. The modification takes the form of a cutoff---as we go to large wavelengths (small $k$) the wave frequency $\omega \rightarrow \pm \kappa$, with the opposing signs corresponding to inward and outward propagating waves. 
Provided that $\kappa^2 > 0$ (as is assuredly the case for a near-Keplerian disk) the wave frequency $\omega$ is always real, so there is no instability.

\subsection{The linear wave equation}
The properties of more general waves in disks can be derived in multiple ways. The derivation given in the review by \citet{balbus03} is quite straightforward and direct, and we follow it here. The starting point is the fluid equations in the absence of magnetic fields or viscosity. Sticking to the inertial rather than shearing sheet frame, the continuity and momentum equations are,
\begin{eqnarray}
  \frac{\partial \rho}{\partial t} + \nabla \cdot \left( \rho {\bf v} \right) & = & 0, \\
  \frac{\partial {\bf v}}{\partial t} + {\bf v} \cdot \nabla {\bf v} & = & 
  - \frac{\nabla P}{\rho} - \nabla \Phi.
\end{eqnarray}
To keep matters (relatively) simple, we adopt a polytropic equation of state,
\begin{equation}
    P = K \rho^{\gamma},
\label{eq_wave_eos}    
\end{equation}
with $K$ and $\gamma$ constants. Although we will not need it in order to look at wave properties, the hydrostatic equilibrium structure implied by this equation of state is,
\begin{equation}
    \rho(z) = \left[ \rho_0^{\gamma-1} - \frac{ (\gamma-1) \Omega_{\rm K}^2 z^2}{2 \gamma K} \right]^{1/(\gamma-1)},
\end{equation}
where $\rho_0$ is the mid-plane density. Unlike in the isothermal case, the disk has a free surface at finite $z$. 

To obtain an equation for linear waves within disks, the procedure is to first linearize the relevant equations, and then convert them from partial differential equations to algebraic equations by specifying a form for the perturbations. We can illustrate the idea by working through the steps for the continuity equation in detail. Adopting cylindrical polar co-ordinates $(r, \phi, z)$, we assume that there is some equilibrium background disk structure $\rho(r,\phi,z)$, ${\bf v} (r,\phi,z)$ in which the fluid velocity is purely azimuthal (i.e. we ignore any inflow). On top of these fields we add perturbations, i.e. we take,
\begin{eqnarray}
  \rho & \rightarrow & \rho + \delta \rho, \\
  {\bf v} & \rightarrow & {\bf v} + \delta {\bf v}, 
\end{eqnarray}
that are assumed to be small. For the density this is as simple as requiring that $| \delta \rho | / \rho \ll 1$. Substituting in the continuity equation, the terms involving the background state drop out and the one second order term in the small quantities is neglected,
\begin{equation}
    \underbrace{\frac{\partial \rho}{\partial t} + \nabla \cdot (\rho {\bf v} )}_{=0} + 
    \frac{\partial \delta \rho}{\partial t} + 
    \nabla \cdot ( \rho \delta {\bf v} + \delta \rho {\bf v} ) + 
    \underbrace{\nabla \cdot ( \delta \rho \delta {\bf v} )}_{\rm neglect} = 0.
\end{equation}
Allowing the density perturbation to have an arbitrary radial and vertical form, but requiring it to be periodic in azimuth, we write,
\begin{equation}
    \delta \rho = \delta \rho (r,z) \exp[ i ( m \phi - \omega t) ],
\end{equation}
where $\omega$ is the frequency of the wave and $m$ is an integer. This implies that,
\begin{equation}
   \frac{\partial \delta \rho}{\partial t} = -i \omega \delta \rho. 
\end{equation}
In the equilibrium state we have purely azimuthal flow ${\bf v} = (0, v_\phi, 0)$. Using the expression for the divergence operator in cylindrical co-ordinates we find that,
\begin{equation}
    \nabla \cdot (\delta \rho {\bf v}) = \frac{1}{r} \frac{\partial}{\partial \phi} \left( v_\phi \delta \rho \right) = i m \frac{v_\phi}{r} \delta \rho = i m \Omega \delta \rho,
\end{equation}
where $\Omega$ is the angular velocity (which is not necessarily Keplerian). Collecting the surviving terms together, the final result is,
\begin{equation}
    - i \bar{\omega} \delta \rho + \nabla \cdot ( \rho \delta {\bf v} ) = 0,
\label{eq_linearized_continuity}    
\end{equation}
where we have defined,
\begin{equation}
    \bar{\omega} \equiv \omega - m \Omega.
\end{equation}
The quantity $\bar{\omega}$, which comes up often in this sort of analysis, is called the ``Doppler-shifted wave frequency".

Turning to the equation of state (equation~\ref{eq_wave_eos}) we define the enthalpy as,
\begin{equation}
    {\cal H} = \int \frac{{\rm d}P}{\rho} = \int K \gamma \rho^{\gamma-2} {\rm d}\rho = \frac{ \gamma P / \rho}{\gamma - 1} = \frac{c_{\rm ad}^2}{\gamma-1},
\end{equation}
where $c_{\rm ad}$ is the adiabatic sound speed. Differentiating, we obtain,
\begin{equation}
    {\delta {\cal H}} = c_{\rm ad}^2 \frac{\delta \rho}{\rho},
\label{eq_perturbed_enthalpy}    
\end{equation}
as the relation between perturbations in density and perturbations in enthalpy for this equation of state.

For the momentum equation, use of equation~(\ref{eq_convective_cylindrical}) gives the leading and first order terms that result from the convective operator in cylindrical co-ordinates as,
\begin{eqnarray}
\left[ ( {\bf v} + \delta {\bf v} ) \cdot \nabla \right] ( {\bf v} + \delta {\bf v}) = \nonumber \\
 \left[ \frac{v_\phi}{r} \frac{ \partial \delta v_r }{\partial \phi} - 
   \frac{v_\phi^2}{r} - \frac{2 v_\phi \delta v_\phi}{r}, 
\delta v_r \frac{\partial v_\phi}{\partial r} \right. \nonumber \\ \left.+ \frac{v_\phi}{r} \frac{\partial \delta v_\phi}{\partial \phi} + \frac{\delta v_r v_\phi}{r}, \,
 \frac{v_\phi}{r} \frac{\partial \delta v_z}{\partial \phi} \right].
\end{eqnarray}
The linearized momentum equation then gives,
\begin{eqnarray}
 -i \bar{\omega} \delta v_r - 2 \Omega \delta v_\phi & = & - \frac{ \partial \delta {\cal H}}{\partial r}, 
\label{eq_linear_mom1} \\
 -i \bar{\omega} \delta v_\phi + \frac{\kappa^2}{2 \Omega} \delta v_r & = & -i \frac{m}{r} \delta {\cal H}, 
\label{eq_linear_mom2} \\
 -i \bar{\omega} \delta v_z & = & - \frac{\partial \delta {\cal H}}{\partial z},
\end{eqnarray} 
where,
\begin{equation}
    \kappa^2 \equiv 4 \Omega^2 + \frac{{\rm d}\Omega^2}{{\rm d}\ln r},
\end{equation}
defines the epicyclic frequency. Solving for $\delta v_r$ and $\delta v_\phi$ between equation~(\ref{eq_linear_mom1}) and equation~(\ref{eq_linear_mom2}) we have,
\begin{eqnarray}
 \delta v_r & = & \frac{i}{\kappa^2 - \bar{\omega}^2} 
   \left[ \bar{\omega} \frac{\partial \delta {\cal H}}{\partial r} - \frac{2 m \Omega}{r} \delta {\cal H} \right], \\
 \delta v_\phi & = & \frac{1}{\kappa^2 - \bar{\omega}^2} 
   \left[ \frac{\kappa^2}{2 \Omega} \frac{\partial \delta {\cal H}}{\partial r} - \frac{m \bar{\omega}}{r} \delta {\cal H} \right], \\
   \delta v_z & = & -\frac{i}{\bar{\omega}} \frac{\partial \delta {\cal H}}{\partial z}.
\end{eqnarray}
Once the rotation profile of the background disk flow is specified (giving us $\Omega$ and $\kappa^2$), these equations relate the velocity and enthalpy perturbations for waves of particular frequency $\omega$ and azimuthal wave number $m$.

The final step in the derivation is to insert the expressions for $\delta {\bf v}$ into the linearized continuity equation (equation~\ref{eq_linearized_continuity}). Using equation~(\ref{eq_perturbed_enthalpy}) the result simplifies to,
\begin{eqnarray}
    \left[ \frac{1}{r} \frac{\partial}{\partial r} \left( \frac{r \rho}{D} \frac{\partial}{\partial r} \right) - \frac{1}{\bar{\omega}^2} \frac{\partial}{\partial z} \left( \rho \frac{\partial}{\partial z} \right) - \frac{m^2 \rho}{r^2 D}  \right. \nonumber \\
    \left. - \frac{1}{r \bar{\omega}} \frac{\partial}{\partial r} \left( \frac{2 m \Omega \rho}{D} \right) - \frac{\rho}{c_{\rm ad}^2} \right] \delta {\cal H} = 0.
\label{eq_wave_equation}    
\end{eqnarray}
Here, we have defined,
\begin{equation}
    D \equiv \kappa^2 - \bar{\omega}^2.
\end{equation}
This is the governing equation for linear hydrodynamic waves in a disk, ignoring magnetic fields and disk self-gravity\footnote{Note that I think there are typos in the signs of the last two terms in \citet{balbus03}. Check yourself before blindly using this equation.}. Even with these simplifications, it's rather complicated! What is easy to see, however, is that there are two ways in which the denominators in the equation can equal zero. These are {\em resonant} locations. When $D=0$ we have the condition for {\em Lindblad} resonance, while $\bar{\omega} = 0$ defines the location of corotation resonances. These resonances are central to the theory of planet-disk interactions, because for a disk perturbed by an external potential angular momentum transfer occurs only in their vicinity \citep{goldreich79}. 

\subsection{Density and inertial waves}
Suppose for now that we are {\em not} in the vicinity of a resonance. Then equation~(\ref{eq_wave_equation}) describes freely propagating waves within the disk. It can be solved numerically, but we can gain analytic insight with the aid of some additional approximations. Assume a WKB solution of the form,
\begin{equation}
    {\cal H} = A(r,z) \exp \left[ \frac{i S (r,z)}{\epsilon} \right],
\end{equation}
where $\epsilon$ is a small parameter that is introduced so that the phase $S/\epsilon$ varies rapidly. The radial and vertical wavenumbers are,
\begin{eqnarray}
    k_r & = & \frac{\partial S}{\partial r}, \\
    k_z & = & \frac{\partial S}{\partial z}.
\end{eqnarray}
We further assume that $k_r \gg m/r$, $k_z \gg m/r$, and that the disk is thin, so that $c_{\rm ad} \ll r \Omega$. Substituting in equation~(\ref{eq_wave_equation}), and keeping only those derivative terms of order $\epsilon^{-2}$, the leading order result is,
\begin{equation}
    \frac{A\rho}{\bar{\omega}^2} e^{iS/\epsilon} \left( \frac{\partial S}{\partial z} \right)^2 - \frac{A \rho}{D} e^{iS/\epsilon} \left( \frac{\partial S}{\partial r} \right)^2 = \frac{A \rho}{c_{\rm ad}^2} e^{iS/\epsilon}.
\end{equation}
The dispersion relation is then,
\begin{equation}
    \frac{k_z^2 c_{\rm ad}^2}{\bar{\omega^2}} + \frac{k_r^2 c_{\rm ad}^2}{\bar{\omega}^2 - \kappa^2} = 1.
\label{eq_wave_dispersion}    
\end{equation}
By taking the WKB limit the general wave equation reduces to a pretty simple dispersion relation.

\begin{figure}
    \includegraphics[width=\columnwidth]{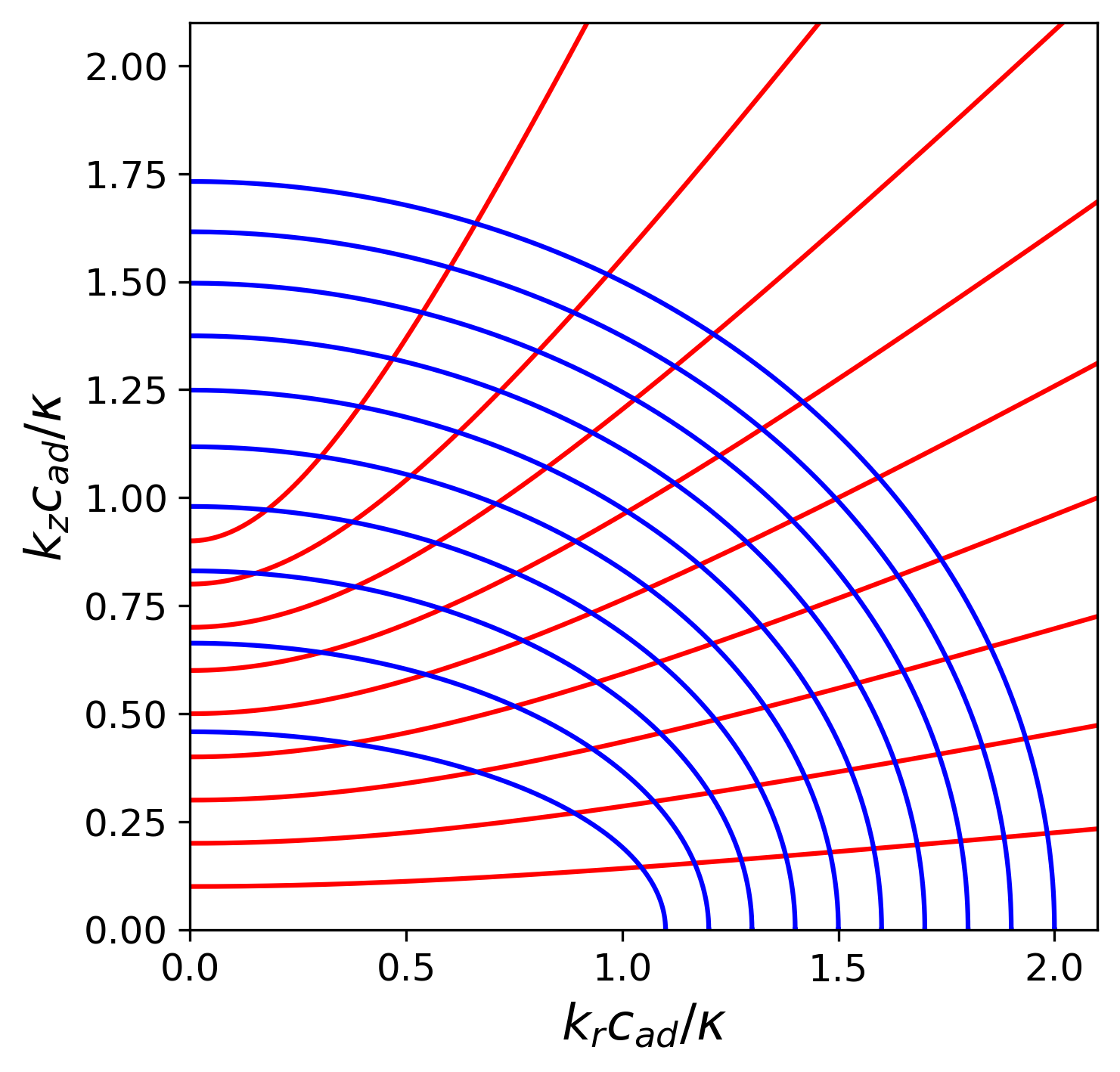}
    \caption{The dispersion relation for disk waves (equation~\ref{eq_wave_dispersion}). Plotted are contours of constant Doppler-shifted wave frequency $\bar{\omega}$, in the $(k_z,k_r)$ plane. The wavenumbers are scaled by $c_{\rm ad}/\kappa$, where $c_{\rm ad}$ is the adiabatic sound speed and $\kappa$ is the epicyclic frequency. The blue ellipses describe high frequency density waves, with $\bar{\omega}^2 > \kappa^2$. The curves are shown for $\bar{\omega} / \kappa$ between 1.1 and 2.0, in intervals of 0.1. The red hyperbolae describe low frequency inertial waves, with $\bar{\omega}^2 < \kappa^2$. The curves are shown for $\bar{\omega} / \kappa$ between 0.1 and 0.9, again in intervals of 0.1.}
    \label{fig_wave_dispersion}
\end{figure}

Recall now some basic properties of conic sections. Ellipses are described by $x^2 / a^2 + y^2 / b^2 = 1$, hyperbolae by $x^2 / a^2 - y^2 / b^2 = 1$. Depending upon the Doppler-shifted wave frequency, solutions to equation~(\ref{eq_wave_dispersion}) in the $(k_r,k_z)$ plane can be of either form. High-frequency waves, with $\bar{\omega}^2 > \kappa^2$, trace out ellipses in this plane, while low-frequency waves, with $\bar{\omega}^2 < \kappa^2$, trace out hyperbolae. These waves evidently have quite different geometric properties, and in fact represent two distinct classes of waves that can be present within disks.

Writing equation~(\ref{eq_wave_dispersion}) in the equivalent form,
\begin{equation}
    \bar{\omega}^4 - (k^2 c_{\rm ad}^2 + \kappa^2 ) \bar{\omega}^2 + \kappa^2 k_z^2 c_{\rm ad}^2 = 0,
\end{equation}
where $k^2 = k_r^2 + k_z^2$, the two types of disk waves can be defined by reference to which terms in the dispersion relation are dominant. {\em Density waves} are what we get if the third term in the dispersion relation can be neglected, either because $k_z = 0$ (or is otherwise small) or because $k c_{\rm ad} \gg \kappa$. In this limit,
\begin{equation}
    \bar{\omega}^2 = \kappa^2 + c_{\rm ad}^2 k^2, 
\end{equation}
and the physics is the same as \S\ref{sec_waves_sound}. These are modified acoustic waves. {\em Inertial waves}, on the other hand, occur when the first term in the dispersion relation is negligible. Taking the limit where the sound speed $c_{\rm ad} \rightarrow \infty$, the dispersion relation takes the form,
\begin{equation}
    \bar{\omega}^2 = \frac{k_z^2}{k_r^2 + k_z^2} \kappa^2.
\end{equation}
Inertial waves are low frequency disk phenomena, and unlike density waves involve fluid motions that are almost incompressible.

\section{Warped and eccentric disks}
It has been known for a very long time that not all accretion disks are circular, planar structures. The X-ray source Hercules X-1 exhibits a 35~dy periodicity that is interpreted as the precession period of a warped disk surrounding the accreting compact object \citep{katz73}. Maser emission from the nuclear region of the Seyfert galaxy NGC~4258 traces a geometrically thin and mildly warped disk down to sub-pc scales from the black hole \citep{miyoshi95}. More recently, several examples of warped (and in some cases ``broken") protoplanetary disks have been observed \citep[e.g.][]{casassus15,kraus20}. Direct evidence for eccentric disks is sparser, but photometric variations known as ``superhumps" in some cataclysmic variables imply excitation of the outer disk into an eccentric, precessing state \citep{whitehurst88}. The formation of an eccentric disk also seems to be an inevitable consequence of tidal disruption events in galactic nuclei \citep[e.g][]{bonnerot16}.

The physics of warped and eccentric disks is subtle, and can be a viperous pit for the unwary. This Section gives a high level overview.

\subsection{Warped disks}
\label{sec_warped_disks}
The state of a warped disk is jointly described by the surface density $\Sigma (r)$ and the unit vector normal to the local disk plane $\hat{\bf l} (r)$\footnote{Some literature makes a distinction between {\em warped} disks, in which the ensemble of tilt vectors for the annuli in the disk lie in a single plane, and {\em twisted} disks, in which they do not. We won't bother.}. We will assume that the disk is geometrically thin, that the orbits at all radii are circular (up to corrections due to the radial gas flow), and that $\hat{\bf l} (r)$ is a smooth, continuous function. It is not immediately obvious whether the evolution of the tilt vector, under the action of internal (and possibly external) torques, preserves continuity, but we defer for now discussion of the possibility that the disk breaks up into disjoint pieces \citep{nixon12}.

\begin{figure*}
    \centering
    \includegraphics[width=\textwidth]{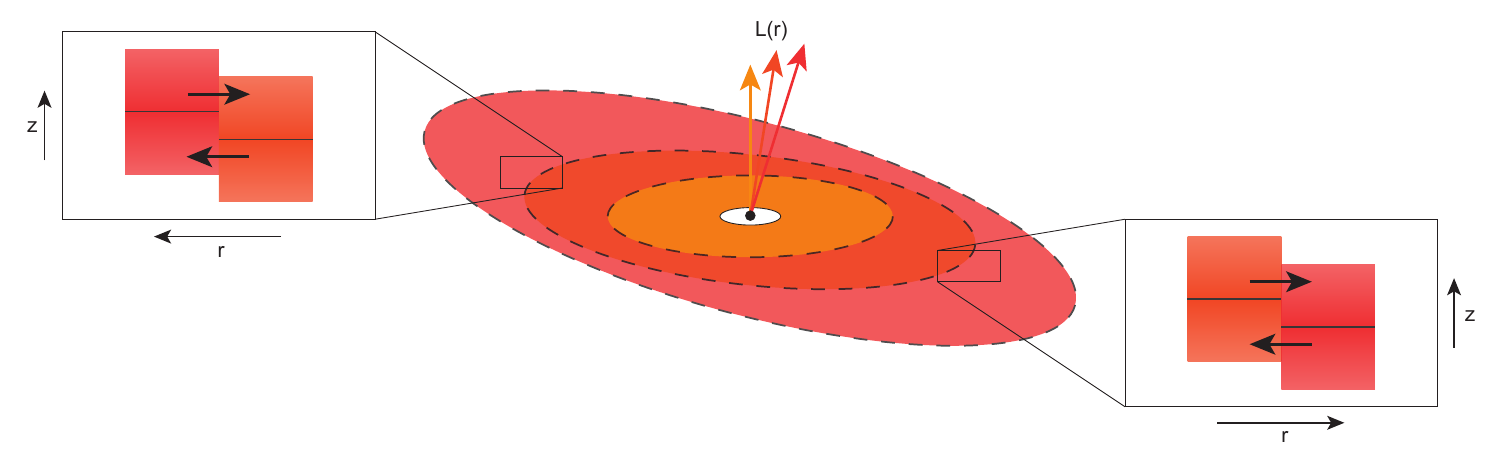}
    \caption{Illustration \citep[after][their Figure~10]{lodato07} of the additional hydrodynamic effects present in a warped disk. Considering two neighboring but misaligned fluid annuli, there is obviously a vertical shear between them that is induced by the warp. As shown in the insets, the relative vertical displacements of the mid-planes of the annuli also result in time- and height-dependent {\em} radial pressure gradients. If the potential is close to Keplerian, the radial forcing is resonant with the disk's epicyclic frequency. This physical effect means that a warped disk is a source of bending waves, which---depending upon the disk viscosity---may damp locally or propagate globally.}
    \label{fig_warp_forcing}
\end{figure*}

Our immediate goal is to derive equations for the time evolution of the surface density and tilt of a warped disk. In addition to the regular planar disk physics there are new effects to consider, illustrated in Figure~\ref{fig_warp_forcing}. If we consider two adjacent disk annuli, any mutual inclination results in vertical as well as radial shear. That vertical shear, in turn, leads to a periodic vertical displacement of the mid-planes of the two fluid annuli, which induces periodic {\em radial} pressure gradients. In a Keplerian potential, the forcing is resonant with the epicyclic response of the disk, such that the misalignment launches a wave.

What happens next depends upon the conditions within the disk. If the disk is sufficiently viscous, the nascent wave is damped locally, and we say that the warp evolution is in the viscous regime. The condition for viscous warp evolution is that \citep{papaloizou83},
\begin{equation}
    \alpha \gtrsim \frac{h}{r}.
\label{eq_warp_regime}    
\end{equation}
In the opposite limit where $\alpha \lesssim h/r$, radial communication of the warp occurs via waves provided that the potential is nearly Keplerian, specifically that,
\begin{equation}
    \left| \frac{ \Omega^2 - \kappa^2 }{\Omega^2} \right | \lesssim \frac{h}{r}.
\end{equation}
The evolution equations for viscous and wave-like warped disks are different, and they predict qualitatively different behavior. Both regimes are physically relevant. Protoplanetary disks are thick and inviscid enough as to fall almost always into the wave-like regime. Disks around black holes and other compact objects, on the other hand, are expected to display a large variation in $h/r$ with radius (Figure~\ref{fig_SS_hr}). The outer regions (at a minimum) are likely to be described by viscous warp dynamics. 

Key references for the analytic theory of the fluid dynamics of warped disks include \citet{papaloizou83}, \citet{papaloizou95}, and \citet{ogilvie99}. The review by \citet{nixon16} provides a good overview, and is worth reading before wading into the technical details of the earlier papers. 

\subsubsection{Viscous limit: heuristic derivation}
Conservation laws strongly constrain how warped disks behave. In the viscous limit, \citet{papaloizou83} and \citet{pringle92} derived an evolution equation for geometrically thin warped disks by combining conservation laws for mass and angular momentum with an assumption that a distinct viscosity acts to damp misalignments. Although this approach does not {\em quite} recover the same equations as a full hydrodynamic analysis \citep{papaloizou83,ogilvie99}, the differences are fairly minor, and the \citet{pringle92} derivation is simple and physically instructive. We follow it closely here.

We consider a geometrically thin disk with orbits that are circular but possibly misaligned with respect to each other. An annulus of the disk at radius $r$ has width $\Delta r$, surface density $\Sigma$, radial velocity $v_r$, and angular velocity $\Omega$. The tilt of the annulus in some inertial frame is described by a unit vector $\hat{\bf l}$ that is normal to the disk (and parallel to the local angular momentum vector).

Conservation of mass gives,
\begin{equation}
    \frac{\partial \Sigma}{\partial t} + \frac{1}{r} 
      \frac{\partial}{\partial r} \left( r \Sigma v_r \right) = 0.
\label{eq_warp_scalar}      
\end{equation}
This equation is identical to the one for a planar disk.

The angular momentum density (i.e. per unit area) is $\Sigma r^2 \Omega {\hat {\bf l}}$, and the total angular momentum content of the annulus is then $2 \pi r \Delta r \times \Sigma r^2 \Omega {\hat {\bf l}}$. The annulus' angular momentum changes due to:
\begin{itemize}
    \item Mass flow into / out of the annulus.
    \item Viscous torques acting within the disk.
    \item External torques, present for example near Kerr black holes, around oblate central objects, and in binary systems.
\end{itemize}
Mathematically \citep{papaloizou83},
\begin{eqnarray}
    \frac{\partial}{\partial t} \left( 2 \pi r^3 \Omega \Sigma {\hat {\bf l}} \Delta r \right) &=& \nonumber \\
    \left. 2 \pi r^3 \Omega \Sigma {\hat {\bf l}} v_r \right|_{r - \Delta r /2} & - & 
    \left. 2 \pi r^3 \Omega \Sigma {\hat {\bf l}} v_r \right|_{r + \Delta r /2} \nonumber \\
    + {\bf G} \left( r + \Delta r/2 \right) & - & {\bf G} \left( r - \Delta r/2 \right) + {\bf T}.
\end{eqnarray}
${\bf G} (r,t)$ is the viscous torque exerted on one annulus by its neighbor, ${\bf T}$ is the external torque. We can ignore ${\bf T}$ for now, as it can easily be added back in to the final evolution equation.

Determining the form for ${\bf G}$ is the crux of the derivation. In a warped disk there are two sources of shear, which we consider independently. First, there is an in-plane component, corresponding (in obvious notation) to the $(r,\phi)$ stress, which acts in the direction of $\hat {\bf l}$. This component has the same form as for a planar disk,
\begin{equation}
    {\bf G}_1 = 2 \pi r \nu_1 \Sigma r \frac{{\rm d}\Omega}{{\rm d}r} r {\hat {\bf l}}.
\end{equation}
Second, there is a component corresponding to the $(r,z)$ stress, which has a $\cos \phi$ azimuthal dependence, and vanishes if $\partial {\hat {\bf l}} / \partial r$ is zero. This component of the stress takes the form \citep{papaloizou83},
\begin{equation}
    {\bf G}_2 = 2 \pi r (\nu_2/2) \Sigma r \Omega \frac{{\partial}{\hat {\bf l}}}{{\partial}r} r.
\end{equation}
The factor of two comes from averaging the oscillatory vertical shear around the orbit. In this telling, $\nu_1$ and $\nu_2$ are separate viscosities that do not have a simple or obvious relationship. In particular, there is {\bf no reason} to assume that $\nu_1 = \nu_2$, and equality of these heuristically defined ``viscosities" is {\bf not} what is expected for a fluid disk described by the Navier-Stokes equations with an isotropic viscosity. We will return to these possible pitfalls later.

Using these forms for the viscous torque, and taking the limit as $\Delta r \rightarrow 0$, the angular momentum conservation equation becomes,
\begin{eqnarray}
  \frac{\partial}{\partial t} \left( \Sigma r^2 \Omega {\hat {\bf l}} \right) + 
  \frac{1}{r}  \frac{\partial}{\partial r} \left( \Sigma v_r r^3 \Omega {\hat {\bf l}} \right) = \nonumber \\
  \frac{1}{r}  \frac{\partial}{\partial r} \left( \nu_1 \Sigma r^3 \frac{{\rm d}\Omega}{{\rm d}r} {\hat {\bf l}} \right) +
  \frac{1}{r}  \frac{\partial}{\partial r} \left( \frac{1}{2} \nu_2 \Sigma r^3 \Omega \frac{\partial {\hat {\bf l}}}{\partial r} \right).
\label{eq_warp_vector}  
\end{eqnarray}
In addition to terms describing advection of angular momentum (and thus tilt), the final term has a diffusive character with a diffusion co-efficient that is $\propto \nu_2$. It expresses the tendency of viscosity to flatten out an initially warped disk.  

The combination of the vector equation~(\ref{eq_warp_vector}) with the scalar equation~(\ref{eq_warp_scalar}) for the surface density completely determines the evolution of viscous warped disks, in the absence of external torques and given the specified assumptions. To combine them, $v_r$ has to be eliminated. We first dot $\hat {\bf l}$ into equation~(\ref{eq_warp_vector}), and make use of the following identities which follow from the fact that $\hat {\bf l}$ is a unit vector,
\begin{eqnarray}
 \hat {\bf l} \cdot \hat {\bf l} & = & 0, \\
 \hat {\bf l} \cdot \frac{\partial \hat {\bf l}}{\partial t}  & = & 0, \\
 \hat {\bf l} \cdot \frac{\partial \hat {\bf l}}{\partial r}  & = & 0.
\end{eqnarray}
The result is,
\begin{eqnarray}
  \frac{\partial}{\partial t} \left( \Sigma r^2 \Omega \right) + 
  \frac{1}{r}  \frac{\partial}{\partial r} \left( \Sigma v_r r^3 \Omega \right) = \nonumber \\
  \frac{1}{r}  \frac{\partial}{\partial r} \left( \nu_1 \Sigma r^3 \frac{{\rm d}\Omega}{{\rm d}r} \right) +
  \frac{1}{2} \nu_2 \Sigma r^2 \Omega {\hat {\bf l}} \cdot 
  \frac{\partial^2 {\hat {\bf l}}}{\partial r^2}.
\end{eqnarray}
We now multiply the continuity equation~(\ref{eq_warp_scalar}) by $r^2 \Omega$ and subtract it to eliminate the time derivative from the left-hand-side. For the final term we note that,
\begin{equation}
    \frac{\partial}{\partial r} \left( {\hat {\bf l}} \cdot \frac{\partial \hat {\bf l}}{\partial r} \right) = 
    {\hat {\bf l}} \cdot \frac{\partial^2 {\hat {\bf l}}}{\partial r^2} + 
    \left| \frac{\partial \hat {\bf l}}{\partial r} \right|^2 = 0,
\end{equation}
so $\hat {\bf l} \cdot \partial^2 \hat {\bf l} / \partial r^2$ can be replaced with $-| \partial \hat {\bf l} / \partial r|^2$. This gets us to,
\begin{eqnarray}
  -r \Omega \frac{\partial}{\partial r} \left( r \Sigma v_r \right) + 
  \frac{1}{r}  \frac{\partial}{\partial r} \left( \Sigma v_r r^3 \Omega \right) = \nonumber \\
  \frac{1}{r}  \frac{\partial}{\partial r} \left( \nu_1 \Sigma r^3 \frac{{\rm d}\Omega}{{\rm d}r} \right) -
  \frac{1}{2} \nu_2 \Sigma r^2 \Omega 
  \left| \frac{\partial \hat {\bf l}}{\partial r} \right|^2.
\end{eqnarray}
The left-hand-side looks as if it involves a spatial gradient of $v_r$, but on application of the chain rule it simplifies considerably to give just $v_r \Sigma {\partial} (r^2 \Omega)/\partial r$. The radial velocity is then,
\begin{equation}
    v_r = \frac{\frac{\partial}{\partial r} \left( \nu_1 \Sigma r^3 \frac{{\rm d}\Omega}{{\rm d}r} \right) -
  \frac{1}{2} \nu_2 \Sigma r^3 \Omega 
  \left| \frac{\partial \hat {\bf l}}{\partial r} \right|^2}{\Sigma r \frac{\partial}{\partial r} \left( r^2 \Omega \right)}.
\end{equation}
The novel piece here is the second term involving $\nu_2$ and the radial gradient of the tilt vector. This term reflects the fact that viscous flattening of local variations in the disk tilt involves energy dissipation, which must be accompanied by radial inflow of gas through the disk.
Using this expression for $v_r$, we can obtain separate equations for the time evolution of the surface density and tilt vector. These can be found in \citet{pringle92}. At this point, physically, we are done. Noting that $\hat {\bf l}$ is a unit vector, however, we still have four equations for only three independent quantities. Some additional simplification is therefore possible. Specializing to a Keplerian potential, straightforward manipulations yield an evolution equation for,
\begin{equation}
    {\bf L} \equiv \left( {GM r} \right)^{1/2} \Sigma {\hat {\bf l}},
\end{equation}
which takes the form,
\begin{eqnarray}
  \frac{\partial {\bf L}}{\partial t} & = & 
    \frac{3}{r} \frac{\partial}{\partial r} 
    \left[ \frac{r^{1/2}}{\Sigma} 
      \frac{\partial}{\partial r} \left( \nu_1 \Sigma r^{1/2} {\bf L} \right) \right] \nonumber \\
      &+& \frac{1}{r} \frac{\partial}{\partial r} 
      \left[ \left( \nu_2 r^2 \left| \frac{\partial \hat {\bf l}}{\partial r} \right|^2 - \frac{3}{2} \nu_1 \right) {\bf L} \right] \nonumber \\
      &+& \frac{1}{r} \frac{\partial}{\partial r} 
      \left[ \frac{1}{2} \nu_2 r | {\bf L} | \frac{\partial \hat {\bf l}}{\partial r} \right].
\label{eq_warp_evolution}      
\end{eqnarray}
If the disk is planar, this equation reduces to the usual evolution equation for the surface density of a thin disk.

\subsubsection{Viscous evolution equation}
\label{sec_viscous_warps}
Imagine a minimal planetary system made up of two planets on stable, circular, but mutually inclined orbits. Gravitational torques will result in precession of the orbital planes, but there is no dissipation in the system and both the energy and the angular momentum stay fixed. Replace the planets with interacting fluid rings, and it becomes obvious that elementary considerations of mass and angular momentum conservation {\em cannot} fully determine how a warped disk evolves, because such considerations cannot capture possible precessional effects. A comprehensive description must instead be based on analysis of the Navier-Stokes equations\footnote{Assuming, as is customary in almost all analytic work, that we can't provide a better simple description of angular momentum transport in a fluid disk.}. A program to do so was started by \citet{papaloizou83} and essentially completed by \citet{ogilvie99}. It leads to both a modified version of equation~(\ref{eq_warp_evolution}) and, more importantly, to an understanding of the relationship between $\nu_1$ and $\nu_2$.

\citet{ogilvie99} is stiff mathematical medicine, which the reader seeking full details will have no choice but to imbibe. A central result is the modified version of equation~(\ref{eq_warp_evolution}). Keeping the notation the same, it reads \citep{nixon16},
\begin{eqnarray}
  \frac{\partial {\bf L}}{\partial t} & = & 
    \frac{3}{r} \frac{\partial}{\partial r} 
    \left[ \frac{r^{1/2}}{\Sigma} 
      \frac{\partial}{\partial r} \left( \nu_1 \Sigma r^{1/2} {\bf L} \right) \right] \nonumber \\
      &+& \frac{1}{r} \frac{\partial}{\partial r} 
      \left[ \left( \nu_2 r^2 \left| \frac{\partial \hat {\bf l}}{\partial r} \right|^2 - \frac{3}{2} \nu_1 \right) {\bf L} \right] \nonumber \\
      &+& \frac{1}{r} \frac{\partial}{\partial r} 
      \left[ \frac{1}{2} \nu_2 r | {\bf L} | \frac{\partial \hat {\bf l}}{\partial r} \right] \nonumber \\
      &+& \frac{1}{r} \frac{\partial}{\partial r} 
      \left[ \nu_3 \Sigma r^3 \Omega {\hat {\bf l}} \times 
      \frac{ \partial {\hat {\bf l}}}{ \partial r} \right].
\label{eq_warp_evolution_ogilvie}      
\end{eqnarray}
As presaged above, the modification takes the form of an additional term which describes precession driven by the presence of a warp. Its magnitude depends on a function $\nu_3$, which has the dimensions of a viscosity. In some cases this extra term is negligible, but there is no reason (other than a rather minor simplification) not to include it in calculations of the evolution of warped, viscous disks.

The relation between the dissipative vertical and horizontal viscosities, $\nu_2$ and $\nu_1$, can be determined from the hydrodynamic theory. For small amplitude (linear) warps the result is \citep{papaloizou83,ogilvie99},
\begin{equation}
    \frac{\nu_2}{\nu_1} \simeq \frac{2(1+7 \alpha^2)}{\alpha^2 (4 + \alpha^2)} \approx \frac{1}{2 \alpha^2},
\label{eq_viscosity_ratio}    
\end{equation}
where the final expression is valid for the usual case where $\alpha \ll 1$. It is worth emphasizing that this normally large ratio between $\nu_2$ and $\nu_1$ occurs {\em even for an intrinsically isotropic} fluid viscosity, and has nothing to do with any anisotropy that might arise, for example, if angular momentum transport originates with the MRI.

The local strength of a warp in an accretion disk depends upon the dimensionless quantity,
\begin{equation}
    | \psi | \equiv r \left| \frac{\partial {\hat {\bf l}}}{\partial r} \right|.
\end{equation}
The analysis of \citet{ogilvie99} yields equations that can be solved to obtain the viscosities ($\nu_1$, $\nu_2$ and $\nu_3$, or $Q_1$ through $Q_3$ in Ogilvie's notation) as $f(\psi)$. They have a non-trivial form, with both $\nu_1$ and $\nu_2$ typically decreasing as the warp amplitude becomes increasingly non-linear.

\subsubsection{Bardeen-Petterson effect}
Returning to equation~(\ref{eq_warp_evolution}), we now consider the effect of external torques ${\bf T}$ on the evolution of warped disks. The most famous example is the Lense-Thirring effect \citep{lense18}, which leads to the precession of orbits around Kerr black holes that are inclined with respect to the equatorial plane. The precession frequency can be written, approximately, as,
\begin{equation}
    \Omega_{\rm LT} = \frac{\omega_{\rm LT}}{r^3},
\end{equation}
where $\omega_{\rm LT} = 2GJ / c^2$, and $J = a GM^2 / c$ is the angular momentum of a black hole with dimensionless spin parameter $a$. \citet{bardeen75} observed that differential precession due to the Lense-Thirring effect, acting on an initially planar but tilted disk, would rapidly warp the disk close to the black hole and drive up $|\psi|$. Viscosity would then damp the disk into the equatorial plane at small orbital radii, while the disk further out remained misaligned. Provided that the disk is in the viscous limit, the same sort of effect can occur when disk precession is driven at small radii by the oblateness of a central body, or at large radii by torque from a binary companion.

Starting from the equations of \citet{papaloizou83}, \citet{kumar85} computed the shape of a warped disk subject to the Bardeen-Petterson effect. A simple analytic solution for the shape can be derived in a limit where the warp is small, the viscosity is constant, and the structure of the disk close to the inner boundary is not important \citep{scheuer96}\footnote{Peter Scheuer taught one of my undergraduate mathematics courses, and this argument is a nice example of a model mathematical physics problem.}. The starting point is equation~(\ref{eq_warp_evolution}). We drop the second order term $|\partial \hat{\bf l} / \partial r|^2$, set the time derivative to zero, and add in a term that represents forced precession due to the Lense-Thirring effect. In steady-state, a warped disk around a spinning black hole obeys, in the viscous limit, the equation,
\begin{eqnarray}
 0 & = & \frac{3}{r} \frac{\rm d}{{\rm d}r} \left[ \frac{r^{1/2}}{\Sigma} \frac{\rm d}{{\rm d}r} \left( \nu_1 \Sigma r^{1/2}\right) {\bf L} \right] + 
 \frac{1}{r} \frac{\rm d}{{\rm d}r} \left[ - \frac{3}{2} \nu_1 {\bf L} \right] \nonumber \\
 & + & \frac{1}{r} \frac{\rm d}{{\rm d}r} \left( \frac{1}{2} \nu_2 r | {\bf L} | 
 \frac{{\rm d} \hat{\bf l}}{{\rm d} r} \right) + \frac{ {\bm \omega}_{\rm LT} \times {\bf L}}{r^3}.
\end{eqnarray}
For now, we pick co-ordinates such that the spin of the black hole is aligned with the $z$-axis, so that the vector ${\bm \omega}_{\rm LT}$ in the above equation is,
\begin{eqnarray}
 {\bm \omega}_{\rm LT} & = & \omega_{\rm LT} \left( 0, 0, 1 \right), \\
 \omega_{\rm LT} & = & \frac{2 GJ}{c^2} = 2 ac \left( \frac{GM}{c^2} \right)^2.
\label{eq_warp_steady}
\end{eqnarray}
Task one is to determine the surface density profile of the disk. For a warped disk this is expressed via the angular momentum density,
\begin{eqnarray}
 {\bf L} & = & L \hat{\bf l}, \\
 L & = & | {\bf L} | = \left( GMr \right)^{1/2} \Sigma.
\end{eqnarray}
In terms of these variables, equation~(\ref{eq_warp_steady}) becomes,
\begin{eqnarray}
 0 = \frac{1}{r} \frac{\rm d}{{\rm d}r} \left[ \left( \frac{3r}{L} \frac{\rm d}{{\rm d}r} \left( \nu_1 L \right) -\frac{3}{2} \nu_1 \right) {\bf L} \right. + \nonumber \\
 \left. \frac{1}{2}\nu_2 r L \frac{{\rm d} \hat{\bf l}}{{\rm d}r} \right] + 
 \frac{{\bm \omega}_{\rm LT} \times {\bf L}}{r^3}.
\label{eq_BP1} 
\end{eqnarray}
To convert this to a scalar equation for $L$ we use the same trick as before. We dot through with $\hat{\bf l}$ and make use of the fact that $\hat{\bf l} \cdot {\rm d}\hat{\bf l} / {\rm d} r = 0$. The result is,
\begin{equation}
    0 = \frac{1}{r} \frac{\rm d}{{\rm d}r} \left[ 3 r \frac{\rm d}{{\rm d}r} \left( \nu_1 L \right) - \frac{3}{2} \nu_1 L \right].
\end{equation}
Assuming that $\nu_1$ is independent of radius, the solution is,
\begin{equation}
    L = c_2 r^{1/2} - 2 c_1,
\label{eq_BP_L}    
\end{equation}
where $c_2 = (GM)^{1/2} \Sigma_\infty$ and $\Sigma_\infty$ is the surface density at large disk radii. $c_1$ is a constant that depends on the inner disk boundary condition. The result, that for a constant viscosity $\Sigma$ tends to a constant as $r \rightarrow \infty$, is the same as for a planar disk (equation~\ref{eq_disk_nuSigma}). {\em Small} warps do not affect the steady state run of the disk surface density.

\begin{figure}
    \centering
    \includegraphics[width=\columnwidth]{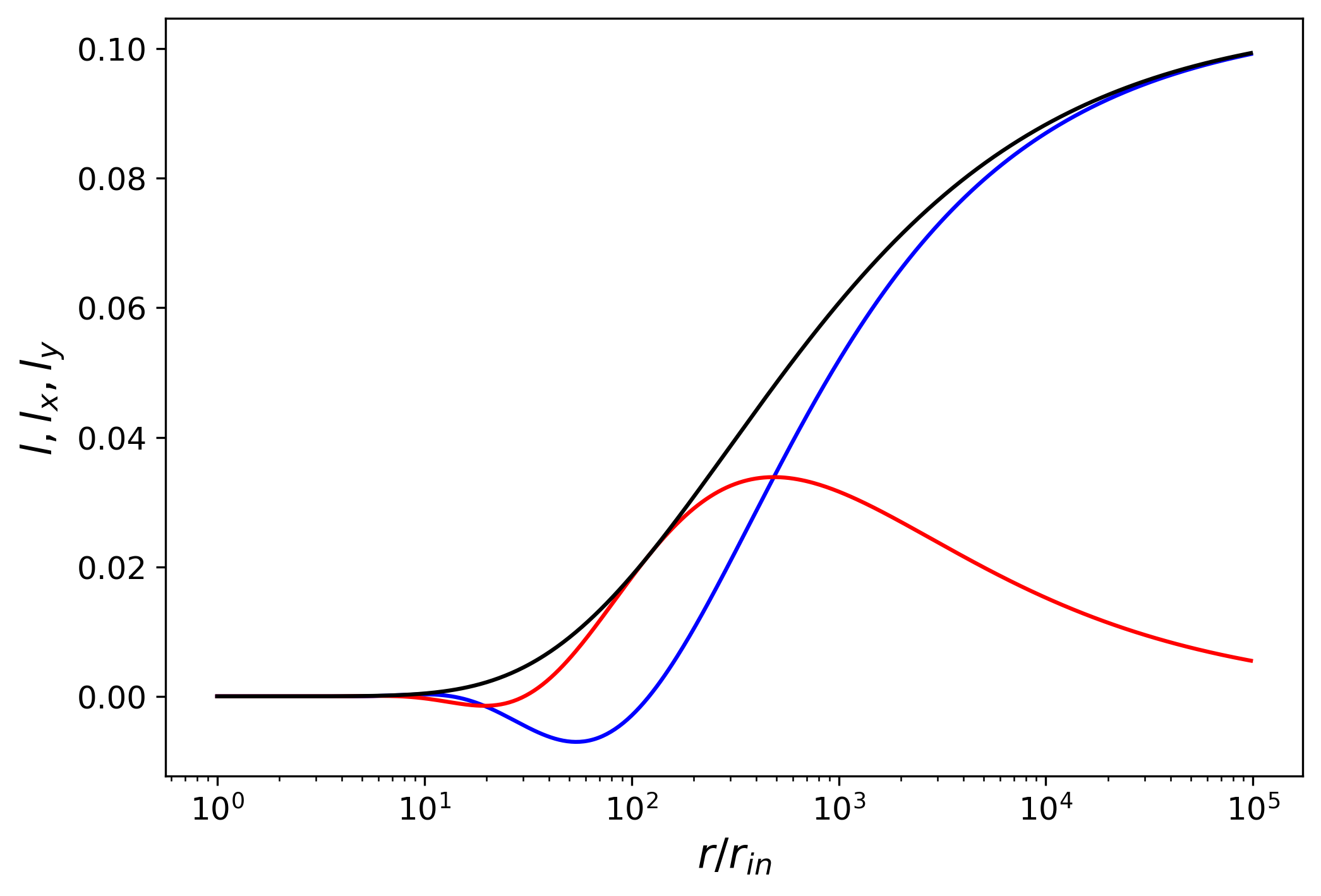}
    \caption{The shape of a warped disk around a Kerr black hole, computed in the viscous limit (with radially constant viscosities) using the \citet{scheuer96} solution. The blue, red and black curves show the components of the tilt, $l_x$ and $l_y$, and the total tilt $l = \sqrt{l_x^2 + l_y^2}$. For this example, $(\omega_{\rm LT}/\nu_2 r_{\rm in})^{1/2} = 300$. The \citet{scheuer96} solution applies to small warps, and hence the values on the $y$-axis are arbitrary. An accretion disk subject to the Bardeen-Petterson effect is predicted to be warped across a broad radial range, and to be modestly twisted.}
    \label{fig_scheuer_feiler}
\end{figure}

We now turn to the crux of the problem: determining the shape of the warp. Using the solution for $L$ (equation~\ref{eq_BP_L}) in equation~(\ref{eq_BP1}) gives,
\begin{eqnarray}
  \frac{\rm d}{{\rm d}r} \left[ 3 c_1 \nu_1 \hat{\bf l} + 
  \frac{1}{2} \nu_2 r \left( c_2 r^{1/2} - 2 c_1 \right) \frac{{\rm d} \hat{\bf l}}{{\rm d}r} \right] = \nonumber \\
  - \frac{ \left( c_2 r^{1/2} - 2 c_1 \right) {\bm \omega}_{\rm LT} \times \hat{\bf l}}{r^2},
\end{eqnarray}
where we have again assumed that $\nu_1$ is not a function of radius. Additionally, assume that the disk region of interest is sufficiently far away from the black hole that the inner disk boundary condition is immaterial. This allows us to drop $c_1$, leaving,
\begin{equation}
    \frac{\rm d}{{\rm d}r} \left[  
  \frac{1}{2} \nu_2 r^{3/2}  \frac{{\rm d} \hat{\bf l}}{{\rm d}r} \right] = 
  - \frac{ {\bm \omega}_{\rm LT} \times \hat{\bf l}}{r^{3/2}}.
\end{equation}
Writing $\hat{\bf l} = (l_x,l_y,l_z)$, $\bm \omega_{\rm LT} \times \hat{\bf l} = \omega_{\rm LT} (-l_y,l_zx,0)$. Because $\hat{\bf l}$ is a unit vector, only two components are independent, and it is convenient to cast the equation in terms of a complex variable,
\begin{equation}
    W \equiv l_x + i l_y.
\end{equation}
We obtain, for constant $\nu_2$, 
\begin{equation}
    r^{3/2} \frac{\rm d}{{\rm d}r} \left( r^{3/2} \frac{{\rm d}W}{{\rm d}r} \right) = 
    - \frac{2 i \omega_{\rm LT} W}{\nu_2}.
\end{equation}
This can simplified further with the substitution $X \equiv 2 r^{-1/2}$, to give,
\begin{equation}
    \frac{{\rm d}^2 W}{{\rm d}X^2} = - \frac{2 i \omega_{\rm LT} W}{\nu_2}.
\end{equation}
The solution is \citep{scheuer96},
\begin{equation}
    W = k \exp \left[ 2 \left( i - 1 \right) \left( \frac{\omega_{\rm LT}}{\nu_2 r} \right)^{1/2} \right],
\end{equation}
with $k$ a constant. As $r \rightarrow \infty$ we have $W \rightarrow k$, representing a flat inclined disk, while $W \rightarrow 0$ at small radius. The solution is well-behaved even though, as we have noted, we have dropped terms that are non-negligible close to the black hole. The solution is plotted as Figure~\ref{fig_scheuer_feiler}.

The shape of a disk subject to the Bardeen-Petterson effect is not something that is easily observed, so a fair bit of the interest in the problem lies in what the warped disk does to the black hole. Given enough time, we expect the angular momentum vectors of the black hole and disk to change, until we reach an end-state with a flat disk in the equatorial plane defined by the (final) black hole spin vector. The disk may be either aligned or counter-aligned with the black hole. General considerations of angular momentum conservation provide constraints on which of these outcomes occurs \citep{king05}, but to get at the time scale for realignment a disk model is indispensable. We sketch out how it's done in the case of the \citet{scheuer96} solution. For the disk, with angular momentum ${\bf J}_{\rm d}$, the rate of change of angular momentum due to the Lense-Thirring effect is,
\begin{eqnarray}
  \frac{{\rm d}{\bf J}_{\rm d}}{{\rm d}t} & = & \int \frac{ {\bm \omega}_{\rm LT} \times {\bf L} }{r^3} 2 \pi r {\rm d}r, \\   
  & = & \int \omega_{\rm LT} \left( -l_y, l_x, 0 \right) 2 \pi c_2 r^{-3/2} {\rm d}r, \\
  & = & 2 \pi i c_2 \omega_{\rm LT} \int W r^{-3/2} {\rm d}r.
\end{eqnarray}
Substituting for $W$ we can do the integral, with the result being,
\begin{equation}
  \frac{{\rm d}{\bf J}_{\rm d}}{{\rm d}t} = - \left( 1 - i \right) \pi k \left( GM \omega_{\rm LT} \nu_2 \right)^{1/2} \Sigma_\infty. 
\label{eq_SF_diskJ}  
\end{equation}
We now consider the black hole's spin. Its angular momentum changes at an equal and opposite rate to that of the disk,
\begin{equation}
\frac{{\rm d}{\bf J}_{\rm BH}}{{\rm d}t} = - \frac{{\rm d}{\bf J}_{\rm d}}{{\rm d}t}.
\end{equation}
Switching to a disk-centered coordinate system, the disk as $r \rightarrow \infty$ lies in the $x$-$y$ plane, rather than that plane coinciding with the equator of the black hole as before. Then $k = - (j_x + i j_y)$, where $j_x$ and $j_y$ are components of the unit vector describing the black hole spin. Equation~(\ref{eq_SF_diskJ}) then gives the time evolution of the black hole spin as \citep{scheuer96},
\begin{equation}
    \frac{{\rm d}(j_x + i j_y)}{{\rm d}t} = 
    - \pi (1 - i) \left( \frac{G \nu_2}{a c M} \right)^{1/2} \Sigma_\infty (j_x + i j_y).
\end{equation}
The imaginary part of this equation describes precession of the black hole, while alignment of the black hole spin with the angular momentum of the disk at large radii is described by the real part. Both precession and alignment occur on the same time scale,
\begin{equation}
    t_{\rm p} = t_{\rm align} = \frac{1}{\pi \Sigma_\infty} \left( \frac{acM}{G \nu_2} \right)^{1/2}
\label{eq_BP_alignment}    
\end{equation}
The dependence on disk parameters comes in via $\nu_2$ and $\Sigma_0$ (which, for a given steady accretion rate $\dot{M}$, would be inversely proportional to $\nu_1$).

This illustrative version of the analysis can be extended to account for radially varying viscosities \citep{natarajan99,martin07}. \citet{gerosa20}, who also study the effect of precession due to a binary companion, provide a detailed analysis. The numerical value of the alignment time scale (equation~\ref{eq_BP_alignment}) depends on the details of the disk model, but generically it works out to be {\em short}---compared in particular to the Salpeter time scale on which accretion would change the magnitude as well as the direction of ${\bf J}$. The ease with which a misaligned disk can change the spin axis of a black hole occurs because of the larger specific angular momentum of disk gas, and because in the viscous regime we expect $\nu_2 \gg \nu_1$. \citet{natarajan98}, for example, estimate that $t_{\rm align}$ may be of the order of one percent of the Salpeter time.

\subsubsection{Wave-like evolution equation}
Warps in protostellar disks, and in the inner regions of AGN disks, are expected (according to equation~\ref{eq_warp_regime}) to evolve in the wave-like regime, and the prior viscous analysis no longer holds. Viewed as $m=1$ bending waves, the linear evolution of warps is a subset of the general problem of disk waves (\S\ref{sec_waves}). It has been treated by \citet{papaloizou95}, \citet{demianski97}, and \citet{lubow00}. These papers use different notation and address distinct scientific questions, but fundamentally solve the same linear problem. As before, \citet{nixon16} is recommended for an accessible introduction.

The derivation of the equations for the linear evolution of warp waves in an accretion disk is given, as compactly as possible, as an Appendix in \citet{lubow00}. We won't repeat it here. If ${\bf G}(r,t)$ is the internal torque in the disk, the evolution of the unit tilt vector $\hat{\bf l}$ in the absence of external torques is given by the coupled equations,
\begin{eqnarray}
 \Sigma r^2 \Omega \frac{\partial \hat{\bf l}}{\partial t} & = & \frac{1}{r} \frac{\partial {\bf G}}{\partial r}, \\
 \frac{\partial {\bf G}}{\partial t} + \alpha \Omega {\bf G} & = & \frac{c_s^2 \Sigma r^3 \Omega}{4} \frac{\partial \hat{\bf l}}{\partial r}.
\end{eqnarray}
The right-hand-side of the second of these equations has some dependence on the disk's vertical structure \citep{lubow00}, with the form given here being true if it is isothermal. For a strictly inviscid disk ($\alpha=0$) we can combine these equations into a single wave equation for the disk tilt,
\begin{equation}
    \frac{\partial^2 \hat{\bf l}}{\partial t^2} = 
    \frac{1}{\Sigma r^3 \Omega} \frac{\partial}{\partial r} \left( 
    \frac{c_s^2 \Sigma r^3 \Omega}{4} \frac{\partial \hat{\bf l}}{\partial r} \right). 
\end{equation}
With a rescaling of the radial co-ordinate \citep[e.g][]{ogilvie06} this equation can be cast into the form of the classical wave equation, and one finds that linear bending waves propagate at speed \citep{papaloizou95},
\begin{equation}
    v_{\rm warp} = \frac{c_s}{2}.
\end{equation}
These waves are non-dispersive.

\subsubsection{Additional dynamical considerations}
Several other dynamical or hydrodynamical effects can come into play for warped disks. We list a few, with brief comments, here.

The dichotomy of having, separately, a non-linear theory for viscous disks with $\alpha > h/r$, and a linear theory (only) for wave-like warp evolution with $\alpha < h/r$, is obviously unsatisfactory. How does a disk with $\alpha \approx h/r$, or one where different radial parts fall into different regimes, evolve? \citet{martin19} proposed an evolution equation that unifies the two regimes, valid in the limit where the warp is small. Their work was extended by \citet{dullemond21}, who used a formalism developed by \citet{ogilvie13} to derive a closely related equation describing warped disk evolution.

We introduced this section by noting that the key physical reason why warped disks behave differently from planar ones is because the warp drives oscillatory, and often resonant, {\em radial} gas flows (Figure~\ref{fig_warp_forcing}). That these secondary flows can be unstable was suggested by \citet{papaloizou95b} and studied in detail by \citet{gammie00}. \citet{deng21} have simulated the resulting parametric instability in a global context, finding that the resulting turbulence strongly damps warps.

A warped disk is ``held together" by internal torques, which may not always be strong enough to prevent the disk breaking up into disjoint annuli. {\em Disk breaking} \citep{nixon12} occurs whenever a disk is subject to particularly strong differential precession, for example due to the Lense-Thirring effect \citep{nixon12b,liska21} or due to torques from a binary companion \citep{nixon13}. It may also be possible for a warped disk, with an initially smooth radial profile of $\hat{\bf l}$, to evolve {\em slowly} (e.g. on a viscous time scale) up to the point where regularity of the solution is lost, and a break develops \citep{dogan18}. The existence of this latter channel for breaking depends upon how the internal torques, $\nu_1$ through $\nu_3$ in equation~(\ref{eq_warp_evolution_ogilvie}), vary with the local warp strength $|\psi|$.

Warped disks are commonly found in binaries. The gravitational potential of a binary gives rise to rich purely gravitational dynamics, most famously the Kozai-Lidov effect \citep{lidov62,kozai62}, which leads to large-amplitude oscillations in eccentricity and inclination for test particles that are sufficiently inclined to the binary plane \citep[for a review see][]{naoz16}. \citet{martin14} identified analogous dynamics in simulations of initially misaligned fluid disks in binary systems. Linear analysis \citep{lubow17,zanazzi17} shows that the fluid version of the Kozai-Lidov effect depends upon $h/r$, and that unlike the free particle version can be present even for low values of the initial misalignment angle.

\subsection{Eccentric disks}
\begin{figure*}
    \centering
    \includegraphics[width=\textwidth]{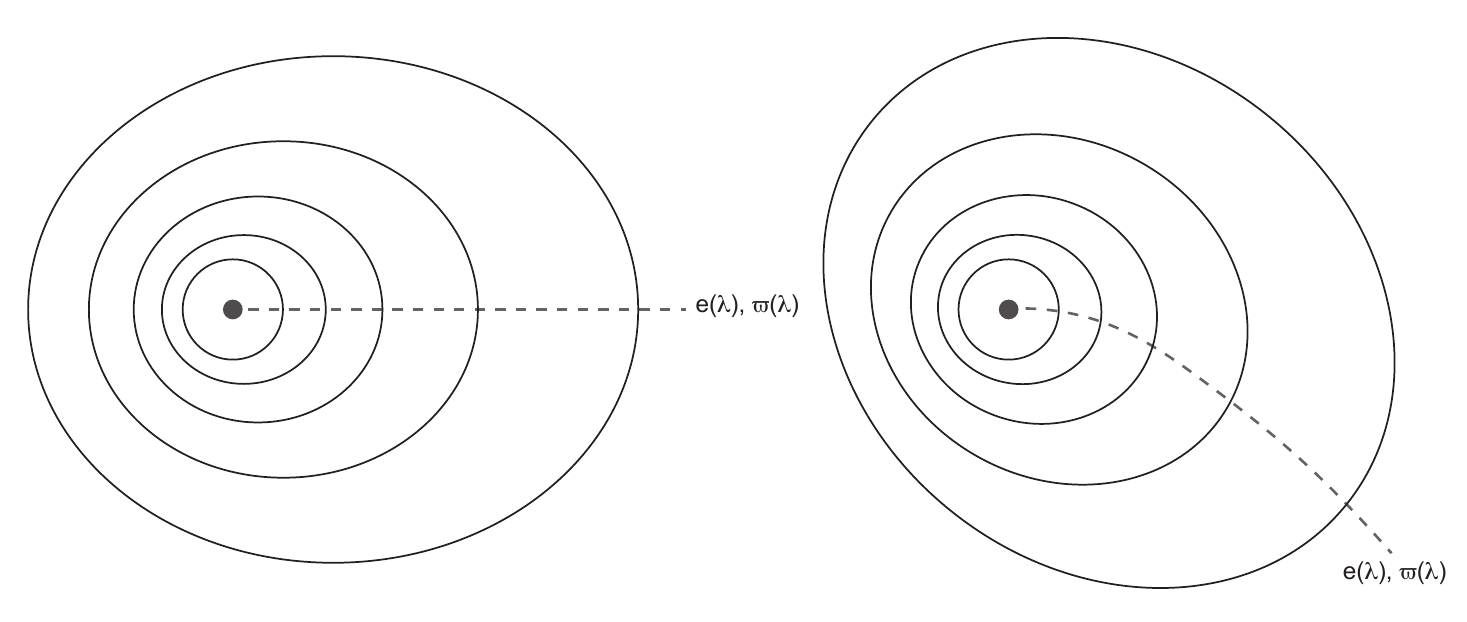}
    \caption{Example streamlines for two eccentric disks. In the left-hand example, the eccentricity $e$ varies radially, but the argument of pericenter $\varpi$ remains constant. In the more general case, shown on the right-hand side, both $e$ and $\varpi$ are functions of distance from the central object. It is often convenient to use the semilatus rectum $\lambda \equiv a (1 - e^2)$ as the radial co-ordinate when describing eccentric disks.}
    \label{fig_eccentric}
\end{figure*}

Returning to planar disks, we drop the prior assumption that orbits in the disk are circular (up to a small correction due to radial inflow) and instead allow them to be eccentric. This leads to new complications:
\begin{itemize}
    \item The time scale for establishing vertical hydrostatic equilibrium in an accretion disk, $t_{\rm hydro} \sim \Omega^{-1}$ (equation~\ref{eq_hydrostatic_equilibrium}), is of the same order as the orbital time scale. In an eccentric disk, the vertical acceleration due to the gravity of the central object changes on this time scale between apocenter and pericenter, dramatically so if $e$ is large. Eccentric disks are therefore never in true vertical hydrostatic equilibrium \citep{ogilvie14}, and the vertical structure couples strongly to the dynamics in situations---such as disks resulting from Tidal Disruption Events---where the eccentricity is large \citep{zanazzi20,ryu21,lynch21}.
    \item Depending upon the nature of angular momentum transport within the disk, an initially circular disk may exhibit {\em viscous overstability} \citep{ogilvie01,lyubarskij94}. If this is the case, the intuitive expectation that an eccentric disk ought to eventually circularize would be violated. Extension of the viscous $\alpha$-model (with a Navier-Stokes viscosity) to eccentric disks is not recommended, and instead we should think about how magnetic fields and the magnetorotational instability interact with disk eccentricity \citep{lynch21b,chan18}. A direct comparison of two-dimensional $\alpha$ disks with three-dimensional MHD simulations, in binary systems with mass ratios in the range where eccentricity is resonantly excited, shows that MHD effects work to damp  eccentricity \citep{oyang21}.
    \item Eccentric disks (in common with warped disks) exhibit a hydrodynamic parametric instability \citep{papaloizou05,wienkers18}. This can generate turbulence, for example in circumbinary disks whose eccentricity is maintained by external forcing \citep{pierens20}.
\end{itemize}
Figure~\ref{fig_eccentric} illustrates example streamlines for fluid in an eccentric disk. Both the eccentricity $e$ and the longitude of pericenter $\varpi$ may vary as functions of some radial co-ordinate used to label streamlines (the semilatus rectum, $\lambda = a (1-e^2)$, where $a$ is the semi-major axis of the orbit, is a good choice). These represent two additional functions whose evolution must be specified, unlike in the case of a thin circular disk where we have only the surface density. 

Although it does not address all of the complexities of eccentric disks, the linear theory of disk eccentricity is already useful. It is the foundation of models for how planets and binary systems affect their surrounding disks. We sketch here the main steps of the derivation by \citet{goodchild06}. 
An incomplete list of related perturbative analyses include those by \citet{kato83}, \citet{lee99}, \citet{tremaine01} and \citet{papaloizou02}.

Following \citet{goodchild06} we consider a two-dimensional inviscid disk model in cylindrical polar coordinates, with density $\rho(r,\phi)$, velocity ${\bf v}=(v_r,v_\phi)$, pressure $p$, and gravitational potential $\Phi$. The potential is assumed to be that of a central object, and to be axisymmetric\footnote{\citet{goodchild06} include an additional term representing a tidal potential, which we ignore here in the interests of thinking about the eccentricity evolution of isolated disks only.}. The fluid equations are,
\begin{eqnarray}
  \frac{\partial v_r}{\partial t} + v_r \frac{\partial v_r}{\partial r} + \frac{v_\phi}{r} \frac{\partial v_r}{\partial \phi} - \frac{v_\phi^2}{r} & = & -\frac{1}{\rho} \frac{\partial p}{\partial r} - \frac{\partial \Phi}{\partial r}, \label{eq_eccentric_fluid1} \\
    \frac{\partial v_\phi}{\partial t} + v_r \frac{\partial v_\phi}{\partial r} + \frac{v_\phi}{r} \frac{\partial v_\phi}{\partial \phi} + \frac{v_r v_\phi}{r} & = & -\frac{1}{r \rho} \frac{\partial p}{\partial \phi}, \label{eq_eccentric_fluid2}  
\end{eqnarray}
\begin{eqnarray}
    \frac{\partial \rho}{\partial t} + v_r \frac{\partial \rho}{\partial r} + \frac{v_\phi}{r} \frac{\partial \rho}{\partial \phi} &=&  \nonumber \\ -\frac{\rho}{r} \left[ \frac{\partial (r v_r)}{\partial r} + \frac{\partial v_\phi}{\partial \phi} \right],
\end{eqnarray}
\begin{eqnarray}
    \frac{\partial p}{\partial t} + v_r \frac{\partial p}{\partial r} + \frac{v_\phi}{r} \frac{\partial p}{\partial \phi} &=&  \nonumber \\ -\frac{\gamma p}{r} \left[ \frac{\partial (r v_r)}{\partial r} + \frac{\partial v_\phi}{\partial \phi} \right].
\label{eq_eccentric_fluid3}    
\end{eqnarray}
The pressure is related to the density with an adiabatic exponent $\gamma$.

The equilibrium state is a circular disk, with $v_r = \partial / \partial \phi = \partial / \partial t = 0$. This gives the usual radial force balance,
\begin{equation}
 \frac{v_\phi^2}{r} = \frac{{\rm d}\Phi}{{\rm d}r} + \frac{1}{\rho}\frac{{\rm d}p}{{\rm d}r}. 
\end{equation}
We now assume, here and subsequently, that the disk is thin such that $c_s \ll v_\phi$. The pressure gradient term is then negligible and,
\begin{equation}
    r \Omega^2 = \frac{{\rm d}\Phi}{{\rm d}r},
\end{equation}
with the angular velocity $\Omega = v_{\phi,0} / r$ being that in the unperturbed disk state.

Our goal is to reduce equations~(\ref{eq_eccentric_fluid1}) through (\ref{eq_eccentric_fluid3}) into a PDE for the time evolution of linear eccentricity perturbations to the base state. To that end, we pursue a variant of the sort of stability analysis we've done before, starting by linearizing and perturbing the fluid equations. We then Fourier analyze in the azimuthal co-ordinate {\em only}, restricting attention to the $m=1$ mode that we will identify with eccentricity. Proceeding informally we make the substitutions,
\begin{eqnarray}
 v_r & \rightarrow & v_r^\prime \exp[-i \phi], \\
 v_\phi & \rightarrow & r \Omega + v_\phi^\prime \exp[-i \phi], \\
 \rho & \rightarrow & \rho + \rho^\prime \exp[-i\phi], \\
  p & \rightarrow & p + p^\prime \exp[-i\phi]
\end{eqnarray}
where the perturbed quantities denoted with primes are understood to be small. Ditching second order terms in the perturbations,
\begin{eqnarray}
  \frac{\partial v_r^\prime}{\partial t} - i \Omega v_r^\prime -2 \Omega v_\phi^\prime & = & 
  -\frac{1}{\rho} \frac{\partial p^\prime}{\partial r} + \frac{\rho^\prime}{\rho^2} \frac{\partial p}{\partial r}, \\
  \frac{\partial v_\phi^\prime}{\partial t} - i \Omega v_\phi^\prime + \frac{v_r^\prime}{r} \frac{\partial}{\partial r} \left( r^2 \Omega \right) & = & \frac{i p^\prime}{r \rho}, \\
  \frac{\partial \rho^\prime}{\partial t} - i \Omega \rho^\prime + v_r^\prime \frac{\partial \rho}{\partial r}  &=& \nonumber \\ - \frac{\rho}{r} \left[ \frac{\partial}{\partial r} \left( r v_r^\prime \right) - i v_\phi^\prime \right], \\
  \frac{\partial p^\prime}{\partial t} - i \Omega p^\prime + v_r^\prime \frac{\partial p}{\partial r}  &=& \nonumber \\ - \frac{\gamma p}{r} \left[ \frac{\partial}{\partial r} \left( r v_r^\prime \right) - i v_\phi^\prime \right].  
\end{eqnarray}
These manipulations have eliminated the azimuthal derivatives, and give equations for the perturbed variables that are just functions of radius and time.

Next, we look for the lowest order solution to the {\em perturbation} equations. It's fine to assume that the time scale for eccentricity evolution is long compared to the orbital time scale, so that $|\partial v_r^\prime / \partial t | \ll |i \Omega v_r^\prime|$. The pressure terms are also negligible at lowest order for a thin disk, so the first two of the equations above both imply,
\begin{equation}
    2 v_\phi^\prime = -i v_r^\prime.
\end{equation}
Accordingly, we can write,
\begin{eqnarray}
 v_r^\prime & = & i r \Omega E(r,t), \\
 v_\phi^\prime & = & \frac{1}{2} r \Omega E (r,t).
\end{eqnarray}
The perturbations to the radial and azimuthal velocity are out of phase, and differ in magnitude by a factor of two. For eccentricities $e \ll 1$ these characteristics describe a two-body eccentric orbit, and we can therefore identify $E(r,t)$ with a complex eccentricity,
\begin{equation}
    E(r,t) = e \exp[i\varpi].
\end{equation}
Both the eccentricity $e$ and the longitude of pericenter $\varpi$ are functions of radius and of time.

At this point physically we are done. The perturbation equations can be combined to give a single equation for the time evolution of the complex eccentricity,
\begin{equation}
    2 r \Omega \frac{\partial E}{\partial t} = \frac{i E}{\rho} \frac{\partial p}{\partial r} + \frac{i}{r^2 \rho} \frac{\partial}{\partial r} \left( \gamma p r^3 \frac{\partial E}{\partial r} \right).
\label{eq_goodchild_ogilvie}    
\end{equation}
The first term on the right-hand-side describes precession in the presence of a radial pressure gradient, while the second describes radial diffusion of the complex eccentricity. We can look for normal mode solutions to this equation by writing,
\begin{equation}
    E(r,t) = E(r) \exp[i \omega],
\end{equation}
where $\omega$ is an eigenvalue determined as the solution of,
\begin{equation}
    2 r \Omega \omega E = \frac{E}{\rho} \frac{{\rm d} p}{{\rm d} r} + \frac{1}{r^2 \rho} \frac{\rm d}{{\rm d} r} \left( \gamma p r^3 \frac{{\rm d} E}{{\rm d} r} \right).  
\end{equation}
In this highly simplified and inviscid model, the eccentricity evolves as a superposition of precessing normal modes, whose structure and amplitude are set by the initial and boundary conditions. 

The reader who wants to delve further into the linear theory of disk eccentricity can find it in \citet{goodchild06}, which contains a lot of interesting stuff. It turns out that including the effects of viscosity leads to only a small modification of equation~(\ref{eq_goodchild_ogilvie}),
\begin{equation}
    2 r \Omega \frac{\partial E}{\partial t} = \frac{i E}{\rho} \frac{\partial p}{\partial r} + \frac{i}{r^2 \rho} \frac{\partial}{\partial r} \left[ \left(\gamma - i \alpha_{\rm b}\right) p r^3 \frac{\partial E}{\partial r} \right].
\end{equation}
Here $\alpha_{\rm b}$ is a Shakura-Sunyaev style {\em bulk} viscosity. It acts straightforwardly in the context of this theory to damp disk eccentricity. Stepping back, however, one might ask whether a Navier-Stokes viscosity (with shear and bulk terms) really applies to a turbulent accretion disk, and if not what sort of effective bulk viscosity is generated by the MRI or some other physical angular momentum transport process. These are assuredly {\em not} straightforward questions.
    
\section{Classical disk instabilities}
Thus far, the instabilities we have discussed have been linear instabilities of the fluid equations (mostly in the inviscid Euler limit), with the addition of magnetohydrodynamics or self-gravity. They are ``disk" instabilities inasmuch as the choice of unperturbed state is appropriate for Keplerian disk flow. Interest in their non-linear evolution centers on their role in angular momentum transport, turbulence, and diffusion.   

These fluid instabilities are not the only class of instability of interest. There is another. The equations describing disk evolution (for example the diffusive equation for the surface density, equation~\ref{eq_1D_disk_evolution}), can also exhibit instabilities. These instabilities are less fundamental than, say, the MRI, because the equations themselves have baked in various assumptions that are only approximate. They are nonetheless of great interest, first because they underly what was historically a great success of accretion disk theory---the identification of {\em thermal disk instability} with observed dwarf nova outbursts---and second because they allow a simplified analysis of complex physical problems such disk warping.

\subsection{Viscous instability}
The condition for viscous stability is determined by first considering a steady-state solution $\Sigma(r)$ to the one-dimensional disk evolution equation~(\ref{eq_1D_disk_evolution}). Following \citet{pringle81} 
we make the substitution $\mu \equiv \nu \Sigma$ and consider perturbations $\mu \rightarrow \mu + \delta \mu$. We assume that $\nu = \nu(\Sigma)$. Substituting in the evolution equation~(\ref{eq_1D_disk_evolution}), the 
perturbation $\delta \mu$ evolves as,
\begin{equation}
 \frac{\partial}{\partial t} \left( \delta \mu \right) = 
 \frac{ \partial \mu}{\partial \Sigma} \frac{3}{r} 
 \frac{\partial}{\partial r} \left[ r^{1/2} 
 \frac{\partial}{\partial r} \left( r^{1/2} \delta \mu \right) \right].
\end{equation} 
The perturbation $\delta \mu$ grows if the diffusion coefficient, proportional to 
$\partial \mu / \partial \Sigma$, is negative. This defines the condition for {\em viscous instability}, which takes the simple form,
\begin{equation}
 \frac{\partial}{\partial \Sigma} \left( \nu \Sigma \right) < 0.
\end{equation}
Toy disk models where $\nu$ is a constant, or a fixed function of radius, are evidently stable, as would be expected given the diffusive nature of the governing equation. In principle, however, there is no reason why some physical mechanism for angular transport could not yield an effective viscosity law that implied viscous instability. Such a disk would have an intrinsic tendency to break up into rings. How viscous instability would saturate is not entirely clear, though the sharpness of ring-like structures in disks is known to be limited by the onset of the Rossby Wave Instability \citep[RWI;][]{lovelace99}\footnote{RWI physics was discussed earlier in a galactic context, under the monikor of the ``negative mass instability" \citep{lovelace78}.}. The RWI, in turn, leads to the formation of non-axisymmetric structure in the form of vortices \citep{li01,richard13}.

\subsection{Thermal instability}
\label{sec_thermal_instability}
For a geometrically thin disk in thermal equilibrium, 
\begin{equation}
    Q_+ = Q_-,
\end{equation}
where $Q_+$ is the heating rate per unit surface area of the disk and $Q_-$ is the corresponding cooling rate. Consider a perturbation to the mid-plane temperature $T_{\rm c}$, on some scale $\lambda \gg h$, so that cooling of the perturbation occurs predominantly vertically rather than radially. Thermal instability occurs if,
\begin{equation}
    \frac{{\rm d} \log Q_+}{{\rm d} \log T_{\rm c}} > \frac{{\rm d} \log Q_-}{{\rm d} \log T_{\rm c}}.
\end{equation}
When the condition for thermal instability is satisfied, either upward or downward perturbations to $T_{\rm c}$ lead to runaway.

The instability condition is readily evaluated for various flavors of $\alpha$-model disks. Starting with the right-hand-side, which represents cooling, $Q_- = 2 \sigma T_{\rm eff}^4$. Using the one-zone relation between the central and effective temperature (equation~\ref{eq_Tc_Teff}), 
\begin{equation}
    Q_- = 2 \sigma T_{\rm eff}^4 \propto \frac{T_c^4}{\tau} \propto \frac{T_c^4}{\Sigma \kappa},
\label{eq_qminus}    
\end{equation}
with $\kappa$ the (mid-plane) opacity. For a {\em constant} opacity, for example that due to electron scattering, we find that ${\rm d} \log Q_- / {\rm d} \log T_c = 4$.

What about the left-hand-side? Using equation~(\ref{eq_qplus}) together with the $\alpha$-prescription, 
\begin{equation}
    Q_+ = \frac{9}{4} \alpha c_s^2 \Sigma \Omega_{\rm K}.
\end{equation}
For a gas pressure dominated fluid $c_s^2 \propto T_c$, and we find that ${\rm d} \log Q_+ / {\rm d} \log T_c = 1$. The simplest limit of a gas pressure dominated disk, with a temperature-independent opacity, is thus thermally stable.

Everything changes if radiation pressure is dominant. For a radiation pressure dominated gas, the sound speed $c_s^2 \propto T_c^4 / \rho$, and
\begin{equation}
    Q_+ \propto T_c^4 h.
\end{equation}
Noting that the scale height $h$ scales with the central pressure, and hence with $T_c^4$, we find that ${\rm d} \log Q_+ / {\rm d} \log T_c = 8$. The inner, radiation pressure dominated / electron scattering opacity zone of the Shakura-Sunyaev model, is analytically expected to be thermally unstable! The presence of such an instability would be expected to drive large-amplitude short-time scale variability of accreting black hole systems \citep{lightman74,shakura76}. Although accreting black holes are indeed variable sources, the character of the observed variability does not match simple expectations for a thermal instability origin. In particular, stellar-mass black holes, whose accretion rates cycle between values where the inner disk would be either gas or radiation pressure dominated, do not show obvious changes in their variability properties at the point where thermal instability would be predicted to set in.

The existence of thermal instability in radiation pressure dominated $\alpha$ disks depends upon multiple assumptions hardwired into the $\alpha$ model, many of which are known to be false, at least when examined closely enough. \citet{agol01}, and many subsequent authors, studied the thermal stability of radiation pressure dominated disks using either viscous or MHD simulations. Current results suggest that weakly magnetized disks {\em would} exhibit thermal instability \citep{jiang13,mishra16}. Disks threaded by enough net flux as to be at least marginally magnetic pressure dominated, however, are stable \citep{begelman07,oda09,sadowski16,jiang19}, and this is a possible resolution of the apparent discrepancy between observations and theory highlighted above.

The radiation pressure version of thermal instability, relevant for very hot disks, is driven by the behavior of the heating term in the thermal equilibrium equation. One can also get instability, at much lower temperatures, from the cooling term. From equation~(\ref{eq_qminus}) it is clear that one can get a small value of ${\rm d} \log Q_- / {\rm d} \log T_c$, and thermal instability, if the opacity $\kappa$ is a sufficiently strongly increasing function of temperature. Physically, this occurs when hydrogen is partially ionized at $T_c \simeq 10^4 \ {\rm K}$, and the opacity is dominated by H$^-$. Analytic fits to the opacity in this regime yield 
$\kappa \propto T_c^{10}$ \citep[e.g.][]{bell94}, which is easily strong enough to source thermal instability under gas pressure dominated disk conditions.  

\subsubsection{The S-curve and limit cycles}
Thermal instability, associated with the ionization of hydrogen, is well-accepted as the root cause of outbursts in dwarf nova systems. Dwarf novae are a subclass of cataclysmic variables \citep{warner95} in which a weakly or non-magnetized white dwarf accretes from a low-mass main sequence star in a Roche-lobe filling mass transfer binary system. The prototypical dwarf nova outburst, seen in systems such as U~Geminorum, is observed as roughly 5~magnitude increases in optical brightness, that last about a week and recur every few months.

\begin{figure*}
    \centering
    \includegraphics[width=\textwidth]{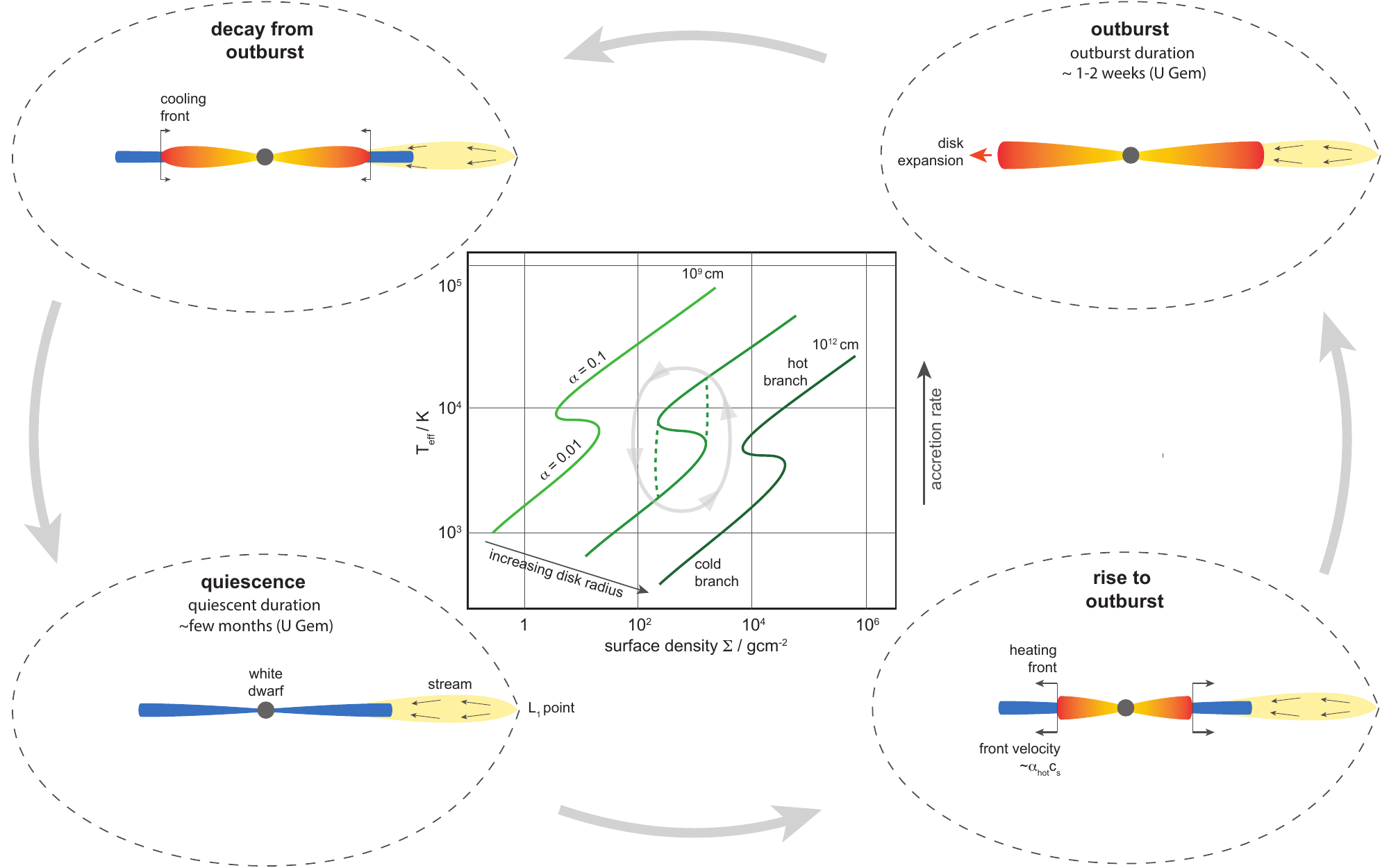}
    \caption{Schematic illustration of how the existence of a {\em local} S-curve leads to {\em global} outbursts in dwarf novae accretion disks. The S-curves shown for different disk radii in the $\Sigma$-$T_{\rm eff}$ plane are simplified from calculations by \citet{bollimpalli18}, for a $1.35 \ M_\odot$ white dwarf. The depicted global evolution, in which outbursts start from the inside-out and decay from the outside-in, is typical for dwarf nova models but not a guaranteed property.}
    \label{fig_scurve}
\end{figure*}

The identification of dwarf nova outbursts with accretion disk thermal instabilities was made in the 1970s \citep{osaki74,hoshi79}, and developed into a working quantitative model by many authors in the early 1980s \citep{meyer81,cannizzo82,minishige83,faulkner83,smak84}. Excellent reviews include those by \citet{lasota01} and \citet{hameury20}.

Dwarf nova outbursts involve both local and global disk processes. At a local level, a thermally unstable annulus of the disk has an unstable thermal equilibrium for some given surface density $\Sigma$ and angular velocity $\Omega_{\rm K}$. Such an annulus would evolve on the thermal time scale $\sim 1/(\alpha \Omega_{\rm K})$ to one of two stable states: a cold state where the central temperature is such that hydrogen is substantially neutral, or a hot state where it is substantially ionized. We can visualize this state of affairs by plotting the thermal equilibria for disk annuli in the $(\Sigma, T_{\rm eff})$ plane. Figure~\ref{fig_scurve}, which is a schematic based on calculations by  \citet{bollimpalli18}, shows what this looks like. The thermal equilibria have a characteristic ``S-curve" appearance, such that two stable equilibria (and an intermediate unstable one) exist across a range of surface densities. The computed S-curves are a function of white dwarf mass, disk radius, and assumed $\alpha$ value(s).

At this point it's worth highlighting an important subtlety of the dwarf nova disk instability model. The existence of a classical thermal instability, here due to the behavior of the opacity near the hydrogen ionization temperature, is not a sufficient condition for producing a clear S-curve. Almost equally important are changes to the disk vertical structure that occur due to vertical energy transport by convection. Even then, the computed S-curves are rather lackluster. Prominent S-curves, of the form shown in Figure~\ref{fig_scurve}, occur if we make the further assumption that $\alpha$ on the hot branch is {\em significantly larger} than on the cold branch. Assumed ratios $\alpha_{\rm hot} /\alpha_{\rm cold} \approx 10$ do the trick.

Figure~\ref{fig_scurve} shows qualitatively the limit cycle that develops in dwarf novae disks as a consequence of the S-curve. During quiescence, gas from the Roche lobe filling secondary streams onto the outer accretion disk at a rate that exceeds what the disk can transport while on the cold branch of the S-curve. Mass accumulates in the disk, until at some radius the surface density exceeds the maximum surface density, $\Sigma_{\rm max}$, permitted on the cold branch. Usually, but not always, the critical point is close to the white dwarf. Once $\Sigma > \Sigma_{\rm max}$ the only allowable local thermal equilibrium solution for the disk lies on the hot branch. The annulus that has been triggered rapidly heats up on the thermal time scale $\sim 1/(\alpha \Omega_{\rm K})$. Radial diffusion of mass and heat can then act to trigger the cold~$\rightarrow$~hot transition in neighboring annuli, and a heating front sweeps through the disk until every annulus is on the hot branch. Once fully in outburst, accretion onto the white dwarf exceeds the rate of mass supply from the secondary, and the surface density drops, triggering a return to quiescence.

Quantitative models for disk instability rely on separate calculations for the vertical structure and radial evolution (``1+1D"). In the radial direction, it is necessary to treat departures from thermal equilibrium, and therefore the usual diffusive equation for the surface density (equation~\ref{eq_1D_disk_evolution}) is supplemented with an equation for the evolution of the central temperature $T_c$,
\begin{equation}
    \frac{\partial T_c}{\partial t} = \frac{Q_+ - Q_- + J}{c_p \Sigma} 
    - \frac{{\cal R}T_c}{\mu c_p} \frac{1}{r} \frac{\partial (r v_r)}{\partial r}
    - v_r \frac{\partial T_c}{\partial r}.
\end{equation}
Here $\cal R$ is the gas constant, $\mu$ is the mean molecular weight, and $c_p$ is the specific heat capacity at constant pressure. $J$ is the radial flux of energy due to basically diffusive processes, which could be radiative diffusion and / or turbulent radial heat transport. Different forms for $J$ have been used in the literature \citep{hameury20,cannizzo93}.

The agreement between disk models and observations of dwarf novae \citep[e.g.][] {cannizzo93} is impressive\footnote{After pages and pages of beating up on the $\alpha$ model, you might have been wondering why anyone takes it seriously. This is a large part of the reason why.}, but leaves open questions such as why the efficiency of angular momentum transport on the hot branch should be much higher than on the cold branch. Local numerical simulations suggest that convection enhances the strength of MRI transport near the tip of the hot branch, while Ohmic diffusion can damp MHD turbulence on the cold branch \citep{hirose14,coleman16,coleman18,scepi18}. As in other accreting systems, net magnetic fields and disk winds may contribute to the observed behavior \citep{scepi19}.

\subsection{Radiation warping instability}
A warped disk that intercepts and re-radiates radiation from a central source experiences a radiative torque that modifies the warp. \citet{pringle96} showed that the simplest model of this physical situation---a razor-thin viscous disk illuminated by a small central point source of radiation---is subject to a linear instability that can warp an initially almost planar disk. The derivation in \citet{pringle96} is clear, and we won't repeat it here. Further linear analysis was given by \citet{maloney96}.

The radiation warping instability sets in beyond a critical radius, which depends upon the luminosity and viscosity of the disk. For a steady, self-luminous disk with radiative efficiency $\eta$, the critical radius is given in terms of the Schwarzschild radius as,
\begin{equation}
    \frac{r}{r_{\rm S}} \gtrsim \frac{8 \pi^2}{\eta^2} \left( \frac{\nu_2}{\nu_1} \right)^2.
\end{equation}
Here $\nu_1$ and $\nu_2$ are the horizontal and vertical viscosities discussed earlier in \S\ref{sec_viscous_warps}.

If the conditions are right to trigger it, calculations show that the non-linear evolution of the instability under X-ray binary conditions leads to evolution that is qualitatively consistent with observations \citep{wijers99}. It is not so obvious, however, whether the conditions are right. The hydrodynamic expectation that $\nu_2 / \nu_1 \approx 1 / (2 \alpha^2) \gg 1$ (equation~\ref{eq_viscosity_ratio}) implies that accretion disks are quite ``stiff", and would only be unstable at unreasonably large radii for typically assumed values of $\alpha$ and $\eta$. For example, even assuming quite large values of $\alpha = 0.3$ and $\eta = 0.3$, instability only sets in for $r \gtrsim 3 \times 10^4 \ r_{\rm S}$. An interesting possibility is that analogous instabilities might be driven by the larger torques from pressure or magnetic forces if there is a wind from the disk \citep{lai03,schandl94}.

\subsection{Other conjectured instabilities}
Many other accretion disk instabilities have been suggested, with varying degrees of theoretical and / or observational evidence. Here are a few.

{\em Gravo-magneto limit cycle instability}. Young Stellar Objects (YSOs) show eruptive behavior, known as FU~Orionis outbursts, that looks very much like a slowed-down version of dwarf nova outbursts \citep{audard14}. Thermal instability models for FUOrs are possible \citep{bell94}, but require high disk masses and unusually low values of $\alpha$. A modified limit cycle, in which non-ideal MHD processes strongly damp turbulence on the cold branch, and the onset of self-gravity triggers the transition to the hot branch, has been suggested for these systems \citep{gammie99b,armitage01,zhu09,martin11}.

{\em Thermal wave or irradiation instability}. A disk whose thermal structure is dominated by irradiation from a central source has a scale height $h(r)$ whose profile is fixed by the {\em shape} of the absorbing surface of the disk. For YSOs and protoplanetary disks, consistent equilibrium disk shapes are modestly flared to large radius \citep{kenyon87,chiang97}. It is easy to imagine how such an equilibrium might be unstable. If the disk surface develops a small ``ripple", the inward-facing part of the perturbation will intercept more radiation from the central source, and the outward-facing part less. The ripple will strengthen---perhaps to the point that an actual shadow is cast---and propagate radially. This is an old idea \citep{cunningham76,dalessio99,dullemond00,watanabe08} that has received renewed study \citep{ueda21,wu21} as a potential cause of the annular structures observed in protoplanetary disks.

{\em Magnetic Prandtl number driven instability}. The microphysical viscosity $\nu$ and resistivity $\eta$ of a plasma have the same dimensions, and from them one can construct a dimensionless ratio, the {\em magnetic Prandtl number},
\begin{equation}
    {\rm Pm} \equiv \frac{\nu}{\eta}.
\end{equation}
Numerical simulations, both of isotropic stirred MHD turbulence \citep{schekochihin04} and of the MRI in shearing boxes \citep{fromang07}, exhibit a dependence on ${\rm Pm}$. The modest value of fluid Reynolds number realized numerically means that these results must be approached cautiously, but they are consistent with a plausible physical argument in which the value of the Prandtl number affects the rate of reconnection and hence large-scale fluid properties. \citet{balbus08} noted that the value of ${\rm Pm}$ in X-ray binary disk models goes from ${\rm Pm} \gg 1$ close to the black hole to ${\rm Pm} \ll 1$ at $10^3 \ r_{\rm S}$, which might feed forward into observable time-dependent behavior \citep{potter14}. A possible confounding factor is the importance of radiation viscosity in the relevant disk regions.

\section{Geometry of accretion}
In several circumstances, the existence of a particular geometry (spherical symmetry, axisymmetry) allows us to calculate how gas flows from large radii toward the accreting object. We reproduce here a couple of well-known results, focusing on the simplest cases.

\subsection{Spherical (Bondi) accretion}
\label{sec_Bondi}

A point mass $M$ is at rest with respect to gas, uniform at infinity, with density $\rho_\infty$ and sound speed $c_s$. The problem is spherically symmetric and the mass accretion rate can be estimated from dimensional analysis. A spatial scale can be constructed as,
\begin{equation}
    R_{\rm B} = \frac{GM}{c_s^2},
\end{equation}
called the {\em Bondi radius}. Multiplying the spherical surface area $4 \pi R_{\rm B}^2$ by the characteristic density and by the characteristic velocity, the accretion rate is,
\begin{equation}
    \dot{M}_{\rm B} \sim 4 \pi \frac{\left(GM\right)^2}{c_s^3} \rho_\infty.
\end{equation}
This is the {\em Bondi accretion rate}.

The dimensional analysis gives the correct estimate, but we can solve the problem exactly \citep{bondi52}. We need the continuity and momentum equations, which for a steady spherically symmetric flow read,
\begin{eqnarray}
 \dot{M} & = & 4 \pi r^2 v_r \rho, \\
 v_r \frac{{\rm d}v_r}{{\rm d}r} & = & - \frac{1}{\rho} 
 \frac{{\rm d}p}{{\rm d}r} - \frac{GM}{r^2}.
\end{eqnarray}
We also need an equation of state. The simplest case is the isothermal one, with $p = \rho c_s^2$. The momentum equation is then,
\begin{equation}
    v_r^2 \frac{{\rm d}\ln v_r}{{\rm d}r} = -c_s^2 
    \frac{{\rm d}\ln \rho}{{\rm d}r} - \frac{GM}{r^2}.
\end{equation}
We take the log of the continuity equation and differentiate with respect to radius to obtain,
\begin{equation}
  0 = \frac{2}{r} + \frac{{\rm d}\ln \rho}{{\rm d}r} 
  + \frac{{\rm d}\ln v_r}{{\rm d}r},
\end{equation}
and use this relation to eliminate density from the momentum equation. The result is,
\begin{equation}
    \left( v_r^2 - c_s^2 \right) \frac{{\rm d}\ln v_r}{{\rm d}r} = 
    \frac{2 c_s^2}{r} \left( 1 - \frac{GM}{2 r c_s^2} \right).
\end{equation}
This equation can describe a number of physical circumstances, but the one that is almost always of interest astrophysically is the one where the flow speed $\rightarrow 0$ at large radius, and approaches some large value $\gg c_s$ near the central object (this {\em must} be true for a black hole, and is usually true even for objects with surfaces). There must therefore be a radius, called the {\em sonic radius} $r_s$, where $v_r^2 = c_s^2$, where the flow makes a transition between subsonic and supersonic motion. Inspection of the above equation pins the sonic radius at,
\begin{equation}
    r_s = \frac{GM}{2 c_s^2}.
\end{equation}
Writing $\rho_s = \rho(r_s)$, the accretion rate for an isothermal equation of state is just $\dot{M} = 4 \pi r_s^2 c_s \rho_s$.

To complete the solution, we need to relate the density at the sonic point $\rho_s$ to the density at infinity $\rho_\infty$. This is accomplished using the Bernoulli constant,
\begin{equation}
    H = \frac{1}{2} v_r^2 + \int \frac{{\rm d}p}{\rho} - \frac{GM}{r}.
\end{equation}
For an isothermal gas at temperature $T$ we have that ${\rm d}p = ( {\cal R}/\mu ) T {\rm d}\rho$, so $H = (1/2)v_r^2 + c_s^2 \ln \rho - GM / r$. The relation between the general conditions in the flow and those at the sonic point is then,
\begin{equation}
    \frac{1}{2} v_r^2 + c_s^2 \ln \rho - \frac{GM}{r} = 
    \frac{1}{2} c_s^2 + c_s^2 \ln \rho_s - \frac{GM}{r_s}.
\end{equation}
On substituting for $r_s$,
\begin{equation}
    v_r^2 = 2 c_s^2 \left[ \ln \left( \frac{\rho_s}{\rho} \right) - \frac{3}{2} \right] + \frac{2 GM}{r}. 
\end{equation}
Noting that as $r \rightarrow \infty$ the radial velocity vanishes, we find that $\ln (\rho_s / \rho_\infty) = 3/2$ and the accretion rate is,
\begin{equation}
    \dot{M} = \pi e^{3/2} \frac{(GM)^2}{c_s^3} \rho_\infty.
\end{equation}
This differs by about 10\% from the answer suggested by dimensional analysis.

The extension to an adiabatic equation of state is not hard, and is covered in many textbooks \citep[e.g.][]{clarke07,frank02}. The same analysis applies equally to the hydrodynamics of inflows and outflows, and in the latter case is the basis of the {\em Parker wind} solution that describes a thermally driven stellar wind \citep{parker58}.

\subsection{Supersonic accretion from an external medium}
A second limit amenable to analytic analysis occurs when a point-mass accretor moves supersonically through a medium that is uniform at infinity. Examples include an isolated neutron star or black hole traversing a molecular cloud, or a compact object accreting from the wind of a massive star in a binary system. This is {\em Bondi-Hoyle-Lyttleton accretion} \citep{bondi44,hoyle39}. Although it is, in principle, at least an axisymmetric fluid problem, a calculation that largely ignores pressure effects is instructive and physically useful. \citet{edgar04} provides a clear review, and we follow his treatment of the basics here.

\begin{figure}
    \centering
    \includegraphics[width=\columnwidth]{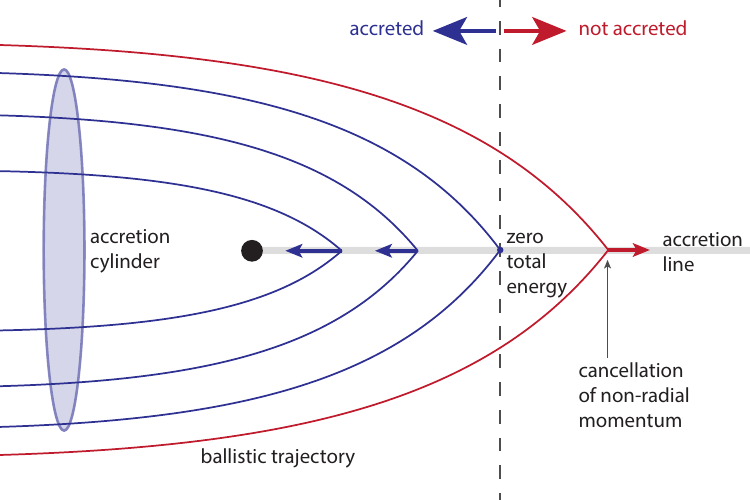}
    \caption{Illustration of the physics of Bondi-Hoyle-Lyttleton accretion. A point mass moving supersonically through a uniform medium focuses streamlines of the fluid toward a trailing {\em accretion line}. Streamlines collide in diametrically opposed pairs when they reach the accretion line, retaining only their radial momentum. Close to the point mass, the total energy after collision is negative, and the gas accretes.}
    \label{fig_BHL1}
\end{figure}

The physical argument for how Bondi-Hoyle-Lyttleton accretion proceeds is illustrated in Figure~\ref{fig_BHL1}. The flow is evidently axisymmetric at large distances from the accreting object, and we assume that there are no instabilities that spoil this symmetry closer in. We further assume that, because the motion is supersonic, fluid follows essentially ballistic trajectories in the gravitational field of the point mass. Given these assumptions, pairs of streamlines that pass by the mass on opposite sides are focused toward collision points that lie on a straight line behind the accreting object. The collision cancels out the non-radial component of the momentum, leaving only a radial component $v(r)$ (which we will calculate shortly). The specific energy, $(1/2)v^2 - GM/r$, is an increasing function of distance $r$ along the accretion line. Gas with negative specific energy after collision flows back along the accretion line and gets accreted, while that further away flows outward. The accretion rate is then given by the mass flux into a cylinder, whose size is defined by the impact parameter of the critical streamline that has zero specific energy after collision on the accretion line.

\begin{figure}
    \centering
    \includegraphics[width=\columnwidth]{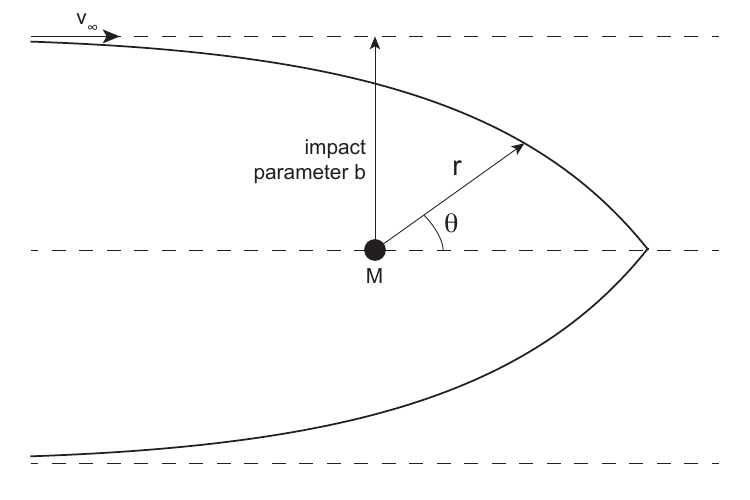}
    \caption{Setup for the calculation of the Bondi-Hoyle-Lyttleton accretion rate.}
    \label{fig_BHL2}
\end{figure}

Quantifying the above argument requires calculating the impact parameter of the hyperbolic orbit that leads to zero energy after collision along the accretion line \citep{edgar04}. The setup for the calculation is shown in Figure~\ref{fig_BHL2}. A point mass $M$ moves with velocity $v_\infty$ with respect to a uniform medium of density $\rho_\infty$. We work in polar co-ordinates $(r,\theta)$, and consider the orbit of a point particle with impact parameter $b$. Ignoring fluid effects, the equations of motion are,
\begin{eqnarray}
 \ddot{r} - r \dot{\theta}^2 & = & -\frac{GM}{r^2}, \\
 r^2 \dot{\theta} & = & b v_\infty.
\end{eqnarray}
The dots denote time derivatives. The specific angular momentum $h=b v_\infty$.

Solving the above equations to determine the shape of the orbit is a standard exercise in celestial mechanics. Substituting $u \equiv r^{-1}$, and repeatedly using the chain rule, one obtains,
\begin{equation}
    \frac{{\rm d}^2 u}{{\rm d} \theta^2} + u = \frac{GM}{h^2}.
\end{equation}
The solution of this ODE can be written in various ways. Often one works toward the real space solution,
\begin{equation}
    r = \frac{h^2 / GM}{1 + e \cos \left( \theta - \varpi\right)},
\end{equation}
describing a general conic section, with $e$ and $\varpi$ being constants that one identifies with the orbital eccentricity and longitude of pericenter. For our purposes, it's a bit simpler to instead write the solution as,
\begin{equation}
    u = c_1 \cos \theta + c_2 \sin \theta + \frac{GM}{h^2},
\end{equation}
with $c_1$ and $c_2$ being constants. The constants are determined by requiring that as $\theta \rightarrow \pi$,
\begin{eqnarray}
  u & \rightarrow & 0, \\
  \dot{r} & \rightarrow & -v_\infty.
\end{eqnarray}
Noting that,
\begin{equation}
    \dot{r} = -h \frac{{\rm d}u}{{\rm d}\theta},
\end{equation}
the solution is,
\begin{equation}
    u = \frac{GM}{h^2} \left[ 1 + \cos \theta \right] - \frac{v_\infty}{h} \sin \theta.
\end{equation}
Finally, setting $\theta = 0$ we find that the streamlines with impact parameter $b$ collide along the accretion line at a distance,
\begin{equation}
    r = \frac{b^2 v_\infty^2}{2 GM},
\end{equation}
downstream behind the point mass. Post-collision, the azimuthal velocity goes to zero, leaving only a radial component,
\begin{equation}
    \dot{r} = v_\infty.
\end{equation}
Note that the radial velocity is independent of the distance of the collision point from the point mass.

Using the above results, we assess whether the post-collision gas is bound to the accretor. The condition for being bound is that,
\begin{equation}
    \frac{1}{2} \dot{r}^2 - \frac{GM}{r} = 
    \frac{1}{2} v_\infty^2 - 2 \left( \frac{GM}{b v_\infty} \right)^2 < 0.
\end{equation}
The critical streamline that yields just-bound gas has an impact parameter,
\begin{equation}
    b_{\rm crit} = \frac{2 GM}{v_\infty^2}.
\end{equation}
All gas flowing into the accretion cylinder with $b < b_{\rm crit}$ will be accreted, so the accretion rate is,
\begin{equation}
    \dot{M} = \pi b_{\rm crit}^2 v_\infty \rho_\infty = 
    4 \pi \frac{(GM)^2}{v_\infty^3} \rho_\infty.
\end{equation}
Up to a numerical factor that depends on the adiabatic index, the Bondi-Hoyle-Lyttelton accretion rate is just the Bondi accretion rate, with the relative velocity of the moving accretor replacing the sound speed that enters into the spherical formula.

Numerical simulations of Bondi-Hoyle-Lyttleton accretion, which is a computationally hard problem, go back a long way \citep{livio86,taam88,matsuda91,ruffert99}. The analytic estimate for the accretion rate is found to be a reasonable estimate, but the flow morphology depends on the adiabatic index, features a bow shock, and is often unsteady. Going beyond the simplest version of the problem, the rates of mass and angular momentum accretion from media with density or velocity gradients, or stochastic inhomogeneities, are of interest. These more complex situations are hard to treat analytically \citep[for a discussion of the physical considerations, see][]{davies80}. Relatively recent works include those by \citet{krumholz06}, \citet{blondin12}, \citet{macleod15}, and \citet{xu19}.

\section*{Acknowledgements}
I'd like to thank my many collaborators, most of all Jim Pringle and Mitch Begelman, for educating me about accretion disks. NASA, the National Science Foundation, and the Simons Foundation have provided research support. Finally, I wish {\em SuperMongo} a happy retirement, and acknowledge the use of {\em matplotlib} \citep{hunter07}, {\em NumPy} \citep{harris20} and {\em Jupyter} \citep{kluyver16}.

\bigskip
\input{accretion.bbl}

\end{document}

%% file: accretion.bbl
%